\documentclass[draft]{agujournal2019}

\usepackage{url} 
\usepackage{lineno}
\usepackage{multirow}
\usepackage{rotating}
\usepackage{booktabs}
\usepackage{graphicx}
\usepackage{caption}
\usepackage{subcaption}
\usepackage{amsmath}
\usepackage{longtable}
\usepackage{rotating,caption,booktabs}
\usepackage{xcolor}

\usepackage{soul}

\journalname{Space Weather}

\begin{document}


\title{Investigating potential benefits of future sub-L1 missions with STEREO-A}


\authors{E. Weiler\affil{1,2}, E.~E.~Davies\affil{1}, C.~M\"ostl\affil{1}, N.~Lugaz\affil{3}, A.~Veronig\affil{2,4}, R.~L.~Bailey\affil{5}, M.~A.~Reiss\affil{6, 7}}

\affiliation{1}{Austrian Space Weather Office, GeoSphere Austria, 
Reininghausstrasse 3, 8020 Graz, Austria}
\affiliation{2}{Institute of Physics, University of Graz, Universit\"atsplatz 5, 8010 Graz, Austria}
\affiliation{3}{Space Science Center and Department of Physics and Astronomy, University of New Hampshire, 8 College Rd, Durham, NH 03824, USA}
\affiliation{4}{Kanzelh\"ohe Observatory for Solar and Environmental Research, University of Graz, Kanzelh\"ohe 19, Treffen am Ossiacher See, 9521, Austria}
\affiliation{5}{Conrad Observatory, GeoSphere Austria, Hohe Warte 38, 1190 Vienna, Austria}
\affiliation{6}{Community Coordinated Modeling Center, NASA Goddard Space Flight Center, 8800 Greenbelt Rd., Greenbelt, MD 20771, USA}
\affiliation{7}{Universities Space Research Association, Washington, DC, USA}


\correspondingauthor{Eva Weiler}{eva.weiler@geosphere.at}

\begin{keypoints}
\item From 2022--2024 STEREO-A acted as a sub-L1 monitor 0.01-0.06~au upstream and within $\pm 15^\circ$ longitude of Earth.

\item A sub-L1 monitor must be placed closer to the Sun than 0.95~au, as 25\% of CMEs measured at both STEREO-A and L1 show negative lead times.

\item 26 of 47 observed geomagnetic storms are correctly identified from STEREO-A data. Intense events are especially well identified.
\end{keypoints}

\begin{abstract}
We present the first statistical study of geomagnetic storm forecasting using in situ data from the STEREO-A spacecraft as a sub-L1 monitor. Between November 2022 and June 2024, STEREO-A crossed the Sun--Earth line, covering longitudinal and radial separations of $\pm15^\circ$ from the Sun--Earth line and 0.01–0.06~au from Earth. This passage provides a unique opportunity to assess future sub-L1 mission concepts by ESA, such as HENON and SHIELD. We identify 32 coronal mass ejections (CMEs) observed by both STEREO-A and L1 spacecraft. Eight of these 32 CME events are first detected at L1, indicating that radial spacecraft separations of up to $\sim0.05$~au do not always yield lead time advantages. Furthermore, we find greater (smaller) gains in lead time when STEREO-A is east (west) of the Sun–Earth line. We develop a baseline methodology for the use of future sub-L1 in situ data to enable time-shifting and real-time modeling of the geomagnetic SYM-H index. This is run continuously over the entire time period, therefore modeling the geomagnetic response of all solar wind structures. Our methodology is empirically motivated and should be considered a first approach in addressing the use of sub-L1 data. Following this methodology, 26 of 47 observed geomagnetic storms are correctly identified from STEREO-A data. Intense events (82\%, SYM-H~$<-100$~nT) are well detected, most of which are also associated with an identified CME event. Most SYM-H minima are predicted later (72\%) and stronger (58\%) than those observed due to biases introduced by our methodology.

\end{abstract}

\section*{Plain Language Summary}
Solar storms can cause disturbances in the Earth's magnetic field, known as geomagnetic storms. If strong enough, these geomagnetic storms can disrupt technological systems on Earth. To enable earlier warnings of solar storms, future missions by ESA, such as HENON and SHIELD, aim to place spacecraft closer to the Sun than the current monitoring L1 point, which gives lead times of only ten to 60 minutes. We use measurements from the STEREO-A spacecraft as it passed in front of the Earth, five times closer to the Sun than the Earth--L1 distance, between November 2022 and June 2024 to test the potential of these future missions. We find that solar storms are not always detected earlier at STEREO-A, meaning that future missions must be placed closer to the Sun than 0.95~au. We develop an empirically based methodology, using STEREO-A measurements for real-time geomagnetic storm prediction. Using this approach, more than half of the observed geomagnetic storms are correctly identified using STEREO-A data. Strong storms are detected particularly well, although their timing and strength are often overestimated. Our results highlight important trade-offs between forecast accuracy and lead time that must be considered in the design of future space weather missions.


\section{Introduction}

One persistent challenge in space weather forecasting is known as the $B_z$ problem \cite<e.g.>[]{kilpua2019review,vourlidas2019review}. This refers to our current inability to predict the north-south component (i.e., $B_z$) of the interplanetary magnetic field (IMF) more than an hour in advance, not least because the magnetic field orientations of large-scale solar wind structures, such as coronal mass ejections (CMEs), cannot be reliably determined from remote observations as of yet. The orientation of $B_z$ plays a critical role in governing the amount of energy and mass transferred from the solar wind into Earth’s magnetosphere. When $B_z$ is directed southward, it can reconnect with the oppositely directed geomagnetic field at the dayside magnetopause, enabling enhanced solar wind–magnetosphere coupling. This enhanced coupling leads to energy and plasma injection into the magnetosphere, driving disturbances known as geomagnetic storms \cite<e.g.>[and references therein]{gonzalez1994}.

To probe the IMF directly and hence estimate any pending geomagnetic disturbances, we use in situ observations from spacecraft positioned at the first Sun--Earth Lagrangian point (L1), located about 0.01~au upstream (sunward) of Earth. Some of these spacecraft, namely the Advanced Composition Explorer \cite<ACE;>[]{stone1998ace} and the NOAA-operated Deep Space Climate Observatory \cite<DSCOVR;>[]{dscovr}, transmit IMF and plasma data in real time, with a delay of only a few minutes. Depending on the speed of the measured structure in the solar wind, measurements taken at L1 give us a lead time of 10 to 60 minutes until the structure reaches Earth. As 10 to 60~minutes is not much time to take any preventive measures, a solution could be to position spacecraft even closer to the Sun than L1, referred to as sub-L1 monitors \cite{lindsay1999dst}. The idea of using spacecraft ahead of L1 to improve the forecast lead time has been around for some time and was addressed in previous (review) papers \cite<e.g.,>{vourlidas2019review,morley2020,palmerio2025_commentary_weiler}. 

Furthermore, although remote imaging observations combined with forward modeling can estimate CME arrivals several days in advance, their predictive accuracy remains limited. By comparing different operational CME arrival models, whose predictions are routinely posted to the CCMC scoreboard, previous studies \cite{kay2024CMEarrivals,riley2018forecasting} show that uncertainties in input parameters and inherent model assumptions typically prevent mean absolute arrival time errors from improving beyond $\sim10$~hours for any model. In addition, even advanced (magneto-)hydrodynamic models such as ENLIL \cite{odstrcil2003enlil}, EUHFORIA \cite{pomoell2018euhforia} or ICARUS \cite{verbeke2022,Baratashvili2022} often struggle to accurately reproduce in situ profiles later measured at L1. The limitations arise because the magnetic structure of a CME at eruption is not directly known and its evolution through the heliosphere toward 1~au is subject to many uncertainties \cite{vourlidas2019review}. Incorporating upstream in situ observations into these forecasting frameworks could help constrain model parameters, potentially improving both arrival time and storm intensity predictions \cite{davies2025realtime}. Direct measurements upstream of L1 may be particularly important for complex events that interact with other solar wind structures \cite{davies2025realtime,weiler2025,liu2024may,laker_2024}.

Keeping sub-L1 monitors in position is not a trivial task due their non-Keplerian nature. The most promising solution so far is the use of spacecraft on stable distant retrograde orbits (DROs), which are Earth-like orbits that differ slightly in their eccentricity from that of the Earth \cite{henon1969,stcyr2000diamond}. Spacecraft in such orbits periodically shift ahead of and behind Earth as they orbit the Sun. By deploying several spacecraft on DROs, it would be possible to ensure that at least one spacecraft is always positioned sunward of L1. From an Earth point of view, these probes would orbit Earth at distances of around 0.1--0.2~au, i.e., roughly 10--20 times closer to the Sun than Earth's distance to L1, and therefore potentially extending lead times for space weather events by 10--20 hours for slow solar wind transients ($\sim 400$~km\,s$^{-1}$) and still several hours for the fastest ones ($>1000$~km\,s$^{-1}$). The ideal orbital distance, and window of longitudinal separation, of a sub-L1 monitor to Earth has not yet been determined, as there is a lack of multi-spacecraft observations at medium separations \cite<1$^\circ$ to 20$^\circ$ in heliospheric longitude,>[]{lugaz2025subL1}. However, recent studies \cite<e.g.>[]{good2016interplanetary,lugaz2024MEwidth,banu2025} have shown that the longitudinal separation between two measuring spacecraft should be less than $15^\circ$ to ensure that both spacecraft measure the same solar wind structure. 
Such a mission concept with four spacecraft is known in the USA as MIIST \cite<Mission to Investigate Interplanetary Space Transients,>[]{Lugaz2024mission}. In Europe, a one-spacecraft CubeSat mission called HENON \cite<Heliospheric Pioneer for Solar and Interplanetary Threats Defence,>[]{Cicalò2025Henon} is planned to be launched on a DRO trajectory at the end of 2026. As recently announced, ESA is also studying a multi-spacecraft sub-L1 DRO mission called SHIELD. This mission just completed the Current Design Facility study and moved on to the next phase of the mission development as of February 2026.  

NASA's twin Solar TErrestrial RElations Observatory \cite<STEREO;>{kaiser2008stereo} was launched in 2006, with two nearly identical observatories being sent into space. One of these, STEREO-ahead (STEREO-A), is slightly ahead of Earth in its orbit, the other, STEREO-behind (STEREO-B), is trailing behind. Unfortunately, communication with STEREO-B was lost in 2014 but STEREO-A continues to collect data to date. In August 2023, STEREO-A crossed the Sun--Earth line for the first time since its launch, which coincided with the solar maximum of the current solar cycle (Solar Cycle 25). From November 2022 to June 2024, STEREO-A was longitudinally separated within $\pm 15^{\circ}$ of the Sun--Earth line, passing approximately 0.01$-$0.05~au ahead of Earth. These longitudinal and radial separations closely match those of future DRO missions, allowing us to utilize this fortunate passage of STEREO-A to quantify the feasibility of sub-L1 monitors. Therefore, one aim of this study is to explore mission requirements for potential operational sub-L1 platforms. 

With STEREO-A acting as a future sub-L1 monitor, we can also investigate how this data can best be used to predict geomagnetic indices. Since solar wind structures evolve from a sub-L1 position to Earth, we can, for example, investigate how to map sub-L1 data to Earth to reproduce observed geomagnetic responses. We approach this specific issue from an empirical perspective by analyzing differences in arrival times and also physical variations of CMEs observed by both STEREO-A and at L1. We focus on CMEs because, on the one hand, they are considered the most geoeffective structures \cite<e.g.>[]{zhang2007,wang2003,gosling1991geo} and, on the other hand, they are relatively easy to identify and link between the two datasets due to the distinct characteristics such as the appearance of shocks, sheath regions and magnetic field rotations \cite{zurbuchen2006situ,kilpua2017coronal}. A statistical understanding of the differences between the CMEs measured at STEREO-A and L1 allows us to create representative ensembles covering slightly different solar wind conditions, resulting in a range of realistic geomagnetic responses. Even though we focus on the effects of CMEs, we apply our pipeline to the whole STEREO-A data set, therefore modeling the geomagnetic effects of every solar wind structure. In this paper, we therefore do not only address the utility of future sub-L1 monitors, but establish our methodology in such a way that it is directly applicable to data from future missions, such as HENON and SHIELD. The goal is to provide a baseline methodology that continuously forecasts the SYM-H index in real time using a spacecraft at a sub-L1 position. 

In the following, we first describe the data we use for our analysis (Section~\ref{sec:data}). We then discuss the differences of CMEs measured by STEREO-A and L1, which are used to establish our methodology in Section~\ref{sec:differences}. In Section~\ref{sec:methods} we introduce our techniques, explaining how L1 and STEREO-A data are shifted to Earth (\ref{sec:time_shift}), and how the geomagnetic indices are modeled (\ref{sec:symh_modeling}). Section~\ref{sec:analysis} is divided into two parts, first dealing with our results from a continuous, statistical perspective on geomagnetic storms (observed and modeled, Section~\ref{sec:continuous_analysis}) and then from a more detailed perspective, considering individual geomagnetic storms in Section~\ref{sec:event_analysis}. Our results are discussed and summarized in Section~\ref{sec:discussion}. A summary with explicit recommendations for future sub-L1 missions can be found in Section~\ref{sec:summary}.


\section{Data}\label{sec:data}
For our analysis, we use the L1 Real Time Solar Wind (RTSW) data provided by NOAA \cite[available at \url{https://www.spaceweather.gov/products/real-time-solar-wind}]{zwickl1998noaartsw}. This data product combines measurements from two spacecraft, namely the Advanced Composition Explorer \cite<ACE;>[]{stone1998ace} and the Deep Space Climate Observatory \cite<DSCOVR;>[]{dscovr}. These spacecraft are positioned around L1 
and transmit magnetic field and plasma data of the solar wind in real time, with a delay of only a few minutes. For DSCOVR, the magnetometer (MAG) and the Faraday Cup (FC) are used for real-time measurements of the local magnetic field and solar wind plasma, respectively. Similarly, we use the Magnetic Field Experiment \cite<MAG;>[]{smith1998ACEmag} and the Solar Wind Electron Proton Alpha Monitor \cite<SWEPAM;>[]{mccomas1998solar} onboard ACE whenever DSCOVR is not observing. We remove spikes in the L1 RTSW density data that are greater than 100~cm$^{-3}$. Furthermore, we calculate the gradient of the density and single out peaks that exceed a prominence threshold of 10~cm$^{-3}$~min$^{-1}$. For readability, we are going to refer to the position of ACE and DSCOVR simply as L1, even though neither the ACE nor the DSCOVR positions exactly correspond to the actual position of L1, but rather orbit around the L1 point with a major and minor axis of about 150,000~km and 75,000~km, respectively \cite{ruedisser2025ARCANE,lotoaniu2022,stone1998advanced}. 

\begin{figure}[ht]
    \centering
    \includegraphics[width=\linewidth]{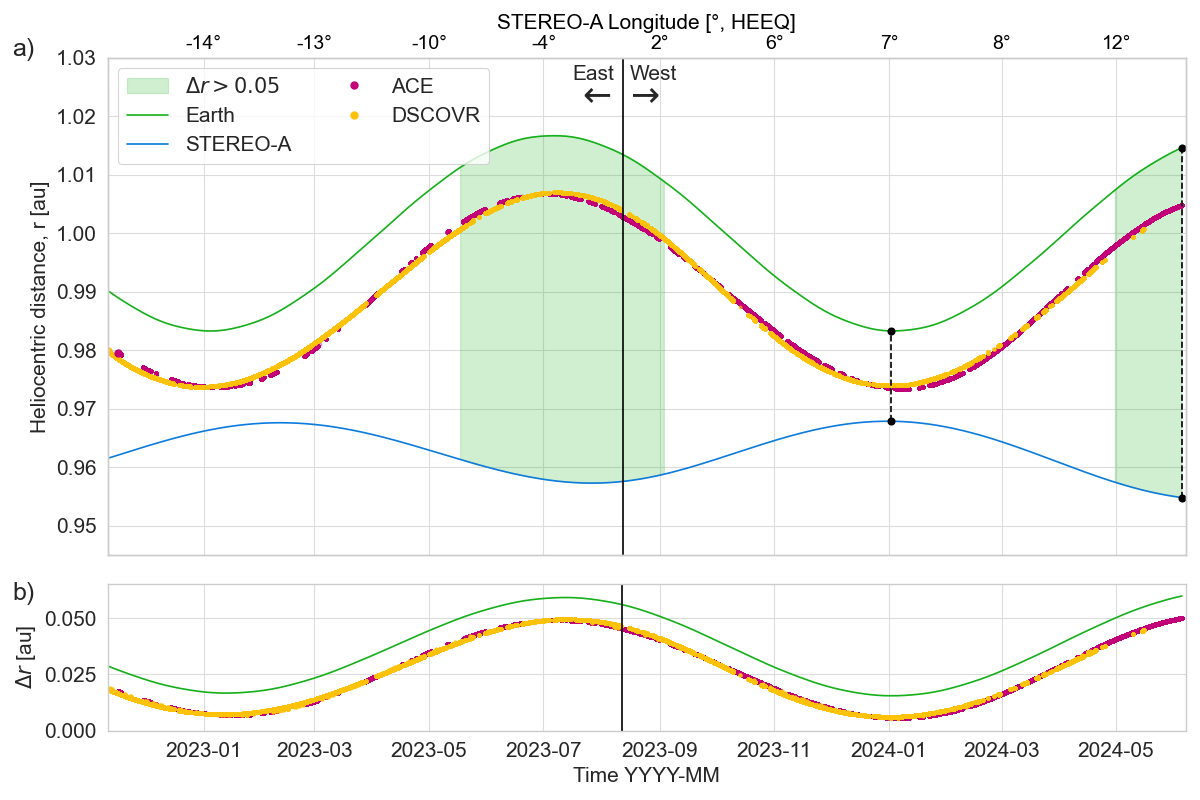}
    \caption{Differences in heliocentric distance for the different spacecraft. (a) Heliocentric distance, $r$, of STEREO-A (blue), ACE (magenta), DSCOVR (orange) and Earth (green). The solid black vertical line indicates the Sun--Earth line crossing of STEREO-A. Periods where STEREO-A and Earth are separated by $\Delta r>0.05$~au are shaded in green. Minimum and maximum radial separations between STEREO-A and Earth are indicated as dashed vertical black lines. (b) Differences in heliocentric distance $\Delta r$, for STEREO-A and Earth (green) and STEREO-A and ACE/DSCOVR (magenta/orange) in au.}
    \label{fig:l1_sta_pos}
\end{figure}

From November 2022 to June 2024, STEREO-A passed in front of L1 within $\pm 15^{\circ}$ longitude (in Heliospheric Earth Equatorial, HEEQ, coordinates), acting as a sub-L1 monitor. The heliocentric distances of STEREO-A, ACE, DSCOVR and Earth are plotted in Figure~\ref{fig:l1_sta_pos}a. The solid black vertical line indicates the time when STEREO-A crossed the Sun--Earth line on 2023 August 12, with STEREO-A being eastward of Earth before and westward after that date. The green shaded areas indicate regions where the radial separation between Earth and STEREO-A, $\Delta r$, is greater than 0.05~au, which corresponds to a spacecraft separation between STEREO-A and L1 of $>0.04$~au. The shaded green region from 2023 May 17 to 2023 September 3 fits the orbital parameters of future mission concepts best and therefore defines our target region, which will become more important later on in our analysis. The dashed black lines indicate the minimum and maximum radial separation between STEREO-A and Earth, with a minimum radial separation on 2024 January 2, coming as close as 0.015~au, and a maximum radial separation of 0.06~au on 2024 June 4.

For evaluating the prediction performance of a potential sub-L1 monitor, we use beacon (produced in real-time) magnetic field data measured by the In situ Measurements of Particles And CME Transients \cite<IMPACT;>{luhmann2008impact} instrument as well as science plasma data measured by the Plasma and Suprathermal Ion Composition \cite<PLASTIC;>{galvin2008plastic} instrument onboard STEREO-A. STEREO-A PLASTIC does not provide plasma parameters in real-time, therefore, we use level 2 science quality plasma data to obtain the speed and density information necessary for establishing our baseline methodology, including a time-shifting procedure and modeling the SYM-H indices. Both the STEREO-A beacon magnetic field as well as L1 magnetic field data are given in Geocentric Solar Magnetospheric (GSM) coordinates.

\section{Differences between STEREO-A and L1 Observations}\label{sec:differences}

Before explaining our methodology for continuously forecasting the SYM-H index in real time from sub-L1 as well as L1 data, it is necessary to discuss differences in the observations by STEREO-A and ACE/DSCOVR. To assess these differences, we focus on CMEs that are observed by both STEREO-A and spacecraft at L1.

CMEs are large expulsions of plasma and magnetic fields from the solar corona. They consist of a magnetic obstacle (MO, sometimes also referred to as magnetic ejecta), often following a shock, if the CME is traveling faster than the local magnetosonic speed, and a sheath region. The MO can sometimes be described as a magnetic flux rope, where the magnetic field lines are wound helically around a central axis. Observationally, this corresponds to a smooth rotation of the magnetic field vector in in situ data. This organized magnetic structure can lead to prolonged and intense southward $B_z$ components, which is why CMEs are known to be the primary drivers of the most severe geomagnetic storms \cite<e.g.>[]{zhang2007,echer2008}. Other solar wind structures such as high-speed streams (HSSs) and their associated stream interaction regions (SIRs) can also lead to significant, albeit typically less intense, geomagnetic disturbances.

In this work, CMEs are mainly identified by typical CME signatures \cite{zurbuchen2006situ,burlaga1981magnetic}, such as enhanced total magnetic field strength and the arrival of a shock or discontinuity followed by the MO. In addition to the magnetic field structure, other typical features of the MO include linearly declining speed profiles and lower proton densities and temperatures compared to the surrounding solar wind \cite{klein1982interplanetary}. The different CMEs are linked between the datasets via matching features such as the similar rotational behavior of the MOs. We also include events where an MO can only be clearly identified at one spacecraft but where the data from the other spacecraft displays similar features that do not fully meet the definition of a MO. In total, we have found 32 CME events from November 2022 to June 2024, some of which are individual CMEs, while others are formed of multiple interacting CMEs. Events involving interacting CMEs are treated as a single event. The HELIO4CAST ICMECAT \cite{moestl2026icmecat} lists 56 CMEs at Wind (L1) and 47 at STEREO-A over the same time period. In contrast to this work, interacting events are, as far as possible, separated into individual MOs in the ICMECAT, explaining the difference between the number of events in the catalog and this study. Nevertheless, it is clear that most of the CMEs that occur from November 2022 to June 2024 are measured by both STEREO-A and L1. For each of the 32 CME events, we have defined a start time, which is either the arrival of a shock or, if there is no shock, the start of the MO, and an end time. To set the start and end time at STEREO-A of CME event 26, we use magnetic field science data, due to data gaps in the beacon data. Table~\ref{tab:CME_events} lists all events and their corresponding start time at STEREO-A and L1. 14 of the identified CME events occur when STEREO-A is eastward of the Sun--Earth line, while we identify 18 events when it is westward. Some of these events have already been studied in \citeA{banu2025}, where the authors investigated the intrinsic variations of 20 CMEs observed by both STEREO-A and L1 from January 2022 to August 2023, when STEREO-A was approaching L1 from $35^\circ$ to $0^\circ$.

\subsection{Arrival Time}\label{sec:differences_arrival}

\begin{table}[htbp]
    \footnotesize
    \caption{Start times of the 32 CME events as measured by STEREO-A and spacecraft at L1. The lead time, $t_s$ is the time of arrival at L1 - time of arrival at STEREO-A. The last column indicates whether the observed CME event would have resulted in a gain ($\Delta t_s > 0$) or loss ($\Delta t_s < 0$) in lead time.}
    \centering
    
    \renewcommand{\arraystretch}{0.9}
    \begin{tabular}{cccc}
        \toprule
        CME event & \multicolumn{2}{c}{Start time [UTC]} & Lead time [h] \\
        \# & STEREO-A & L1 & Gain $(+)$, Loss $(-)$ \\
        \midrule
            1  &  2022-12-18T10:48Z  &  2022-12-18T19:51Z  &  +9.0 \\
            2  &  2023-01-04T12:34Z  &  2023-01-03T20:58Z  &  -15.6 \\
            3  &  2023-01-17T14:36Z  &  2023-01-17T21:03Z  &  +6.4 \\
            4  &  2023-03-15T01:16Z  &  2023-03-15T03:47Z  &  +2.5 \\
            5  &  2023-03-22T21:46Z  &  2023-03-23T07:31Z  &  +9.8 \\
            6  &  2023-04-20T04:30Z  &  2023-04-19T08:15Z  &  -20.2 \\
            7  &  2023-04-23T14:29Z  &  2023-04-23T16:58Z  &  +2.5 \\
            8  &  2023-05-09T19:06Z  &  2023-05-09T22:06Z  &  +3.0 \\
            9  &  2023-05-12T01:06Z  &  2023-05-12T05:48Z  &  +4.7 \\
            10 &  2023-07-14T08:36Z  &  2023-07-14T15:24Z  &  +6.8 \\
            11  &  2023-07-16T12:50Z  &  2023-07-16T18:35Z  & +5.8 \\
            12  &  2023-07-25T18:51Z  &  2023-07-25T21:52Z  &  +3.0 \\
            13  &  2023-08-01T06:08Z  &  2023-08-01T11:06Z  &  +5.0 \\
            14  &  2023-08-04T02:09Z  &  2023-08-04T06:52Z  &  +4.7 \\
            15  &  2023-09-12T06:52Z  &  2023-09-12T10:24Z  &  +3.5 \\
            16  &  2023-09-18T09:04Z  &  2023-09-18T12:55Z  &  +3.8 \\
            17  &  2023-09-24T17:35Z  &  2023-09-24T19:53Z  &  +2.3 \\
            18  &  2023-10-20T08:25Z  &  2023-10-20T08:04Z  &  -0.4 \\
            19  &  2023-11-05T03:39Z  &  2023-11-05T08:06Z  &  +4.4 \\
            20  &  2023-11-12T03:36Z  &  2023-11-12T05:30Z  &  +1.9 \\
            21  &  2023-11-30T20:50Z  &  2023-11-30T23:34Z  &  +2.7 \\
            22  &  2023-12-15T12:12Z  &  2023-12-15T10:56Z  &  -1.3 \\
            23  &  2023-12-29T10:10Z  &  2023-12-29T11:37Z  &  +1.4 \\
            24  &  2024-01-03T15:37Z  &  2024-01-03T14:20Z  &  -1.3 \\
            25  &  2024-02-11T00:52Z  &  2024-02-11T01:19Z  &  +0.4 \\
            26  &  2024-03-03T06:00Z  &  2024-03-03T08:47Z  &  +2.8 \\
            27  &  2024-03-21T04:16Z  &  2024-03-21T02:24Z  &  -1.9 \\
            28  &  2024-03-24T14:23Z  &  2024-03-24T14:07Z  &  -0.3 \\
            29  &  2024-04-19T09:30Z  &  2024-04-19T04:53Z  &  -4.6 \\
            30  &  2024-04-30T00:29Z  &  2024-04-30T22:32Z  &  +22.0 \\
            31  &  2024-05-02T11:26Z  &  2024-05-02T13:16Z  &  +1.8 \\
            32  &  2024-05-10T14:03Z  &  2024-05-10T16:38Z  &  +2.6 \\
        \bottomrule
    \end{tabular}
    \label{tab:CME_events}
\end{table}

\begin{figure}[htbp]
    \centering
    \includegraphics[width=\linewidth]{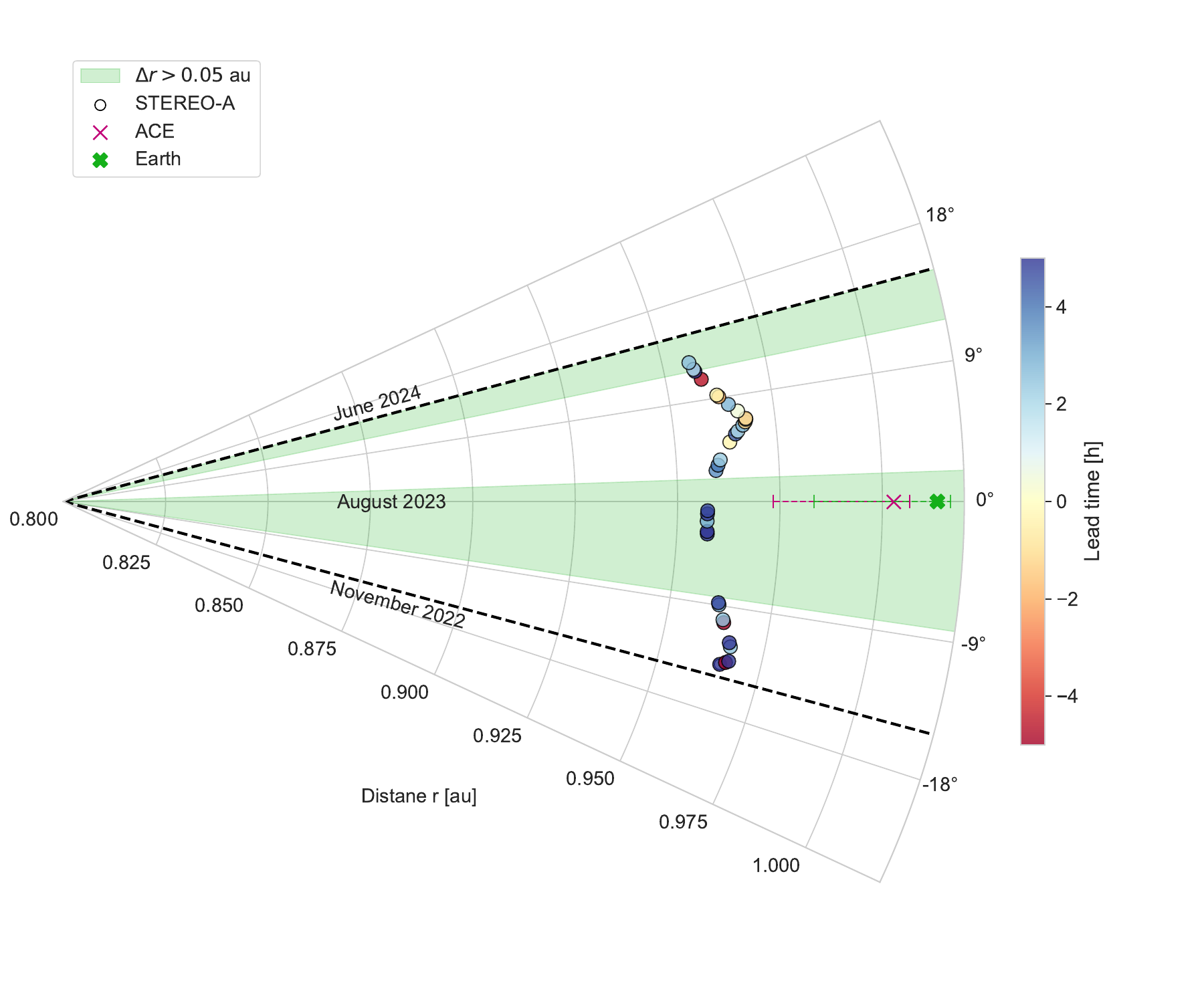}
    \caption{Visualization of the 32 CME events observed by both STEREO-A and L1. STEREO-A (color-coded), ACE (magenta), and Earth (green) positions are plotted with the coordinates given in Heliospheric Earth Equatorial (HEEQ) coordinates. The colored horizontal dashed lines indicate the heliospheric distance range for spacecraft at L1 (magenta) and Earth (green) from November 2022 to June 2024. At STEREO-A positions, the difference in event start time for the identified 32 CME events are color-coded corresponding to the gain/loss in lead time. The shaded green areas indicate regions where STEREO-A and Earth are radially separated by more than 0.05~au.}
    \label{fig:diff_arrivals}
\end{figure}

First, we look at the differences in arrival time at STEREO-A and L1. Whenever the start of an event is observed at STEREO-A first, we declare this as a gain in lead time ($\Delta t_{s} > 0$), whereas a loss in lead time ($\Delta t_s < 0$) would indicate that the CME was first observed at L1 and later at STEREO-A. For obvious reasons, the latter would not be advantageous for space weather forecasting. Figure~\ref{fig:diff_arrivals} presents the differences in arrival time ($\Delta t_s$) between STEREO-A and L1 for the 32 CME events listed in Table \ref{tab:CME_events}. Figure~\ref{fig:diff_arrivals} shows the spacecraft positions of STEREO-A (circles) in  HEEQ coordinates at the observed time of the events with the losses (negative values) and gains (positive values) in lead time color-coded by the scale on the right of the figure. For simplicity, we only plot one representative position of ACE (magenta cross) and Earth (green cross) on 2023 August 12, the day when STEREO-A crossed the Sun--Earth line. The dashed lines in magenta and green indicating the corresponding distance ranges for ACE and Earth, respectively, from November 2022 to June 2024.

\begin{figure}[htbp]
    \centering
    \includegraphics[width=\linewidth]{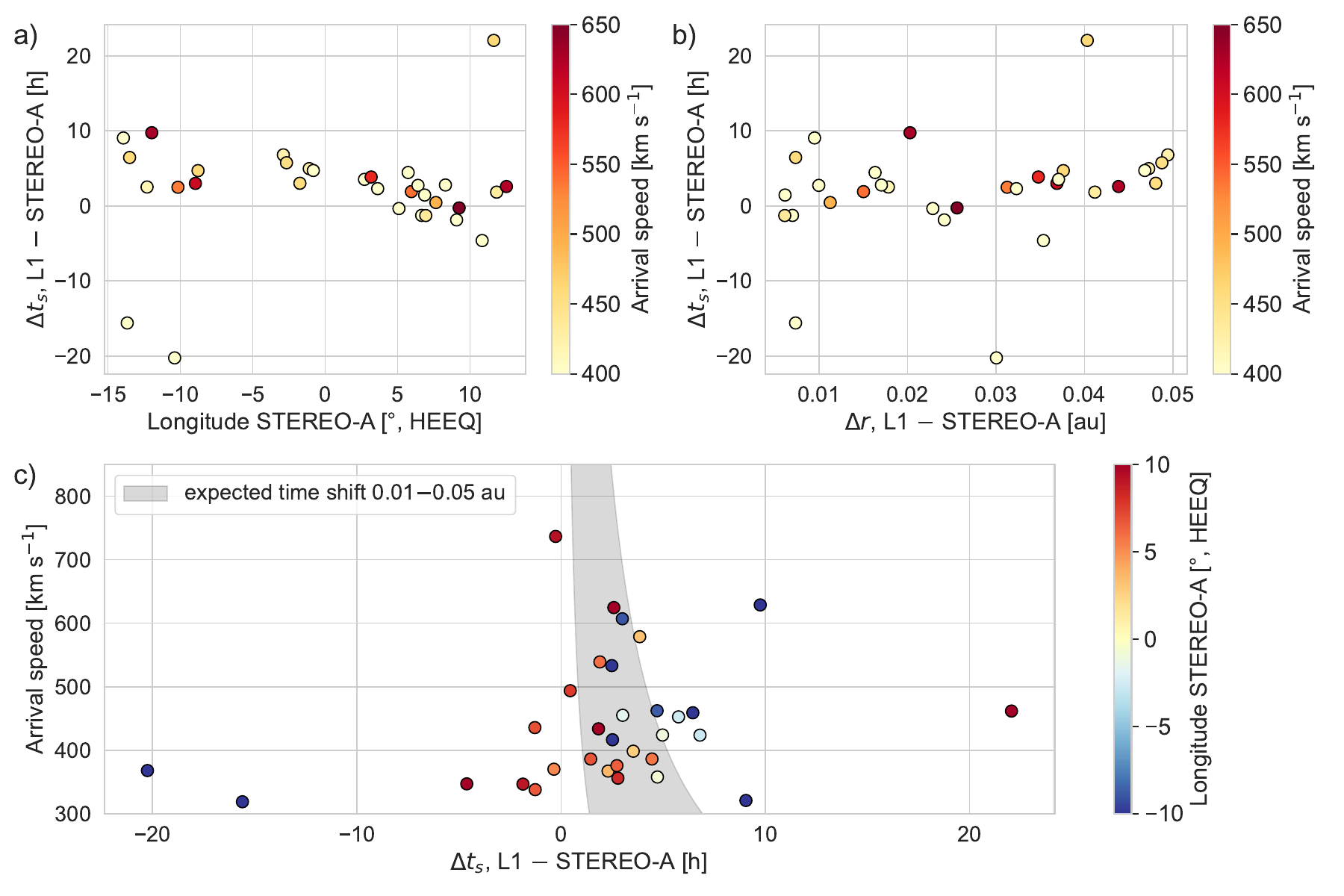}
    \caption{Differences in arrival time ($\Delta t_{s}$) between STEREO-A and L1 for the 32 CME events listed in Table~\ref{tab:CME_events}. (a) $\Delta t_s$ as a function of STEREO-A's longitude, given in HEEQ. The colorbar indicates the arrival speed of the CME events at STEREO-A. (b) $\Delta t_s$ as a function of radial separation from STEREO-A to L1. As in panel a, the colorbar indicates the arrival speed of the CME events at STEREO-A. (c) $\Delta t_s$ is plotted against the arrival speed of the 32 CME events. The gray area represents the expected time shift for spacecraft separations of 0.01--0.05~au, assuming a purely radially propagating CME front. The colors indicate the longitude of STEREO-A in HEEQ coordinates.}
    \label{fig:diff_arrivals1}
\end{figure}

Interestingly, 8 of the 32 (25\%) CME events identified show a loss in lead time ($\Delta t_s < 0$), meaning that the shock was first observed at L1 and then at STEREO-A. In addition, there appears to be a dependence on longitude, with events east of the Sun--Earth line showing a greater gain in lead time than events west of it. This is also illustrated in Figure~\ref{fig:diff_arrivals1}a, where $\Delta t_s$ is plotted against the longitude of STEREO-A. In contrast, the lead time trend seems to be less apparent when considering the radial separation of STEREO-A and L1 ($\Delta r$), plotted in Figure~\ref{fig:diff_arrivals1}b. Quantitatively, ignoring the three CME events with $\lvert \Delta t_s \rvert > 10$~h, which deviate the most from the expected time shift, the Pearson correlation coefficient ($p$-value) between $\Delta t_s$ and longitude is 0.70 ($2.72\times10^{-5}$), while it is 0.17 (0.37) for $\Delta t_s$ and $\Delta r$. Previous studies \cite<e.g.>{wang2002,zhang2003} suggest that CMEs are influenced by the Parker spiral as they propagate toward Earth, which could be a physical explanation for this asymmetry. It could, however, also be a consequence of STEREO-A's trajectory, which is not symmetric with respect to the Sun--Earth line (see also Figure~\ref{fig:l1_sta_pos}). We also note that the source regions have not been investigated in this work, which could also contribute to an apparent asymmetry. For the two outliers eastward of the Sun--Earth line in Figure~\ref{fig:diff_arrivals1}a with $\lvert \Delta t_s \rvert > 10$~hours we found corresponding entries in the CCMC Space Weather Database Of Notifications, Knowledge, Information (DONKI, \url{https://kauai.ccmc.gsfc.nasa.gov/DONKI/}). Both show a much more extended structure with higher magnetic field strengths at L1 than at STEREO-A (see Table~\ref{tab:CME_events}, events 2 and 6), hinting at a propagation direction directed more toward L1. However, only event 6 with a propagation direction of $-14^\circ$ agrees with that statement whereas Event 2, with $\mathrm{lon}=+14^\circ$, propagates more toward the west. Unfortunately, there is no DONKI entry for the outlier on the west side (event 30), which has a longer duration at STEREO-A but a weaker total magnetic field strength. Without knowledge of the source region, the propagation direction cannot be determined with certainty for this event.

Considering Figure~\ref{fig:diff_arrivals1}b, we note that for spacecraft separations of $\Delta r > 0.037$~au, the lead time is always positive. This can be also seen Figure~\ref{fig:diff_arrivals}, where all events that occurred in the highlighted green regions, indicating a separation of STEREO-A and Earth of more than 0.05~au, show a gain in lead time, five of which occurred in our defined target region (events 10--14 in Table~\ref{tab:CME_events}), and three in the second shaded area (events 30--32 in Table~\ref{tab:CME_events}). In Figure~\ref{fig:diff_arrivals1}c, we plot the differences in arrival time between STEREO-A and L1 against the arrival speed at STEREO-A, which is calculated by averaging the speed in a 30~minute window after arrival, defined by either a shock arrival or the start of the MO in the in situ data. The gray area represents the expected time shift for spacecraft separations of 0.01 to 0.05~au, assuming that the fronts of the CMEs propagate radially. Beyond the two large negative outliers both occurring for slow events, we find that the lead time does not seem to be strongly dependent on the arrival speed. The color coding corresponds to the longitudinal separation between STEREO-A and Earth and again shows the trend in lead time, with reddish colors indicating CME events on the west side, usually associated with smaller $\Delta t_s$ values.

\subsection{Physical Properties}\label{sec:differences_parameters}

As a CME propagates from one location in the heliosphere to another, its inherent properties, such as magnetic field strength, speed, density, and temperature change. This happens because the CME expands as a response to internal and external pressure balance, deforms, or gets deflected due to the surrounding ambient solar wind, and possibly interacts with other large-scale structures \cite<e.g.,>{manchester2017physical,alhaddad2025}. The identified 32 events, observed by both STEREO-A, and spacecraft at L1, allow us to not only investigate differences in arrival time but also to study intrinsic variations from one location to the other.

\begin{figure}[htbp]
    \centering
    \includegraphics[width=\linewidth]{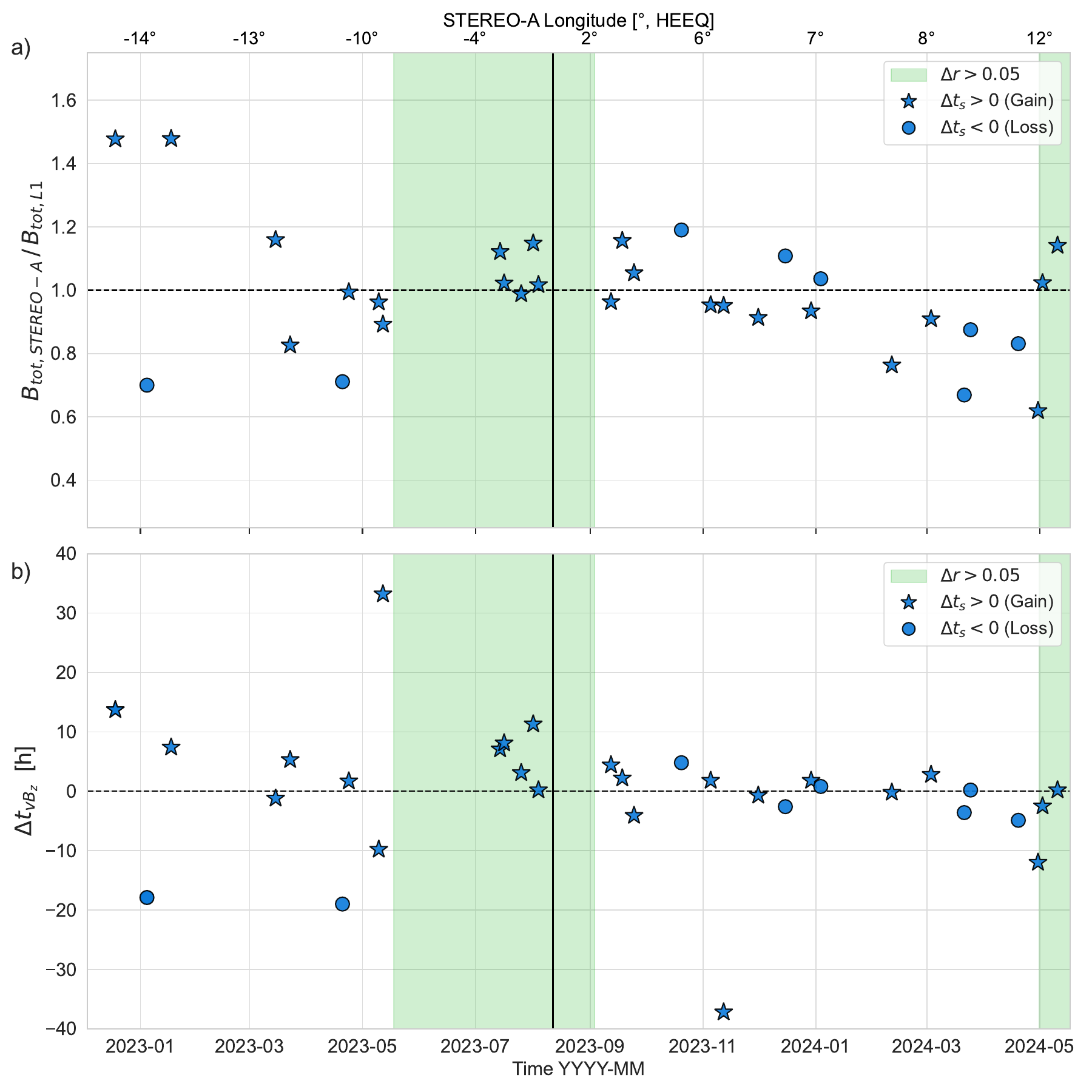}
    \caption{Differences in properties between STEREO-A and L1 for the 32 CME events. (a) Relative change in mean magnetic field strength $B_{tot}$ from STEREO-A to L1 plotted against the start time of the event at STEREO-A. Data points marked as stars indicate a gain in lead time ($\Delta t_s > 0$), while circles indicate a loss in lead time ($\Delta t_s < 0$), where the lead time is again defined via the shock/start of a MO as in Figure~\ref{fig:diff_arrivals} and Figure~\ref{fig:diff_arrivals1}. (b) Time difference in measured minimum $vB_z, \Delta t_{vB_z}$, between STEREO-A and L1. The symbols indicate again, whether the start of the event was measured first at STEREO-A (stars) or L1 (circles).}
    \label{fig:diff_min_bv}
\end{figure}

To do this, we first calculate the mean magnetic field strength for all 32 CME events using STEREO-A and L1 data. The mean value is calculated using the entire time range of an event, both single CMEs and interacting ones from the defined start (either the arrival of a shock or the start of the MO, as explained above) to the defined end. We note that interacting events can consist of more than one MO. We define one start and end time for these events also, averaging the magnetic field strength over the whole complex structure or complex ejecta \cite<i.e., MOs that cannot be differentiated anymore due to merging processes, see>{burlaga2002}. We then derive the relative change in mean $B_{tot}$ ($B_{tot, \, STEREO-A} / B_{tot, \, L1}$) from STEREO-A to L1, shown in Figure~\ref{fig:diff_min_bv}a. On average, the mean magnetic field strength changes by $\sim16\%$ from STEREO-A to L1, with the mean difference and mean absolute difference being 1.8~nT and -0.17~nT, respectively. Interestingly, 18 events have a higher mean $B_{tot}$ at L1 than at STEREO-A, which cannot be explained by statistical global expansion laws $B \propto r^{-\alpha}, \, \alpha=-1.66$ \cite{davies2021catalogue} since STEREO-A is always positioned sunward of L1. Furthermore, the relative change of mean $B_{tot}$ varies less closer to the Sun--Earth line, which is counterintuitive to the expected global expansion as the radial separation between STEREO-A and L1 is largest at this time (see Figure~\ref{fig:l1_sta_pos}). Specifically, the mean relative change in $B_{tot}$ for the five events in our target region is 6\%, while it is roughly thrice that (18\%) eastward and westward of the Sun--Earth line. While a relative change of 6\% at a radial distance of 0.05~au agrees with $\alpha=-1.3$ quite well with established global expansion laws, a relative change of 18\% requires much steeper slopes with $\alpha=-4.1$ for similar radial separations. This suggests that as STEREO-A deviates further from the Sun--Earth line, longitudinal effects become more relevant for the variations in $B_{tot}$ than global expansion. Furthermore, the change in mean $B_{tot}$ is also independent of whether the arrival of the event was first measured at STEREO-A (events marked as stars) or at L1 (events marked as circles), with 54\% (13/24) of the events that are first measured at STEREO-A having a higher mean $B_{tot}$ at L1, while the remaining 46\% have a higher mean $B_{tot}$ at STEREO-A. This also supports our finding that the longitudinal variations and possibly apex direction (not investigated) determine the magnetic field strength rather than global expansion, at least for spacecraft with radial separations of up to 0.05~au.

We perform a similar analysis of the mean speed and density parameters for each event (not shown). The speed differs on average by roughly 5\% from STEREO-A to L1, and the density by 28\%. The density varies quite a lot due to the lower quality of the near real-time plasma data of ACE/DSCOVR and the density values being smaller in general, leading to a large relative change. Both the mean speed and density show less dependence on longitude than the mean $B_{tot}$. The derived variations of $B_{tot}$, speed, and density will become important in Section~\ref{sec:symh_modeling} when we build our ensemble to model the geomagnetic indices from the shifted STEREO-A and L1 data.

Furthermore, we investigate the time difference of the product of the southward magnetic field component and speed, $vB_z $, between the two spacecraft ACE/DSCOVR and STEREO-A. The so-called dawn-to-dusk electric field, $vB_z$, is considered an important indicator when it comes to the production of geomagnetic storms and is highly correlated with the $Dst$ index \cite{gonzalez1998, pokharia2018}. 
The strength of a geomagnetic storm does not only depend on the minimum of $vB_z$, but also on negative $B_z$ values themselves---especially periods where $B_z$ stays negative for several hours---and on the dynamic pressure \cite{gopalswamy2022}. However, as investigated by \citeA{pokharia2018}, $vB_z$ is better correlated with the $Dst$ profile than $B_z$ and $v$ separately. Identifying the time difference of $min(vB_z)$ between STEREO-A and L1, $\Delta t\;min(vB_z)$ (hereafter $\Delta t_{vB_z}$), will therefore give us clues as to when we can expect the strongest geomagnetic response and how this might affect the modeled SYM-H indices, bearing in mind that not every CME identified is geoeffective.

Figure~\ref{fig:diff_min_bv}b shows $\Delta t_{vB_z}$ against the start time of the events at STEREO-A. On average, $min(vB_z)$ is measured 4.1 hours earlier at STEREO-A than at L1 when STEREO-A is east of the Sun--Earth line, and 2.6 hours later when STEREO-A is west of it. We can see that for nine events, the minimum of $vB_z$ at STEREO-A is measured after the corresponding measurement at L1, even though the shock is observed at STEREO-A first (stars below zero-line). In contrast, we can identify three events, for which the start of the event was first observed at L1, but the $vB_z$-minimum is nonetheless first observed at STEREO-A (circles above zero-line). Since $vB_z$ is an important driver for geomagnetic activity, this is naturally also going to affect the modeled SYM-H index. Hence, when talking about a potential gain in lead time, we should compare not only the arrivals of CMEs at the different spacecraft ($\Delta t_s$), but also the temporal differences in the geoeffective solar wind parameters measured by each ($\Delta t_{vB_z}$). In this case, a sub-L1 monitor probably would have been most advantageous for only the 15 events in Figure~\ref{fig:diff_min_bv}b, where the shock and $min(vB_z)$ are measured at STEREO-A first (stars above zero-line). Of these 15 events, 10 events (67\%) occurred east of the Sun--Earth line. In comparison with the previous Section~\ref{sec:differences_arrival}, where 24 events are identified to have a gain in lead time (i.e. the start of the event is observed at STEREO-A first), this number is reduced by nine when comparing the timings of $min(vB_z)$. For the five events in our target region (see also Table~\ref{tab:spec_CMEs} in Section~\ref{sec:event_analysis}), there is one event, where $min(vB_z)$ is observed at L1 before STEREO-A, with the difference being only 12~minutes. For the other four events within the same area, $min(vB_z)$ is observed first at STEREO-A, as is the shock/start of the MO.

\section{Methods}\label{sec:methods}

In the previous section, we have analyzed differences between CMEs as measured by STEREO-A and spacecraft at L1, since they are important drivers for geoeffectiveness. If Earth-directed CMEs exhibit strong southward magnetic fields, reconnection with Earth's magnetic field can occur, allowing for energy and mass transfer into the magnetosphere. The ring current is an equatorial electric current carried by trapped, drifting charged particles within the inner magnetosphere, typically at distances of 2--8~R$_E$ \cite<e.g.>{desslerparker1959,sckopke1966}. It is sustained by ions and electrons undergoing gradient and curvature drifts around the Earth. As more plasma is injected during geomagnetic storms, the ring current intensifies and generates a magnetic field that opposes Earth’s main field, leading to a measurable depression in the global magnetic field strength, especially at low and mid-latitudes.

These variations in Earth’s magnetic field are measured by ground observatories, located at low and mid-latitudes, and are quantified by geomagnetic indices such as the Disturbance Storm Time ($Dst$) index and the SYM-H index \cite{sugiura1964,iyemori1990,kyoto_dst}. The $Dst$ index provides an hourly measure of the horizontal component of the magnetic field, while SYM-H offers a higher time resolution at one-minute intervals. Although derived from different sets of observatories, SYM-H can be considered a high-resolution counterpart of the $Dst$ index \cite{wanliss2006}. The indices become more negative, the more geoeffective the storm is, with a minimum $Dst$/SYM-H index of less than -50, -100, and -250~nT indicating moderate, intense, and geomagnetic superstorms, respectively \cite<e.g.>{gonzalez1994,echer2008}.

In this section, we explain how we model geomagnetic indices from sub-L1 as well as L1 data using the Temerin and Li model \cite<hereafter TL model>{temerin2002model, Temerin2006}, which is explained in Section~\ref{sec:symh_modeling}. Applying the TL model requires some data preparation first, which includes time shifting the data from a sub-L1 and L1 position to Earth (Section~\ref{sec:time_shift}) to compare modeled with observed SYM-H indices.



\subsection{Mapping L1 and Sub-L1 Data to Earth}\label{sec:time_shift}

Here, we first explain the time-shifting approach for L1 data using Equations~\ref{eq:time_shift} and \ref{eq:time_shift_1}, which are given in the OMNI low resolution documentation \cite<\url{https://omniweb.gsfc.nasa.gov/html/ow_data.html}, Section 12,>{king2025OMNI}. The time-shifting for the STEREO-A data then follows this approach closely. The time shift $\Delta t$ (in seconds) is calculated using

\begin{equation}
\begin{aligned}
\Delta t &= \frac{X}{V} \left[\left(1+\frac{Y \cdot 
W}{X}\right)\right]\cdot\left[\left(1-\frac{V_e \cdot 
W}{V}\right)\right]^{-1}\,\mathrm{with} 
\label{eq:time_shift}
\end{aligned}
\end{equation}

\begin{equation}
\begin{aligned}
W &= \tan [n\, \cdot \,\arctan (V/V_{SW})].
\label{eq:time_shift_1}
\end{aligned}
\end{equation}

\noindent $X$ and $Y$ in Equation~\ref{eq:time_shift} are components of the spacecraft position vector in GSE coordinates given in km, $V$ is the observed solar wind speed in km\,s$^{-1}$ (assumed radial), and $V_e$ is the speed of the Earth's orbital motion, set to 30 km\,s$^{-1}$. $W$, explicitly given in Equation~\ref{eq:time_shift_1}, relates to the assumed orientation of the phase front relative to the Sun--Earth line (normal to the ecliptic), where $V_{SW}$ is the ambient solar wind speed set to 428~km\,s$^{-1}$. The parameter $n$ governs the orientation of the phase front and is set to 0.5. Therefore, for wind speeds around 400~km\,s$^{-1}$ the phase front is close to 22.5$^{\circ}$, i.e., intermediate corotation geometry ($45^\circ$) and radially expanding fronts ($0^\circ$). The assumption that the phase fronts of solar wind structures are not radial but rather tilted toward the average IMF direction has previously been discussed in \citeA{richardson1998}. 

When using this formula, one has to be careful that data points with higher speeds do not pass data points with slower speeds. One approach to deal with this rather unphysical behavior, namely, particles passing each other without any interaction, would be down-sampling the resolution of the measurements, as is often done when shifting data from L1 to Earth. However, since we want to compare modeled indices to observed SYM-H indices given in minute resolution, we instead take $V$ to be the average of the observed speed over the whole time range, hence $V=V_{mean_{L1}}=433$~km\,s$^{-1}$, and insert this mean speed at L1 into Equation~\ref{eq:time_shift}. To also consider higher and lower speeds that deviate from the mean speed in our time shifting approach, we randomly vary the free parameters $V$, $n$, and $V_{SW}$ along a normal distribution. The mean and standard deviation for $V$ are those of the observed speed over the entire time range. Similarly, $n$ is set to 0.5 and is randomly varied with $\sigma = 0.1$, while $V_{SW}$ is set to 428~km\,s$^{-1}$ and varied by 10\%, hence 42.8~km\,s$^{-1}$. Our ensemble for the time shift has 2,500 members. We then get a mean time shift and its expected $1\sigma$ spread. The advantage of using an ensemble is that on the one hand, we avoid down-sampling the high resolution of the dataset. On the other hand, since the ensemble considers a spread in speeds as well as phase front orientations, we do not necessarily have to differentiate between ambient solar wind conditions and fast solar wind structures. This proves to be advantageous in real-time settings as one does not need any additional user input or software to identify CMEs in real time. The error of the time shift is also considered when modeling the SYM-H index, resulting in a forecast window defining when to expect disturbed geomagnetic conditions.

As mentioned, the time shifting of STEREO-A data to Earth practically follows the same approach as described above. However, since we want to put particular emphasis on forecasting the geomagnetic effects of CMEs, being the most geoeffective structures in the solar wind, we adapt Equation~\ref{eq:time_shift} to better fit the expected time shift for CMEs. To do so, we take into account the 32 CME events measured by both STEREO-A and L1 and investigated in Section~\ref{sec:differences_arrival}: instead of inserting the mean speed in Equation~\ref{eq:time_shift}, we average the arrival speeds of the 32 CMEs as measured by STEREO-A (see also Figure~\ref{fig:diff_arrivals1}c) and use this value as $V=444$~km\,s$^{-1}$. As explained in 
Furthermore, we take the observed arrivals of the CME events at L1 and shift them to Earth, employing the procedure explained above, to get the time difference between arrivals at Earth and STEREO-A. We consider this to be the ``true'' time shift for CMEs measured at STEREO-A and Earth, where we introduce an error of about 15~minutes corresponding to the $1\sigma$ uncertainty from the ensemble. When comparing this true time shift with the calculated one, we notice that a phase orientation of $n=0.5$ would overestimate the true time shift eastward of Earth, while underestimating it westward, showing even expected negative time shifts for a long period (not shown). We therefore set the parameter $n$ in Equation~\ref{eq:time_shift_1} to 0.2 to better fit observed arrivals. We derive this value minimizing the mean absolute error (MAE) between the observed and calculated time shifts, varying only $n$ in Equation~\ref{eq:time_shift}, keeping $V$ constant at 444~km\,s$^{-1}$, and excluding the three events that deviate the most from the expected time shift. Interestingly, this is less than $n=0.5$ used for the OMNI low resolution data set, but still assumes a tilted front rather than a purely radial propagation. Similarly to the L1 time shift, we use an ensemble, varying $n$, $V_{SW}$, and $V$  by 0.04, 42.8~km\,s$^{-1}$, and 101~km\,s$^{-1}$, respectively, where the latter is the standard deviation of the arrival speeds for the different CME events.   

\begin{figure}[htbp]
    \centering
    \includegraphics[width=\linewidth]{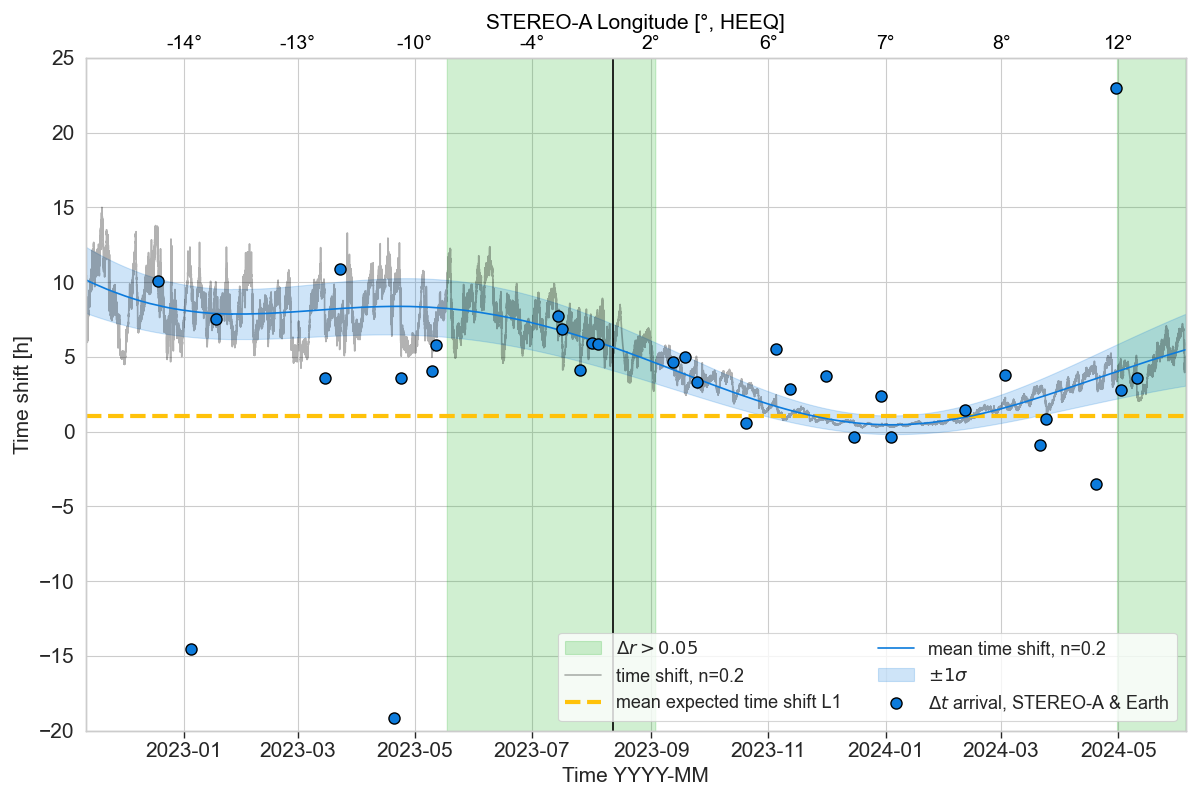}
    \caption{Calculated time shift for STEREO-A data when shifted to Earth. The blue line shows the expected shift when the parameters $V$ and $n$ in Equation \ref{eq:time_shift_1} are set to 444~km\,s$^{-1}$ and 0.2, respectively, with the corresponding $1\sigma$ spread from the ensemble. The gray line corresponds to the time shift when inserting the actual measured speed instead of the averaged into Equation~\ref{eq:time_shift}. The actual difference in arrival of the 32 CME events between STEREO-A and Earth is given by the blue data points. For comparison, the dashed yellow line indicates the mean expected time shift for L1 data as approximately one hour.} 
    \label{fig:time_shift_noaa_sta}
\end{figure}

The calculated time shift for STEREO-A data as well as the true differences in arrival for STEREO-A and Earth (blue data points) are plotted in Figure~\ref{fig:time_shift_noaa_sta}. In contrast to Figure~\ref{fig:l1_sta_pos}, we see that there are now only six instead of eight CME events with a loss in lead time, with two CME events achieving $\Delta t_s$~(Earth\,$-$\,STEREO-A)~$> 0$ after shifting the L1 data to Earth. For comparison, we also indicate the mean expected time shift for L1 data (dashed yellow line in Figure~\ref{fig:time_shift_noaa_sta}), which is roughly one hour. As can be seen, for this specific spacecraft constellation and with $n$ = 0.2, the expected time shift for STEREO-A from December 2023 to February 2024 is $<1$~hours and hence smaller than for L1. Setting $V$ to a constant value introduces a bias of 2.2~hours, which results in an overestimation of the calculated time shift compared to the true shift. This overestimation could result in predicting the onset of a geomagnetic storm later than actually observed. Concerning our target region, the time shift fits particularly well with the observed arrival times for CME events 11, 13 and 14. Our time shifting approach slightly underestimates the arrival time difference for CME event 10 and significantly overestimates it for CME event 12 arriving at STEREO-A on 25 July 2023.

\subsection{Modeling Geomagnetic Indices}\label{sec:symh_modeling}

After shifting the STEREO-A and L1 data to Earth, we apply the TL model to produce a 1-minute SYM-H index. The TL model consists of six terms, for which the parameters were empirically determined using solar wind data from 1995–2002. It requires the interplanetary magnetic field and dynamic pressure as input, and therefore solely depends on in situ measurements of the solar wind. A detailed explanation of the model can be found in \citeA{Temerin2006}, and has also been been introduced in \citeA{weiler2025}. We will therefore only highlight the differences to our previous paper, resulting mainly from handling the time shifting differently in this statistical analysis. 
Since the time step $dt$ is a variable in the TL model, it can be applied straightforwardly to any time resolution, as long as the time steps are equidistant. As the calculated time shift is a variable of spacecraft position, this is not the case for the time shifted datasets, and we have to resample the data to an equidistant time grid with a one minute resolution before applying the TL model.

\subsubsection{Ensemble Method}
As discussed in Section~\ref{sec:differences_parameters}, the properties of the solar wind or transients measured at STEREO-A are not the same as those measured as they pass L1. Specifically, for the 32 identified CME events, the mean total magnetic field strength, speed, and density vary by about 16\%, 5\%, and 28\%, respectively, from STEREO-A to L1. To take this variation into account when modeling geomagnetic indices, we use an ensemble that randomly varies the observed magnetic field, speed, and density data by 16\%, 5\%, and 28\% along a normal distribution. The intrinsic variations of CME parameters from one location in the heliosphere to another therefore do not depend on any theoretical assumptions, like the global expansion, but rather takes the data as is and considers various outcomes by utilizing the statistics analyzed in Section~\ref{sec:differences_parameters}. The ensemble established from these statistics contains 1,000 members. Since the STEREO-A dataset is big, covering one and a half years in minute resolution, we have to bin the dataset when applying the TL model. We divide the dataset into equal bins of 10 days. The TL model needs at least one day of data to initialize; therefore, we start the model always three days into the previous bin, and skip this overlapping part later when stitching the data back together. While this empirical approach might not be the most accurate, it fits our forecasting purpose as it takes possible deformation, expansion, and longitudinal differences of the transients into account, represented by a spread of possible SYM-H values at each time step. Furthermore, ensemble forecasts are fast and hence well suited for real-time applications. In this case, using the entire time range of one and a half years, calculating the SYM-H value takes about half an hour for 1,000 ensemble members. In contrast, it takes only 6 seconds to generate SYM-H indices when the model is applied to only 10 days instead. This is more in line with a real-time setting, where we would only apply the model to the last few days.

\section{Results}\label{sec:analysis}
In this section, we analyze the prediction performance of sub-L1 monitors comparing modeled SYM-H indices from STEREO-A and L1 data to observed indices. In Section~\ref{sec:continuous_analysis}, we present our results in a statistical manner taking the whole time range from November 2022 to June 2024 into account (including all solar wind structures and transients, e.g., CMEs, HSSs and SIRs), comparing modeled with observed geomagnetic storms. In Section~\ref{sec:link_CME_GS} we link the 32 identified CME events to observed and modeled geomagnetic storms, while we provide a more detailed analysis of individual CME events and geomagnetic storms within our target region in Section~\ref{sec:event_analysis}. 

\subsection{Continuous Analysis}\label{sec:continuous_analysis}

First, we apply a peak finding algorithm to the modeled SYM-H indices from STEREO-A and L1 data as well as to the observed ones, taken from the OMNI high resolution data set \cite{king2025OMNI}. We then check if a modeled minimum from STEREO-A (L1) data is found within two days (one day) of an observed minimum, i.e., we match modeled and observed geomagnetic storms whether the modeled SYM-H minimum occurs within $\pm 24$~hours ($\pm 12$~hours) of the observed SYM-H minimum. We use a larger search window for SYM-H minima modeled from STEREO-A data because $\Delta t_{B_vz}$ is large for some events ($\sim \pm 30$ h), even after the data has been time-shifted to Earth. We set the threshold for the peak finding algorithm to SYM-H~$<-50$~nT, hence filtering for geomagnetic storms that are at least of moderate strength.

\begin{figure}[htbp]
    \centering
    \includegraphics[width=\linewidth]{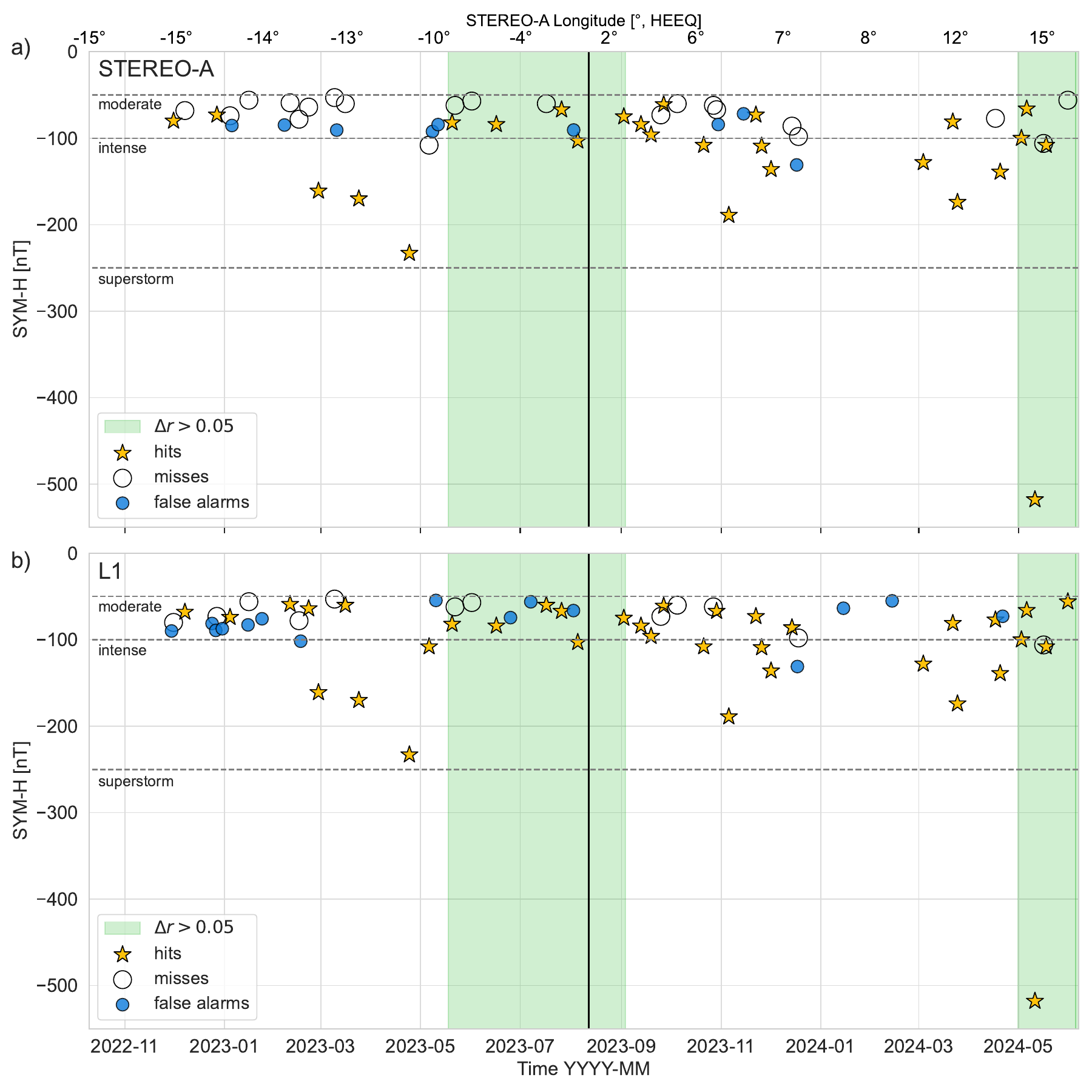}
    \caption{SYM-H minima of 47 observed geomagnetic storms from November 2022 to June 2024. (a) Overview of hits, misses, and false alarms when using STEREO-A data to model the geomagnetic response. Events that are detected by indices modeled from STEREO-A data are indicated as yellow stars. Missed geomagnetic storms are marked as empty circles, whereas filled blue circles correspond to false alarms, i.e., geomagnetic storms that are forecasted but not observed. (b) Corresponding results for SYM-H indices modeled from L1 data, with stars, open circles, and filled blue cirlces again indicating hits, misses, and flase alarms, respectively.}
    \label{fig:hits_misses_false_alarms}
\end{figure}

In Figure~\ref{fig:hits_misses_false_alarms}a we plot the minima of 47 observed geomagnetic storms and indicate whether they are found in the modeled SYM-H indices from STEREO-A data as well, labeling them as ``hits" (marked stars). The ``misses", being the observed but not modeled geomagnetic storms, are marked as black circles, whereas ``false alarms" (i.e., modeled but not observed) are marked as filled blue circles. 

Figure~\ref{fig:hits_misses_false_alarms}a shows that out of the 47 observed geomagnetic storms, 26 are correctly identified when using STEREO-A data. These 26 geomagnetic storms (hereafter referred to as GS events) are listed in Table~\ref{tab:GS_events}, with the differences to observed values (time of $min$(SYM-H) and $min$(SYM-H)) given in separate columns. Hence, 55\% of the modeled storms occur within $\pm24$~hours of a corresponding negative peak in the observed SYM-H values, while the other observed geomagnetic storms (45\%) are missed. It is striking that the missed events occur mainly at the beginning of the time series with eight missed geomagnetic storms up to March 15, 2023. These missed geomagnetic storms make up more than a third (38\%) of all misses for only 22\% of the entire time series. This is also the time period where the longitudinal separation of STEREO-A from Earth is largest ($\Delta \mathrm{lon}<-12^{\circ}$), and the gain in lead time is expected to be highest, as shown by Figures~\ref{fig:diff_arrivals}a, \ref{fig:diff_arrivals1}a, and Figure~\ref{fig:time_shift_noaa_sta}. It is worth mentioning that there are also six missed geomagnetic storms in the period from September 2023 to January 2024. However, the detection rate (i.e., 7 hits versus 6 misses) is 53\%, which is higher than in the time range up to April 2023, where the detection rate is 33\% (4 hits versus 8 misses). 

Furthermore, Figure~\ref{fig:hits_misses_false_alarms}a shows that our pipeline performs well for intense geomagnetic storms (SYM-H $\leq-100$~nT) with the ten strongest events that occurred within the investigated time range correctly captured. In fact, 90\% of all missed events are of moderate strength ($-100$~nT~$<$~SYM-H~$<-50$~nT), with moderate events accounting for 66\% of all observed geomagnetic storms. Only two intense geomagnetic storms, on 2023 May 6 and 2024 May 16, out of a total of 16, are missed when modeling SYM-H from STEREO-A data. Of these 14 correctly identified intense geomagnetic storms, 11 are also modeled as geomagnetic storms with $min$(SYM-H)\,~$\leq-100$~nT. The three events, for which the observed strength is underestimated, correspond to GS events 14, 19 and 26 in Table~\ref{tab:GS_events}. Event 19 deviates the most from the observed intense storm, with a modeled $min$(SYM-H)$=-59$~nT and $min$(SYM-H)$=-128$~nT. In contrast, seven observed moderate geomagnetic storms (SYM-H~$\,\geq-50$~nT) are modeled as intense events (1, 6, 10, 11, 12, 13, 14 in Table~\ref{tab:GS_events}), with GS~event 12 overestimating the observed strength of $min$(SYM-H)$=-96$~nT the most.

The false alarms in Figure~\ref{fig:hits_misses_false_alarms}a consist mainly of modeled moderate storms, since only one intense geomagnetic storm is falsely forecasted on 2023 December 17. Within $\pm5^\circ$ there is only one false alarm, which is also the only false alarm in our target region. The period up to April 2023 stands out again, with 3 false alarms and only 4 hits.

In comparison, when modeling the geomagnetic response from L1 data, 35 observed geomagnetic storms are identified, which are 9 more hits than when modeling SYM-H from STEREO-A data, as shown in Figure~\ref{fig:hits_misses_false_alarms}. Using L1 data, the peak finding algorithm matches ten more moderate storms and one more intense storm, while two moderate storms are missed, which are, however, identified using STEREO-A data (GS events 1 and 2 in Table~\ref{tab:GS_events}). Of the ten additionally identified moderate storms, five are observed before March 15, 2023, where STEREO-A, in contrast, missed eight storms. Most misses (83\%), similar to the indices modeled from STEREO-A data, are again moderate geomagnetic storms. 

As can be seen in Table~\ref{tab:GS_events}, the strength of most geomagnetic storms as calculated from STEREO-A data (58\%) tends to be overestimated. On average, the strength of geomagnetic storms modeled from STEREO-A data, $\langle \Delta \mathrm{SYM\!-\!H}\rangle = -9.8$~nT. Compared to L1 (not shown), the strength of geomagnetic storms with $\langle\Delta \mathrm{SYM\!-\!H}\rangle = -2.0$~nT is overestimated on average as well, albeit less so than from STEREO-A data. This overestimation of geoeffectiveness could be related to the TL model itself, as it tends to overestimate ambient solar wind conditions, leading to an offset, which also affects the modeled SYM-H minima. This bias arises from the fact that the model was originally trained on data from 1995 to 2002 only \cite<see also, e.g.,>{bailey2020prediction}.

Furthermore, Table~\ref{tab:GS_events} shows the time difference between the correctly identified SYM-H minima from STEREO-A data and the observed minima. For most events (73\%) the SYM-H minimum is predicted later than observed. On average, the minima are predicted 2.4~hours too late, which corresponds well to the bias of 3.1~hours introduced by our time shift approach. The event on 16 June 2023, being within our target region, deviates the most from the timing of observed minimum by $-19.4$~hours. For comparison, the timing of the SYM-H minima modeled from L1 data are better on average, but tend to be predicted later than the corresponding observations as well, with the mean difference between observed and modeled minima being $\langle \Delta t_{SYM-H}\rangle \sim-48$ minutes.

\begin{table}[htbp]
    \caption{26 geomagnetic storms that are correctly identified by the SYM-H indices modeled from STEREO-A data. The time of $min$(SYM-H) as well as the $min$(SYM-H) are observed values, whereas $\Delta t_{SYM-H}$ and $\Delta$~SYM-H gives the difference between modeled and observed indices.}
    \centering
    \begin{tabular}{ccccc}
        \toprule
        GS event & time $min$(SYM-H) & $\Delta t_{SYM-H}$ [h] & $min$(SYM-H) [nT] & $\Delta$~SYM-H [nT] \\
        \midrule
            1 & 2022-11-30T19:19Z & +10.8 & -80.0 & -27.8 \\
            2 & 2022-12-27T08:11Z & -9.8 & -73.0 & -20.0 \\
            3 & 2023-02-27T12:13Z & +6.6 & -161.0 & +24.5 \\
            4 & 2023-03-24T05:22Z & +4.6 & -170.0 & +6.1 \\
            5 & 2023-04-24T04:03Z & +7.5 & -233.0 & -27.4 \\
            6 & 2023-05-20T07:11Z & +2.3 & -82.0 & -32.0 \\
            7 & 2023-06-16T09:32Z & -19.4 & -84.0 & +10.6 \\
            8 & 2023-07-26T07:54Z & +2.6 & -67.0 & -24.1 \\
            9 & 2023-08-05T05:49Z & +7.2 & -103.0 & -10.6 \\
            10 & 2023-09-02T10:20Z & +12.0 & -75.0 & -34.3 \\
            11 & 2023-09-12T22:03Z & +1.4 & -84.0 & -20.1 \\
            12 & 2023-09-19T02:50Z & +13.4 & -96.0 & -76.0 \\
            13 & 2023-09-26T20:12Z & -11.2 & -61.0 & -46.1 \\
            14 & 2023-10-21T07:28Z & -1.2 & -108.0 & +9.3 \\
            15 & 2023-11-05T16:55Z & +1.0 & -189.0 & +51.7 \\
            16 & 2023-11-22T06:43Z & +4.8 & -73.0 & -18.4 \\
            17 & 2023-11-25T19:11Z & +2.6 & -109.0 & -68.6 \\
            18 & 2023-12-01T13:30Z & +7.1 & -136.0 & -52.3 \\
            19 & 2024-03-03T18:06Z & -2.1 & -128.0 & +69.0 \\
            20 & 2024-03-21T19:36Z & +5.7 & -81.0 & +21.6 \\
            21 & 2024-03-24T16:21Z & +7.3 & -174.0 & +20.5 \\
            22 & 2024-04-19T19:21Z & +9.5 & -139.0 & +19.1 \\
            23 & 2024-05-02T19:43Z & +4.7 & -100.0 & -41.7 \\
            24 & 2024-05-06T01:13Z & -2.9 & -66.0 & -38.6 \\
            25 & 2024-05-11T02:14Z & -3.0 & -518.0 & +27.4 \\
            26 & 2024-05-17T23:05Z & +0.7 & -108.0 & +10.7 \\
        \bottomrule
    \end{tabular}
    \label{tab:GS_events}
\end{table}

\subsection{Link to Identified CME Events}\label{sec:link_CME_GS}

In this Section we will investigate whether the observed geomagnetic storms are related to the identified CMEs from Section~\ref{sec:differences}. For this purpose, we define that a relation between CME events and observed geomagnetic storms exists, if the SYM-H minimum occurs within the duration of a CME event or within 10 hours of its end time, which is the L1 end time propagated to Earth. We find that of the 47 observed geomagnetic storms, 20 are caused by CME events listed in Table~\ref{tab:CME_events}. CME event 17 led to two identified minima in the observed SYM-H profile, resulting in 19 CME events being linked to 20 observed geomagnetic storms. Of these 19 CME events, 7 happen as STEREO-A is eastward of the Sun--Earth line and 12 as it is westward. In contrast, of the 32 CME events in Table~\ref{tab:CME_events}, 13 events do not cause a geomagnetic storm, with the east-west ratio being fairly balanced at 7:6.

Focusing on correctly identified geomagnetic storms only, corresponding to the 26 GS events listed in Table~\ref{tab:GS_events}, 16 are caused by a CME event in Table~\ref{tab:CME_events}. As the modeled SYM-H indices miss one out of two minima that are caused by CME event 17, we arrive at 16 CME events being associated with 16 GS events, which are listed in Table~\ref{tab:CME_all_GS}. This means that four geomagnetic storms that are associated with CME events are missed from STEREO-A data. These occur on 2023 January 1 2023, 2023 March 15, 2023 July 16 2023, and 2023 September 25, all of which are of moderate strength. In contrast, using STEREO-A data, moderate geomagnetic storms are forecasted for CME events 2, 8, 13, and 20 but not observed (corresponding to the false alarms in Figure~\ref{fig:hits_misses_false_alarms}, occurring on 2023 January 5, 2023 May 11, 2023 August 2, and 2023 November 14). This results in ten true negative detections, i.e., CME events that cause neither an observed nor modeled geomagnetic storm. Table~\ref{tab:CME_all_GS} provides a summary of the number of hits, misses, false alarms, and true negatives, where we also indicate whether or not a modeled geomagnetic storm from Table~\ref{tab:GS_events} is linked to a CME event from Table~\ref{tab:CME_events}.

\begin{table}[htbp]
    \caption{\footnotesize Identified CME events (Table~\ref{tab:CME_events}) linked to correctly modeled geomagnetic storms (i.e., GS event in Table~\ref{tab:GS_events}). The column ``Outcome'' indicates whether a CME event caused an observed geomagnetic storm that is also correctly identified (hit) or missed (miss) from modeled indices using STEREO-A data. We also indicate whether a false alarm from Figure~\ref{fig:hits_misses_false_alarms} is associated with a CME event. CME events that neither cause an observed nor modeled geomagnetic storm are labeled as true negatives. $^{*}$CME event 2 causes an observed geomagnetic storm, that is missed when modeling the SYM-H index from STEREO-A data. A modeled minimum occurs 27~hours later, and is therefore not matched with the observed minimum, but still within the time range of the CME which is why CME event 2 is linked to both a miss and a false alarm. $^{\dagger}$CME event 17 causes two observed SYM-H minima, of which only one is correctly identified from modeled indices.}
    \scriptsize
    \centering
    \begin{tabular}{ccc}
        \toprule
        CME event & GS event & Outcome  \\
        \midrule
            1 & -- & true negative \\
            2$^{*}$ & -- & miss \\
            2$^{*}$ & -- & false alarm \\
            3 & -- & true negative \\
            4 & -- & miss \\
            5 & 4 & hit \\
            6 & -- & true negative \\
            7 & 5 & hit \\
            8 & -- & false alarm \\
            9 & -- & true negative \\
            10 & -- & true negative \\
            11 & -- & miss \\
            12 & 8 & hit \\
            13 & -- & false alarm \\
            14 & 9 & hit \\
            15 & 11 & hit \\
            16 & 12 & hit \\
            17$^{\dagger}$ & -- & miss \\
            17$^{\dagger}$ & 13 & hit \\
            18 & 14 & hit \\
            19 & 15 & hit \\
            20 & -- & false alarm \\
            21 & 18 & hit \\
            22 & -- & true negative \\ 
            23 & -- & true negative \\
            24 & -- & true negative \\
            25 & -- & true negative \\
            26 & 19 & hit \\
            27 & 20 & hit \\
            28 & 21 & hit \\
            29 & 22 & hit \\
            30 & -- & true negative \\
            31 & 23 & hit \\
            32 & 25 & hit \\
        \bottomrule
    \end{tabular}
    \label{tab:CME_all_GS}
\end{table}

\begin{figure}[htbp]
    \centering
    \includegraphics[width=\linewidth]{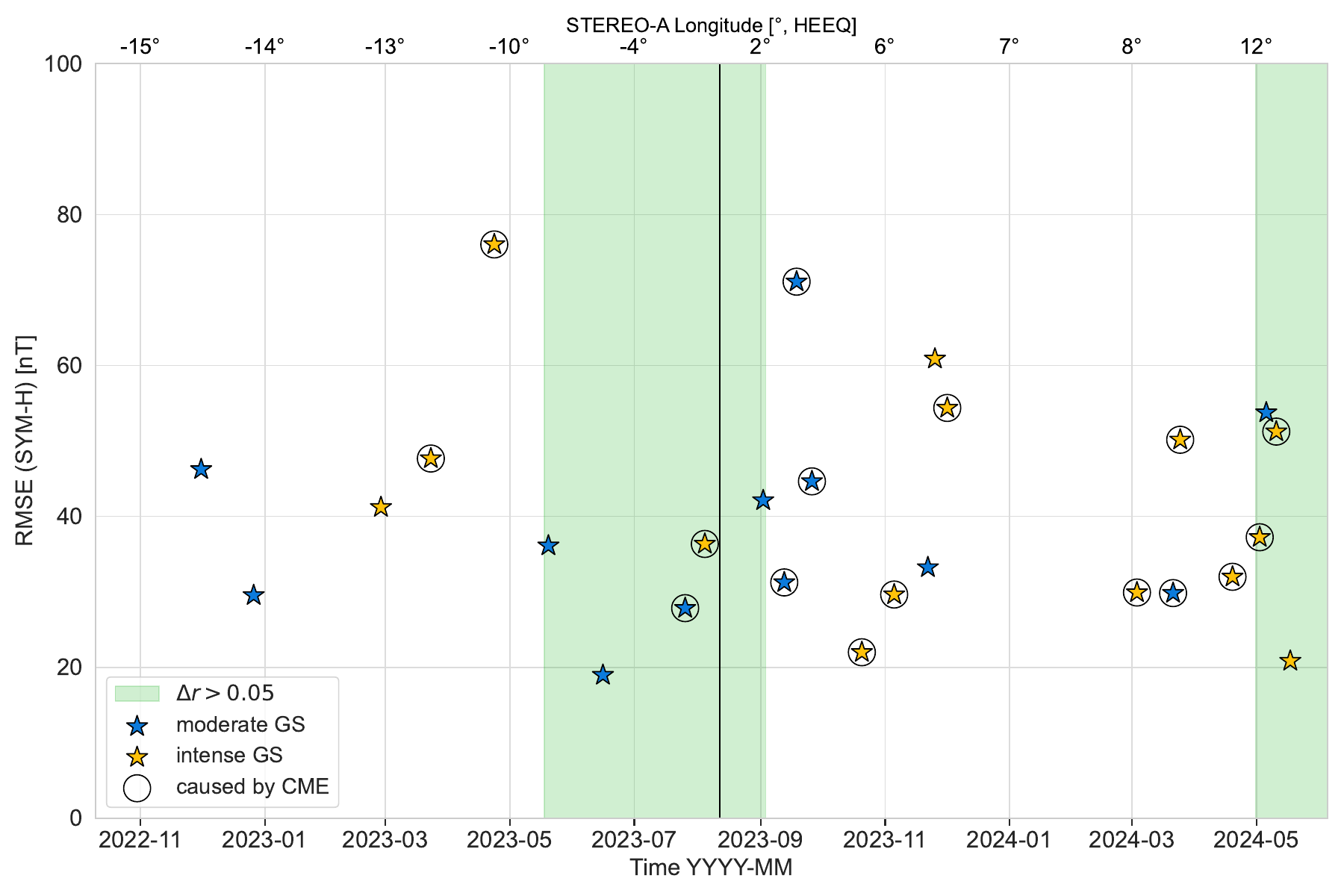}
    \caption{RMSE between observed and modeled SYM-H indices for the 26 GS events. Blue stars indicate geomagnetic storms of observed moderate strength, yellow ones storms of at least intense strength. GS events that are highlighted with circles are associated with a CME event from Table~\ref{tab:CME_events}.}
    \label{fig:rmse}
\end{figure}

Figure~\ref{fig:rmse} shows the RMSE of the 26 GS events between modeled and observed geomagnetic indices, where blue stars indicate observed moderate geomagnetic storms and yellow stars storms of at least intense strength (SYM-H~$\leq -100$~nT). The RMSE is calculated for the duration of the observed geomagnetic storm, where we identify the start mainly due to a sudden increase in the SYM-H profile (likely corresponding to a sudden storm commencement) and the end when the SYM-H values return to pre-storm levels. Furthermore, in Figure~\ref{fig:rmse}, we highlight whether the geomagnetic storm is caused by a CME event from Table~\ref{tab:CME_events} and \ref{tab:CME_all_GS} by circling these events. As can be seen, most of the intense geomagnetic storms (79\%) are caused by a CME event observed by both STEREO-A and L1. With more CMEs being geoeffective as STEREO-A is west of the Sun--Earth line, we also get a higher rate of correctly identified geomagnetic storms there. 

Looking at the in situ data for the geomagnetic storms not associated with a CME event, we notice that most of the other correctly identified geomagnetic storms are caused by HSSs/SIRs which are observed by both STEREO-A and L1. This is especially true for storms that happen within $\pm 10^\circ$ of the Sun--Earth line, where the HSS causing GS event 17 (observed $min$(SYM-H)~$=-109$~nT) shows a particularly strong negative $B_z$ component of $\sim-15$~nT for an extended period of time at both STEREO-A and L1. For the first two events in Figure~\ref{fig:rmse} (GS events 1 and 2 in Table~\ref{tab:GS_events}) we believe that the modeled and observed SYM-H minima are probably matched by chance, as the in situ magnetic field structures look different at STEREO-A and L1, caused by possibly different structures. 

Taking all of this into account, we would like to highlight the following at this point: If the same structure (CME event or HSS/SIR) is observed by both spacecraft, the prediction performance, as measured by the RMSE, does not vary significantly with the longitudinal separation of the spacecraft. This is not true for the time of arrival, $\Delta t_s$, and the timing of minimum ($vB_z$), $\Delta t_{vB_z}$, which show a stronger dependence on longitudinal separation, as analyzed in Section~\ref{sec:differences}.

If we align the modeled SYM-H minima with observed ones, thereby eliminating the error we make when shifting the data temporally to Earth, and add $+10$~nT to the entire modeled SYM-H profile, which corresponds to the bias introduced by the TL model (see Section~\ref{sec:continuous_analysis}), we noticeably improve the mean RMSE for the 26 GS events from $\sim41$~nT to $\sim33$~nT. The best improvement is achieved for the intense GS event 5 on 2023 April 24, where the large RMSE mainly stems from overestimating the time shift to Earth, but the SYM-H profile being otherwise well reproduced. After correcting for the time shift and bias introduced by the TL model, the RMSE improves by 33~nT for this GS event.

\subsection{Target Region}\label{sec:event_analysis}

In this section, we focus on GS as well as CME events within our target region, defined by a radial separation of STEREO-A and Earth of more than 0.05~au and a longitudinal separation of $-8^\circ$ to $+2^\circ$. This area is considered most interesting for future missions such as HENON and SHIELD. HENON will send one CubeSat into a DRO with an orbital distance of roughly 0.1~au to Earth and will therefore be 10 times closer to the Sun than the Earth--L1 distance. SHIELD currently aims for a larger orbital distance to Earth, namely 0.14~au, using nine spacecraft instead of just one. For this configuration, at least one spacecraft would be within $\pm 10^{\circ}$ of the Sun--Earth line at any given time.

Concerning geomagnetic storms in our target region, we see from Figure~\ref{fig:hits_misses_false_alarms} that there are five hits, three misses, and one false alarm when modeling indices from STEREO-A data. As discussed in Section~\ref{sec:link_CME_GS}, of the five hits, two are caused by CME events given in Table~\ref{tab:CME_events}, while the other three are caused by HSSs. For the five hits in our target region, we get a mean RMSE of 32~nT. The SYM-H minima are predicted later and stronger for all five geomagnetic storms, except for the storm that occurred on 2023 June 16 (event 7, see Table~\ref{tab:GS_events}), where the modeled minimum is predicted 19.4~hours earlier and +10~nT weaker than the observed minimum. This geomagnetic storm is caused by a HSS, which arrives roughly 16~hours earlier at STEREO-A and causes a steeper drop in the modeled SYM-H profile than in the observed one.

The RMSE for the three missed geomagnetic storms on 2023 May 22, June 1, and July 17, on the other hand, is also moderate with RMSE~$=32,\, 12,\,\mathrm{and}\, 15$~nT, respectively. The SYM-H profile is mostly well reproduced for these missed storms, however, since the modeled SYM-H profile is rather spiky and shallow, especially for the first missed geomagnetic storm on 2023 May 22, the peak finding algorithm fails to single out a minimum and therefore cannot identify the modeled signal as a geomagnetic storm. Interestingly, the false alarm on 2023 August 2 is associated with a CME event (CME event 13, see Table~\ref{tab:CME_all_GS}). This geomagnetic storm is not only falsely modeled as moderate event using STEREO-A data but also using L1 data. The false detection is therefore probably also due to the model introducing a bias that tends to push the modeled SYM-H index to lower values.

Concerning CMEs, in Section~\ref{sec:differences} we have identified five CME events, observed by both STEREO-A and L1, within the target region. These CME events 10, 11, 12, 13, and 14 are explicitly listed again in Table~\ref{tab:spec_CMEs}. The CME event numbers in Table~\ref{tab:spec_CMEs} correspond to those in Table~\ref{tab:CME_events}. As can be seen from Table~\ref{tab:spec_CMEs}, CME events 11 and 12 cause a geomagnetic storm of moderate strength, while CME event 14 causes the intense GS event 9. CME events 10 and 13, on the other hand, do not cause a geomagnetic storm. As in Table~\ref{tab:CME_all_GS} we indicate whether the geomagnetic storms caused by the CME events are identified correctly (i.e., hit) or missed using STEREO-A data. Again, false alarms are CME events that are falsely associated with a geomagnetic storm, while true negatives are CME events that are (correctly) not associated with a geomagnetic storm.

\begin{table}[ht]
\caption{Overview of the five CMEs occurring in our defined target region, when STEREO-A and Earth are separated by more than 0.05~au. The column ``Strength'' indicates whether the observed geomagnetic storm is of moderate (M) or intense (I) strength.}
\footnotesize
\label{tab:spec_CMEs} 
\begin{tabular*}{\textwidth}{@{\extracolsep\fill}c|c|c|cc|cc}
\toprule
         Event & Strength & Outcome & \multicolumn{2}{c}{Differences GS} & \multicolumn{2}{c}{Differences CME} \\
         CME / GS &  &  & $\Delta t_{SYM-H}$ [h] & $\Delta$~SYM-H [nT] & $\Delta t_s$ [h] & $\Delta t_{vB_z}$ [h] \\
\midrule
        10 / - & -- & true negative & -- & -- & 6.8 & 9.0 \\
        11 / - & M & miss  & -- & -- & 5.8 & 8.1 \\
        12 / 8 & M & hit & +2.6 & -24.1 & 3.0 & 4.0 \\
        13 / - & -- & false alarm & -- & -- & 5.0 & 11.3 \\
        14 / 9 & I & hit & +7.2 & -10.5 & 4.7 & -0.2 \\
\bottomrule
\end{tabular*}
\end{table}

\begin{figure}[htbp]
    \centering
    \includegraphics[width=\textwidth]{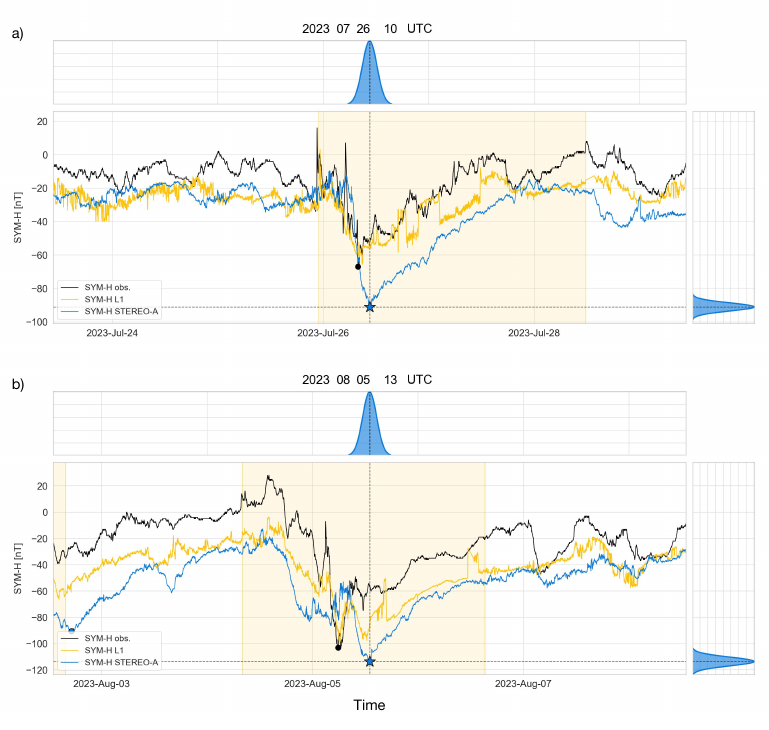}
    \caption{Observed (black) and modeled SYM-H indices using STEREO-A (blue) and L1 (yellow data). Observed SYM-H minima are indicated by black circles. In the panels above and right of the SYM-H plot we show the error distribution for the time and SYM-H indices, respectively. (a) GS event 8, with the associated modeled minimum highlighted as blue star. The duration of CME event 12 is indicated via the shaded yellow area. (b) GS event 9 with the duration of CME event 14 being again highlighted as yellow region.}
    \label{fig:CME_events_GS}
\end{figure}

In Figure~\ref{fig:CME_events_GS}, we show the modeled SYM-H response for the correctly identified GS events on 2023 July 26 and 2023 August 5 (GS events 8 and 9) that are caused by CME events 12 and 14. Observed indices are plotted in black, modeled indices from STEREO-A and L1 data in blue and yellow, respectively. The CME events, with their corresponding start and end times shifted to Earth, are highlighted in yellow. The resulting error distribution for STEREO-A, for both the time shift and the geomagnetic response, is plotted above and right of the SYM-H time series, respectively. We display our results chronologically, with Figure~\ref{fig:CME_events_GS}a showing GS event 8 (CME event 12) and Figure~\ref{fig:CME_events_GS}b showing GS event 9 (CME event 14).

As can be seen in Figures~\ref{fig:CME_events_GS}a and b, both correctly identified minima are predicted later and stronger than the observed minimum (marked by the black circle), which is consistent with the general trend when applying the time shift and TL model (see Table~\ref{tab:GS_events}). However, the overall profile of the geomagnetic storms is well reproduced, especially for the one that occurred on 2023 July 26 (GS event 8, CME event 12), although its strength is overestimated by 24.1~nT. The fact that the modeled geomagnetic storm is stronger may be because the $B_z$-profile at STEREO-A remains negative for a longer period of time than at L1. The steep drop in the SYM-H index five hours after the start of the CME is, however, well captured by both STEREO-A and L1. For this storm, there is a gain in lead time considering both the arrival as well as the timing of $min(vB_z)$. In contrast, for GS event 9 (CME event 14), the gain in lead time is greater when the arrival of the CME is taken into account, but the timing difference of $min(vB_z)$ is negative and occurs in the magnetic obstacle 12 minutes later than its corresponding measurement at L1. Hence, in terms of a potential sub-L1 monitor, we would have known earlier that a CME is arriving, but the SYM-H profile would have started approaching its minimum at about the same time as for L1. Additionally, the modeled SYM-H profile from STEREO-A data shows two broad dips for this event, which is less pronounced in the observed one. This CME event is an interaction of two CMEs, where the second drop in SYM-H is associated with the arrival of another shock. At STEREO-A, the $B_z$-values are more consistently negative before this second shock arrival than at L1, possibly explaining the differences in the shape of the SYM-H profile.

\section{Discussion}\label{sec:discussion}

In this paper, we aim to quantify the feasibility and performance of future sub-L1 monitors for space weather forecasting, such as the ESA missions HENON and SHIELD. HENON is planned to start already by the end of year 2026, while SHIELD is currently under study and will launch earliest in the mid-2030s. With the passage of STEREO-A through the Sun--Earth line in August 2023, while being up to 0.06~au ahead of Earth, such a statistical study is possible for the first time. The passage also coincided with the higher solar activity of the current solar cycle, resulting in many CMEs being observed at both STEREO-A and spacecraft at L1. In this study, we investigate these CMEs, allowing us to derive statistics on how CMEs vary from one spacecraft to the other. We also want to investigate whether future spacecraft at a position similar to STEREO-A are advantageous for forecasting the geoeffectiveness of solar wind structures. In the following, we discuss our results in different sections, where Section~\ref{sec:diss_lead_time} explores the lead time we can expect for distances defined by STEREO-A's trajectory, whereas Section~\ref{sec:diss_methods} assesses the applicability of our methodology for real-time predictions and discusses its limitations. Section~\ref{sec:diss_symh} investigates the performance of the modeled geomagnetic indices from STEREO-A data in comparison with observed indices. 

\subsection{Lead Time}\label{sec:diss_lead_time}
Even though STEREO-A is always radially ahead of L1, there are 8 out of 32 events (25\%), where the start of the event is measured at L1 first and at STEREO-A second. One main finding of this study is therefore that future sub-L1 monitors need to be put closer to the Sun than 0.95~au, to always ensure a gain in lead time. While the ESA mission HENON will be already closer and at an orbital distance of 0.90~au to the Sun, SHIELD is currently planned to be closer still at 0.86~au.\\
Furthermore, the difference in arrival time seems to depend on the longitude of STEREO-A, as the gain in lead time tends to be greater eastward than westward of the Sun--Earth line. We note that within the scope of this study it is not possible to differentiate whether this east-west asymmetry in lead time is a physical effect of the CMEs adapting to the IMF or whether it is an effect of STEREO-A's trajectory, which is not symmetric with respect to the Sun--Earth line. Additionally, we did not investigate the source region and propagation direction of the identified CMEs, which could affect the arrivals at the different spacecraft as well. If the asymmetry, however, stems from CMEs adapting to the Parker spiral, this asymmetry would generalize to any DRO, and should therefore be systematically studied with HENON once sufficient data is available.

In addition to the difference in arrival time, we also investigate the timing when the minimum value of the product $vB_z$ appears within the CME events measured by STEREO-A and spacecraft at L1. In contrast to the arrival time, which affects the predicted onset of the geomagnetic storm, the differences in the timing of $min(vB_z)$, which strongly correlates with $Dst$, affects the predicted SYM-H profile and especially the timing of $min$(SYM-H). We notice that the timing of $min(vB_z)$ follows the same trend as the arrival time, with $min(vB_z)$ being measured on average 4.1~hours earlier at STEREO-A on the east side and $-2.6$~hours later on the west side. Furthermore, with sheaths and especially MOs lasting several hours, the time difference of $min(t_{vB_z})$ between STEREO-A and L1 can be quite large and mostly exceeds the differences in arrival time, $\Delta t_s$. This is also the reason why nine CME events are first observed at STEREO-A, but $min(vB_z)$ is nevertheless observed at L1 first. As argued in Section~\ref{sec:differences_parameters}, when talking about lead time, it is important to take CMEs as a whole into account: It is essential to consider not only the arrival time of CMEs at different spacecraft but also the timings at which geoeffective solar wind parameters arise within the CMEs. The result of $\Delta t_s$ and $\Delta t_{vB_z}$ being so different means that the time shift to Earth cannot be assumed constant over the duration of the CME event. This introduces major difficulties on how time series should be shifted from a sub-L1 position to Earth, which is addressed but not solved in this study, as will be discussed in the following Section~\ref{sec:diss_methods}. 

For the five CME events in our target region, defined by a radial separation of STEREO-A and Earth of more than 0.05~au and a longitudinal separation of $-8^\circ$ to $+2^\circ$, we only see gains in lead time when considering the arrival time, however, one negative lead time of $-12$~minutes for CME event 14 when considering $\Delta t_{vB_z}$.

One additional finding discussed in Section~\ref{sec:differences_parameters} is that for spacecraft separations of up to 0.05~au, longitudinal differences seem to dominate differences in intrinsic properties rather than radial and global effects, such as expansion.

\subsection{Baseline Real-Time Pipeline}\label{sec:diss_methods}
Our approach for setting up a real-time pipeline for sub-L1 data is empirically motivated, using derived CME statistics to build representative ensembles that cover physical differences between the CMEs measured at STEREO-A and L1. We also want to note that we use STEREO-A plasma science data, as we need speed information for both shifting the data to Earth and as input for the TL model. We assume that our methodology would perform much worse if we used beacon plasma data and emphasize the need for high quality plasma data in real time.

For shifting the sub-L1 data temporally to Earth, we use Equation~\ref{eq:time_shift} and \ref{eq:time_shift_1}, given in the documentation of the OMNI low resolution data set \cite{king2025OMNI}. For STEREO-A, the parameters $V$ and $n$ are set to 444~km\,s$^{-1}$, which is the averaged arrival speed of all 32 CME events, and 0.2, respectively. These parameters are assumed constant for the entire period, which prevents faster data points from passing slower ones ahead. Therefore, the resulting time shift and its uncertainty depend mainly on STEREO-A's spatial separation to Earth (see Figure~\ref{fig:time_shift_noaa_sta}). We are aware that by keeping $V$ constant, we introduce a bias, systematically underestimating the time shift for solar wind speeds $<444$~km\,s$^{-1}$ and overestimating it for faster structures. We therefore vary $V$ and $n$ randomly at each time step, employing an ensemble when calculating the time shift. This should account for different structures traveling at different speeds and, furthermore, defines a forecast window for geoeffective disturbances. 

One advantage of this approach is that we do not need to distinguish CMEs from other solar wind structures in real time. However, when comparing the calculated time shift with the observed difference in arrival time between STEREO-A and L1, our time shift approach results in a bias of 2.2 hours, which affects the modeled SYM-H. We emphasize therefore that our approach serves as a baseline methodology that should be further improved. In the future, one could consider implementing automated in-situ CME detection frameworks such as ARCANE \cite{ruedisser2025ARCANE} in sub-L1 data processing pipelines to improve the time shift for these geoeffective structures. One could also improve the time shift by taking the actual measured solar wind speed and including, e.g., inelastic collisions between the plasma parcels, to inhibit unphysical behavior \cite<for different mapping strategies see, e.g.,>{riley2011mapping,milosic2025}. A general investigation of OMNI time-shifted data in comparison with actual in situ observations by spacecraft near Earth's bow shock can be found in \citeA{bluethner2026}.

To produce a 1-minute SYM-H index we use the TL model. The TL model is biased, due to being trained on data from 1995 to 2002 and being applied with an expected offset to data from November 2022 to June 2024 (see also, e.g., \citeA{bailey2020prediction} where the model was applied in a similar manner). Despite offset correction, there is still a deviation of about $-10$ to $-20$~nT from the observed SYM-H value, which affects the strength of the modeled geomagnetic storms. One way to improve this would be to use a different model for the SYM-H prediction \cite<e.g.>{Liemohn2018}. 

As shown, shifting sub-L1 data to Earth is not a trivial task and will become increasingly important as potential sub-L1 monitors are positioned further away from Earth. In our case, where STEREO-A was radially separated from Earth by less than 0.06~au, the longitudinal differences dominated over the radial ones anyway, so we only had to shift the data in time, while our ensemble covered the differences in longitude. However, the further away from Earth future upstream spacecraft are positioned, the more we need to pay attention to the radial and global evolution of large-scale solar wind structures and take into account how their intrinsic parameters change over greater distances. To answer this question, future studies could use data from other spacecraft such as Solar Orbiter and Parker Solar Probe, which have passed the Sun--Earth line at varying distances from Earth in the past. \citeA{laker_2024, davies2021solo} have shown that it is possible to use data from Solar Orbiter at a location far upstream of Earth (at a heliocentric distance of 0.8 and 0.5~au, respectively) to predict the magnetic field profile and geomagnetic indices. \citeA{davies2025realtime} has recently shown that far upstream monitoring is even possible in real time, using Solar Orbiter observations at heliocentric distances of 0.53 and 0.60 au. They found that good predictions of the geomagnetic indices can be made with longitudinal separations up to 10$^{\circ}$ from the Sun--Earth line for separations where radial evolution effects dominate over longitudinal effects. However, further research is needed to determine the best methodology for spacecraft that are widely separated in the radial direction.

\subsection{SYM-H Accuracy}\label{sec:diss_symh}
When SYM-H indices are modeled from the STEREO-A data, 26 out of 47 observed geomagnetic storms are correctly detected, where the detection rate is especially high for stronger geomagnetic storms \cite<SYM-H~$\leq-100$~nT, see also>{lugaz2025subL1}. Quantitatively, 14 out of 16 intense geomagnetic storms (82\%) are identified using STEREO-A data, where 11 are also modeled as events with SYM-H~$\leq -100$~nT. In contrast, considering the 21 missed geomagnetic storms in Figure~\ref{fig:hits_misses_false_alarms}, 90\% are of moderate strength. We consider the high detection rate of intense events to be a very positive result of our analysis, as these events are of utmost interest to us as space weather forecasters.

The biases from our time shifting approach and from the TL model also translate into the prediction performance of the SYM-H indices, with the SYM-H minima of the 26 correctly identified GS events in Table~\ref{tab:GS_events} being often predicted too late (73\%) and too strong (58\%). This also leads to seven geomagnetic storms being modeled as intense events, which are, however, only of moderate strength in the observed SYM-H data. 
Furthermore, we notice that the detection rate is also high if a geomagnetic storm is related to a CME event, given in Table~\ref{tab:CME_events}, which is the case for 20 observed geomagnetic storms and 16 modeled storms (see also Table~\ref{tab:CME_all_GS}). Of the 20 observed geomagnetic storms caused by a CME event, 11 are of at least intense strength. All of these 11 stronger geomagnetic storms associated with a CME event are also modeled from STEREO-A data (circled yellow stars in Figure~\ref{fig:rmse}). 

Interestingly, the performance of the SYM-H modeling does not depend strongly on the longitudinal separation of STEREO-A to Earth (see Figure~\ref{fig:rmse}). If STEREO-A measures the same large-scale structure (i.e., CMEs or HSSs/SIRs) as spacecraft at L1, the prediction performance does not vary significantly for the different correctly identified geomagnetic storms, where the chance of STEREO-A measuring the same structure as spacecraft at L1 increases with STEREO-A's proximity to the Sun--Earth line \cite<see also>{banu2025}.

\begin{figure}[htbp]
    \centering
    \includegraphics[width=\textwidth]{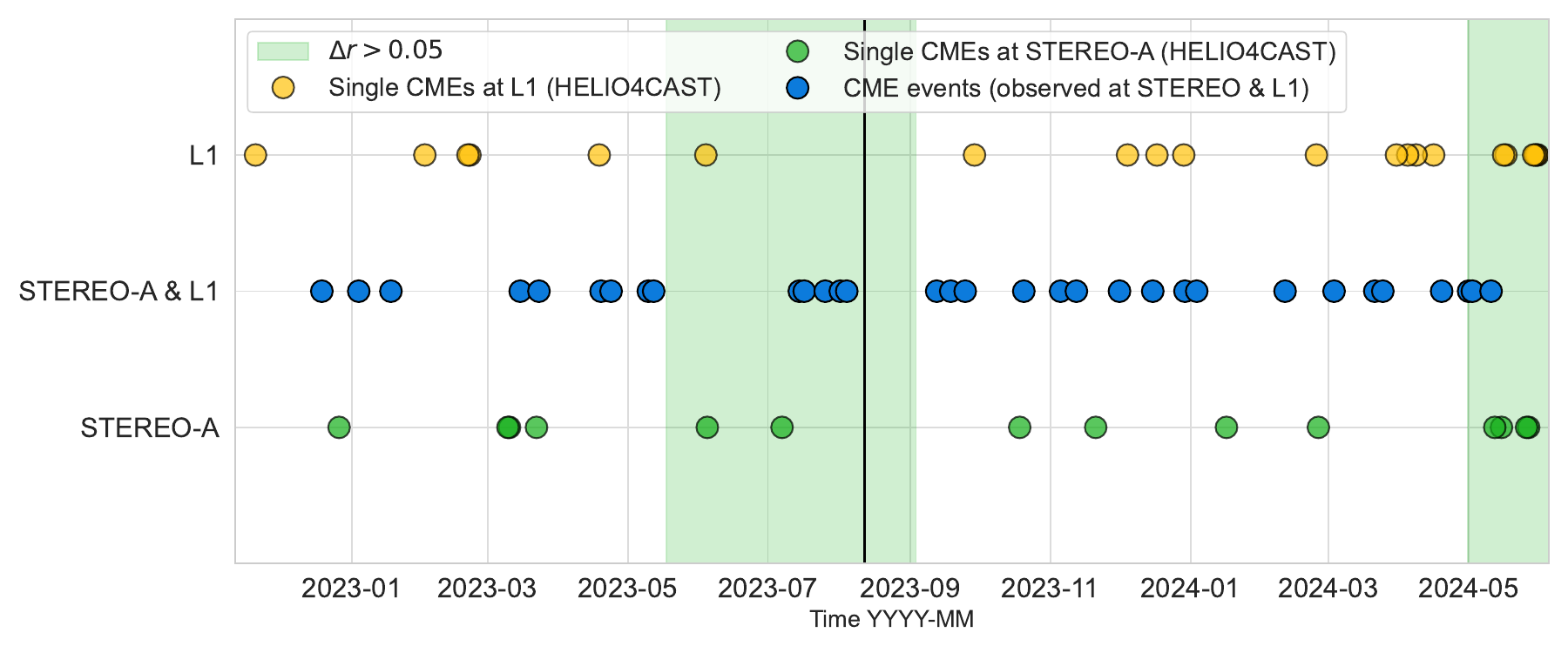}
    \caption{Distribution of CMEs given in the HELIO4CAST ICMECAT, that are either only observed at L1 (yellow) or STEREO-A (green). Blue data points indicate CME events in Table~\ref{tab:CME_events}, which are observed at both STEREO-A and spacecraft at L1.}
    \label{fig:HELIO4CAST}
\end{figure}

Figure~\ref{fig:HELIO4CAST} shows the distribution of CMEs measured at L1 and STEREO-A that are given in the HELIO4CAST ICMECAT \cite{moestl2026icmecat} as yellow and green circles, respectively. As these CMEs are not associated with any CME event given in Table~\ref{tab:CME_events}, they are considered ``single'' CME events, which are either measured only at STEREO-A or at L1. Additionally, we also plot the distribution of our 32 CME events that are, in contrast, identified at both STEREO-A and L1. It is striking that there are no ``single'' CME events close to the Sun--Earth line (marked by the vertical line on 2023 August 12) and that the rate of ``single'' CMEs increases as STEREO-A's longitude to the Sun--Earth line increases. We note that for the single CMEs at L1 and STEREO-A, we plot every MO listed in ICMECAT, i.e., we represent multiple MOs within the same complex structure as a single data point, since each of them has its own entry in ICMECAT. However, we have also checked the data and counted individual MOs from the ICMECAT within the same complex structure as one event (corresponding to MOs really close together in Figure~\ref{fig:HELIO4CAST}), arriving at 12 CME events that are observed at STEREO-A but not L1, while 17 CME events are observed only at L1 but not STEREO-A.

From the CMEs measured by STEREO-A and L1, we know that the longitudinal differences are significant and that they dominate over radial ones, at least for the investigated spacecraft distances of up to 0.05~au. In addition, the probability that the two spacecraft will measure the same structure increases the closer STEREO-A is to the Sun--Earth line. We therefore emphasize that it is important to limit the longitudinal separation of a future sub-L1 monitor and Earth to a maximum of 15$^\circ$ (corresponding to 0.27~au), which is in line with past studies \cite<e.g.>[]{good2016interplanetary,lugaz2024MEwidth,banu2025}. We are assuming that this really is an upper limit, since at longitudinal distances of $>\!12^\circ$, some smaller solar wind structures that cause moderate geomagnetic storms are unlikely to be detected by STEREO-A, whereas more intense geomagnetic storms, likely triggered by more longitudinally extended structures, can be correctly reproduced even at these distances (see also Figure~\ref{fig:hits_misses_false_alarms}).


\section{Summary and Future Mission Recommendations}\label{sec:summary}
Our analysis uses the passage of STEREO-A through the Sun--Earth line up to 0.06~au ahead of Earth as a good test case for future space weather missions such as the ESA missions HENON and SHIELD. In our analysis, we find that future sub-L1 missions must be placed closer than 0.95~au to the Sun to ensure exclusive lead time gains for space weather forecasting. We also note that when STEREO-A measures the same solar wind structure (i.e., CME or HSS/SIR) as L1 monitors, the prediction performance of geomagnetic indices does not vary significantly with longitude for these structures, where the chance of a structure being measured by both spacecraft increases with STEREO-A's proximity to the Sun--Earth line. We therefore recommend that a future sub-L1 monitor and Earth should be separated by less than $15^\circ$ (corresponding to 0.27~au) in the longitudinal direction, or even $12^\circ$ if one wants to catch smaller, but still potentially geoeffective, structures as well. Furthermore, at radial distances of up to 0.05~au the differences in magnetic field strength appear to be determined more by the longitudinal distances between the spacecraft than by global expansion. The analysis of CMEs measured at both STEREO-A and L1 also shows that the lead time and also the timing of $vB_z$ vary with longitude. However, we cannot say with certainty whether this is due to the CMEs adapting to the IMF or because of STEREO-A's asymmetric trajectory. This needs to be investigated with HENON on a DRO that, unlike STEREO-A's trajectory, is symmetric east and west of the Sun--Earth line. Since the ESA SHIELD mission is currently aiming for an orbit of 0.14~au from Earth and plans to deploy nine spacecraft on DROs, at least one spacecraft would be within $10^\circ$ of the Sun--Earth line at any given time, which is well in line with our recommendations. In contrast, HENON will deploy only one spacecraft in a DRO, which is why it will only periodically pass in front of Earth at a distance of 0.1~au. Nevertheless, it will be an excellent precursor mission for SHIELD, addressing important issues such as possible east-west asymmetries in lead time, refining the methodology for sub-L1 data, and further investigating the trade-off between prediction accuracy and lead time regarding heliocentric distance.


\section*{Open Research Section}
The data files as well as python scripts used in this study and the animations that we produced can be found on figshare \cite{Weiler2026_figshare}. This work made use of the following Python packages: sunpy \cite{sunpy_2020}, astropy \cite{astropy:2022}, numpy \cite{harris2020numpy}, pandas \cite{reback2020pandas}, spacepy \cite{niehof2022spacepy}, scipy \cite{scipy2020}, plotly \cite{plotly2026} and matplotlib \cite{hunter2007matplotlib}.
All in situ measurements used in this research are publicly available through NASA SPDF (\url{https://spdf.gsfc.nasa.gov}). STEREO-A beacon and science data was downloaded from the STEREO science center: \url{https://stereo-ssc.nascom.nasa.gov/data/beacon/ahead/impact/} and \url{https://spdf.gsfc.nasa.gov/pub/data/stereo/ahead/l2/impact/magplasma/}. The ACE and DSCOVR data is taken from the NOAA real time solar wind data product:  \url{https://services.swpc.noaa.gov/products/solar-wind/}. The OMNI dataset provides the SYM-H index \cite[\url{https://spdf.gsfc.nasa.gov/pub/data/omni/high_res_omni/}]{omnimin}. For our analysis we use the HELIO4CAST ICMECAT \cite{moestl2026icmecat} which lists CMEs measured in situ at different spacecraft, see also \url{https://helioforecast.space/icmecat}. Here, version 24 on figshare was used \cite{moestl_icme_figshare}.

\acknowledgments
We sincerely thank the reviewers for their valuable input towards refining the structure of the paper and enhancing the clarity of key arguments. E.~W., E.~D., and C.~M. are funded by the European Union (ERC, HELIO4CAST, 101042188). Views and opinions expressed are however those of the author(s) only and do not necessarily reflect those of the European Union or the European Research Council Executive Agency. Neither the European Union nor the granting authority can be held responsible for them. N.~L. acknowledges support from NASA grants 80NSSC24K1245 and 80NSSC25K7409. M.~R. thanks the Austrian Science Fund (FWF) [P 34437].

\section*{Conflict of Interest Statement}
The authors have no conflicts of interest to disclose.

\bibliography{bibliography}

\begin{thebibliography}{}

\bibitem [\protect \citeauthoryear {%
{Al-Haddad}%
\ \BBA {} {Lugaz}%
}{%
{Al-Haddad}%
\ \BBA {} {Lugaz}%
}{%
{\protect \APACyear {2025}}%
}]{%
alhaddad2025}
\APACinsertmetastar {%
alhaddad2025}%
\begin{APACrefauthors}%
{Al-Haddad}, N.%
\BCBT {}\ \BBA {} {Lugaz}, N.%
\end{APACrefauthors}%
\unskip\
\newblock
\APACrefYearMonthDay{2025}{{\APACmonth{06}}}{}.
\newblock
{\BBOQ}\APACrefatitle {{CME Magnetic Field Structure: A Comprehensive Review}} {{CME Magnetic Field Structure: A Comprehensive Review}}.{\BBCQ}
\newblock
\BIn{} \APACrefbtitle {SHINE 2025 Workshop} {Shine 2025 workshop}\ (\BPG~128).
\PrintBackRefs{\CurrentBib}

\bibitem [\protect \citeauthoryear {%
{Astropy Collaboration}%
\ \protect \BOthers {.}}{%
{Astropy Collaboration}%
\ \protect \BOthers {.}}{%
{\protect \APACyear {2022}}%
}]{%
astropy:2022}
\APACinsertmetastar {%
astropy:2022}%
\begin{APACrefauthors}%
{Astropy Collaboration}%
, {Price-Whelan}, A\BPBI M.%
, {Lim}, P\BPBI L.%
, {Earl}, N.%
, {Starkman}, N.%
, {Bradley}, L.%
\BDBL {}{Astropy Project Contributors}%
\end{APACrefauthors}%
\unskip\
\newblock
\APACrefYearMonthDay{2022}{{\APACmonth{08}}}{}.
\newblock
{\BBOQ}\APACrefatitle {{The Astropy Project: Sustaining and Growing a Community-oriented Open-source Project and the Latest Major Release (v5.0) of the Core Package}} {{The Astropy Project: Sustaining and Growing a Community-oriented Open-source Project and the Latest Major Release (v5.0) of the Core Package}}.{\BBCQ}
\newblock
\APACjournalVolNumPages{\apj}{935}{2}{167}.
\newblock
\begin{APACrefURL} \url{http://www.astropy.org} \end{APACrefURL}
\newblock
\begin{APACrefDOI} \doi{10.3847/1538-4357/ac7c74} \end{APACrefDOI}
\PrintBackRefs{\CurrentBib}

\bibitem [\protect \citeauthoryear {%
{Bailey}%
\ \protect \BOthers {.}}{%
{Bailey}%
\ \protect \BOthers {.}}{%
{\protect \APACyear {2020}}%
}]{%
bailey2020prediction}
\APACinsertmetastar {%
bailey2020prediction}%
\begin{APACrefauthors}%
{Bailey}, R\BPBI L.%
, {M{\"o}stl}, C.%
, {Reiss}, M\BPBI A.%
, {Weiss}, A\BPBI J.%
, {Amerstorfer}, U\BPBI V.%
, {Amerstorfer}, T.%
\BDBL {}{Leonhardt}, R.%
\end{APACrefauthors}%
\unskip\
\newblock
\APACrefYearMonthDay{2020}{{\APACmonth{05}}}{}.
\newblock
{\BBOQ}\APACrefatitle {{Prediction of Dst During Solar Minimum Using In Situ Measurements at L5}} {{Prediction of Dst During Solar Minimum Using In Situ Measurements at L5}}.{\BBCQ}
\newblock
\APACjournalVolNumPages{Space Weather}{18}{5}{e02424}.
\newblock
\begin{APACrefDOI} \doi{10.1029/2019SW002424} \end{APACrefDOI}
\PrintBackRefs{\CurrentBib}

\bibitem [\protect \citeauthoryear {%
Banu%
\ \protect \BOthers {.}}{%
Banu%
\ \protect \BOthers {.}}{%
{\protect \APACyear {2025}}%
}]{%
banu2025}
\APACinsertmetastar {%
banu2025}%
\begin{APACrefauthors}%
Banu, S\BPBI A.%
, Lugaz, N.%
, Zhuang, B.%
, Al-Haddad, N.%
, Farrugia, C\BPBI J.%
\BCBL {}\ \BBA {} Galvin, A\BPBI B.%
\end{APACrefauthors}%
\unskip\
\newblock
\APACrefYearMonthDay{2025}{mar}{}.
\newblock
{\BBOQ}\APACrefatitle {Investigating Coronal Mass Ejections through Multispacecraft Measurements: STEREO-A and L1 in 2022–2023} {Investigating coronal mass ejections through multispacecraft measurements: Stereo-a and l1 in 2022–2023}.{\BBCQ}
\newblock
\APACjournalVolNumPages{The Astrophysical Journal}{982}{1}{47}.
\newblock
\begin{APACrefURL} \url{https://dx.doi.org/10.3847/1538-4357/adb60c} \end{APACrefURL}
\newblock
\begin{APACrefDOI} \doi{10.3847/1538-4357/adb60c} \end{APACrefDOI}
\PrintBackRefs{\CurrentBib}

\bibitem [\protect \citeauthoryear {%
{Baratashvili}%
, {Verbeke}%
, {Wijsen}%
\BCBL {}\ \BBA {} {Poedts}%
}{%
{Baratashvili}%
\ \protect \BOthers {.}}{%
{\protect \APACyear {2022}}%
}]{%
Baratashvili2022}
\APACinsertmetastar {%
Baratashvili2022}%
\begin{APACrefauthors}%
{Baratashvili}, T.%
, {Verbeke}, C.%
, {Wijsen}, N.%
\BCBL {}\ \BBA {} {Poedts}, S.%
\end{APACrefauthors}%
\unskip\
\newblock
\APACrefYearMonthDay{2022}{{\APACmonth{11}}}{}.
\newblock
{\BBOQ}\APACrefatitle {{Improving CME evolution and arrival predictions with AMR and grid stretching in Icarus}} {{Improving CME evolution and arrival predictions with AMR and grid stretching in Icarus}}.{\BBCQ}
\newblock
\APACjournalVolNumPages{\aap}{667}{}{A133}.
\newblock
\begin{APACrefDOI} \doi{10.1051/0004-6361/202244111} \end{APACrefDOI}
\PrintBackRefs{\CurrentBib}

\bibitem [\protect \citeauthoryear {%
Blüthner%
\ \protect \BOthers {.}}{%
Blüthner%
\ \protect \BOthers {.}}{%
{\protect \APACyear {2026}}%
}]{%
bluethner2026}
\APACinsertmetastar {%
bluethner2026}%
\begin{APACrefauthors}%
Blüthner, G\BPBI H.%
, Volwerk, M.%
, Schmid, D.%
, Nakamura, R.%
, Temmer, M.%
, Roberts, O\BPBI W.%
\BDBL {}Varsani, A.%
\end{APACrefauthors}%
\unskip\
\newblock
\APACrefYearMonthDay{2026}{}{}.
\newblock
{\BBOQ}\APACrefatitle {Evaluating the OMNI Database: Statistical Analysis of Time-Shifted L1 Data Versus Direct Near-Earth Solar Wind Observations} {Evaluating the omni database: Statistical analysis of time-shifted l1 data versus direct near-earth solar wind observations}.{\BBCQ}
\newblock
\APACjournalVolNumPages{Journal of Geophysical Research: Space Physics}{131}{3}{e2025JA034781}.
\newblock
\begin{APACrefURL} \url{https://agupubs.onlinelibrary.wiley.com/doi/abs/10.1029/2025JA034781} \end{APACrefURL}
\newblock
\APACrefnote{e2025JA034781 2025JA034781}
\newblock
\begin{APACrefDOI} \doi{https://doi.org/10.1029/2025JA034781} \end{APACrefDOI}
\PrintBackRefs{\CurrentBib}

\bibitem [\protect \citeauthoryear {%
{Burlaga}%
, {Sittler}%
, {Mariani}%
\BCBL {}\ \BBA {} {Schwenn}%
}{%
{Burlaga}%
\ \protect \BOthers {.}}{%
{\protect \APACyear {1981}}%
}]{%
burlaga1981magnetic}
\APACinsertmetastar {%
burlaga1981magnetic}%
\begin{APACrefauthors}%
{Burlaga}, L.%
, {Sittler}, E.%
, {Mariani}, F.%
\BCBL {}\ \BBA {} {Schwenn}, R.%
\end{APACrefauthors}%
\unskip\
\newblock
\APACrefYearMonthDay{1981}{{\APACmonth{08}}}{}.
\newblock
{\BBOQ}\APACrefatitle {{Magnetic loop behind an interplanetary shock: Voyager, Helios, and IMP 8 observations}} {{Magnetic loop behind an interplanetary shock: Voyager, Helios, and IMP 8 observations}}.{\BBCQ}
\newblock
\APACjournalVolNumPages{\jgr}{86}{A8}{6673-6684}.
\newblock
\begin{APACrefDOI} \doi{10.1029/JA086iA08p06673} \end{APACrefDOI}
\PrintBackRefs{\CurrentBib}

\bibitem [\protect \citeauthoryear {%
Burlaga%
, Plunkett%
\BCBL {}\ \BBA {} St.~Cyr%
}{%
Burlaga%
\ \protect \BOthers {.}}{%
{\protect \APACyear {2002}}%
}]{%
burlaga2002}
\APACinsertmetastar {%
burlaga2002}%
\begin{APACrefauthors}%
Burlaga, L\BPBI F.%
, Plunkett, S\BPBI P.%
\BCBL {}\ \BBA {} St.~Cyr, O\BPBI C.%
\end{APACrefauthors}%
\unskip\
\newblock
\APACrefYearMonthDay{2002}{}{}.
\newblock
{\BBOQ}\APACrefatitle {Successive CMEs and complex ejecta} {Successive cmes and complex ejecta}.{\BBCQ}
\newblock
\APACjournalVolNumPages{Journal of Geophysical Research: Space Physics}{107}{A10}{SSH 1-1-SSH 1-12}.
\newblock
\begin{APACrefURL} \url{https://agupubs.onlinelibrary.wiley.com/doi/abs/10.1029/2001JA000255} \end{APACrefURL}
\newblock
\begin{APACrefDOI} \doi{https://doi.org/10.1029/2001JA000255} \end{APACrefDOI}
\PrintBackRefs{\CurrentBib}

\bibitem [\protect \citeauthoryear {%
{Burt}%
\ \BBA {} {Smith}%
}{%
{Burt}%
\ \BBA {} {Smith}%
}{%
{\protect \APACyear {2012}}%
}]{%
dscovr}
\APACinsertmetastar {%
dscovr}%
\begin{APACrefauthors}%
{Burt}, J.%
\BCBT {}\ \BBA {} {Smith}, B.%
\end{APACrefauthors}%
\unskip\
\newblock
\APACrefYear{2012}.
\newblock
\APACrefbtitle {{Deep Space Climate Observatory: The DSCOVR mission}} {{Deep Space Climate Observatory: The DSCOVR mission}}.
\newblock
\begin{APACrefDOI} \doi{10.1109/AERO.2012.6187025} \end{APACrefDOI}
\PrintBackRefs{\CurrentBib}

\bibitem [\protect \citeauthoryear {%
Cical{\`o}%
\ \protect \BOthers {.}}{%
Cical{\`o}%
\ \protect \BOthers {.}}{%
{\protect \APACyear {2025}}%
}]{%
Cicalò2025Henon}
\APACinsertmetastar {%
Cicalò2025Henon}%
\begin{APACrefauthors}%
Cical{\`o}, S.%
, Alessi, E\BPBI M.%
, Provinciali, L.%
, Amabili, P.%
, Saita, G.%
, Calcagno, D.%
\BDBL {}Khan, M.%
\end{APACrefauthors}%
\unskip\
\newblock
\APACrefYearMonthDay{2025}{Aug}{13}.
\newblock
{\BBOQ}\APACrefatitle {Mission analysis for the HENON CubeSat mission to a large Sun-Earth distant retrograde orbit} {Mission analysis for the henon cubesat mission to a large sun-earth distant retrograde orbit}.{\BBCQ}
\newblock
\APACjournalVolNumPages{Astrophysics and Space Science}{370}{8}{83}.
\newblock
\begin{APACrefURL} \url{https://doi.org/10.1007/s10509-025-04473-0} \end{APACrefURL}
\newblock
\begin{APACrefDOI} \doi{10.1007/s10509-025-04473-0} \end{APACrefDOI}
\PrintBackRefs{\CurrentBib}

\bibitem [\protect \citeauthoryear {%
{Cyr}%
\ \protect \BOthers {.}}{%
{Cyr}%
\ \protect \BOthers {.}}{%
{\protect \APACyear {2000}}%
}]{%
stcyr2000diamond}
\APACinsertmetastar {%
stcyr2000diamond}%
\begin{APACrefauthors}%
{Cyr}, O\BPBI C\BPBI S.%
, {Mesarch}, M\BPBI A.%
, {Maldonado}, H\BPBI M.%
, {Folta}, D\BPBI C.%
, {Harper}, A\BPBI D.%
, {Davila}, J\BPBI M.%
\BCBL {}\ \BBA {} {Fisher}, R\BPBI R.%
\end{APACrefauthors}%
\unskip\
\newblock
\APACrefYearMonthDay{2000}{{\APACmonth{09}}}{}.
\newblock
{\BBOQ}\APACrefatitle {{Space Weather Diamond: a four spacecraft monitoring system}} {{Space Weather Diamond: a four spacecraft monitoring system}}.{\BBCQ}
\newblock
\APACjournalVolNumPages{Journal of Atmospheric and Solar-Terrestrial Physics}{62}{14}{1251-1255}.
\newblock
\begin{APACrefDOI} \doi{10.1016/S1364-6826(00)00069-9} \end{APACrefDOI}
\PrintBackRefs{\CurrentBib}

\bibitem [\protect \citeauthoryear {%
{Davies}%
, {Forsyth}%
, {Winslow}%
, {M{\"o}stl}%
\BCBL {}\ \BBA {} {Lugaz}%
}{%
{Davies}%
, {Forsyth}%
\BCBL {}\ \protect \BOthers {.}}{%
{\protect \APACyear {2021}}%
}]{%
davies2021catalogue}
\APACinsertmetastar {%
davies2021catalogue}%
\begin{APACrefauthors}%
{Davies}, E\BPBI E.%
, {Forsyth}, R\BPBI J.%
, {Winslow}, R\BPBI M.%
, {M{\"o}stl}, C.%
\BCBL {}\ \BBA {} {Lugaz}, N.%
\end{APACrefauthors}%
\unskip\
\newblock
\APACrefYearMonthDay{2021}{{\APACmonth{12}}}{}.
\newblock
{\BBOQ}\APACrefatitle {{A Catalog of Interplanetary Coronal Mass Ejections Observed by Juno between 1 and 5.4 au}} {{A Catalog of Interplanetary Coronal Mass Ejections Observed by Juno between 1 and 5.4 au}}.{\BBCQ}
\newblock
\APACjournalVolNumPages{\apj}{923}{2}{136}.
\newblock
\begin{APACrefDOI} \doi{10.3847/1538-4357/ac2ccb} \end{APACrefDOI}
\PrintBackRefs{\CurrentBib}

\bibitem [\protect \citeauthoryear {%
{Davies}%
, {M{\"o}stl}%
\BCBL {}\ \protect \BOthers {.}}{%
{Davies}%
, {M{\"o}stl}%
\BCBL {}\ \protect \BOthers {.}}{%
{\protect \APACyear {2021}}%
}]{%
davies2021solo}
\APACinsertmetastar {%
davies2021solo}%
\begin{APACrefauthors}%
{Davies}, E\BPBI E.%
, {M{\"o}stl}, C.%
, {Owens}, M\BPBI J.%
, {Weiss}, A\BPBI J.%
, {Amerstorfer}, T.%
, {Hinterreiter}, J.%
\BDBL {}{Harrison}, R\BPBI A.%
\end{APACrefauthors}%
\unskip\
\newblock
\APACrefYearMonthDay{2021}{{\APACmonth{12}}}{}.
\newblock
{\BBOQ}\APACrefatitle {{In situ multi-spacecraft and remote imaging observations of the first CME detected by Solar Orbiter and BepiColombo}} {{In situ multi-spacecraft and remote imaging observations of the first CME detected by Solar Orbiter and BepiColombo}}.{\BBCQ}
\newblock
\APACjournalVolNumPages{\aap}{656}{}{A2}.
\newblock
\begin{APACrefDOI} \doi{10.1051/0004-6361/202040113} \end{APACrefDOI}
\PrintBackRefs{\CurrentBib}

\bibitem [\protect \citeauthoryear {%
Davies%
\ \protect \BOthers {.}}{%
Davies%
\ \protect \BOthers {.}}{%
{\protect \APACyear {2025}}%
}]{%
davies2025realtime}
\APACinsertmetastar {%
davies2025realtime}%
\begin{APACrefauthors}%
Davies, E\BPBI E.%
, Weiler, E.%
, Möstl, C.%
, Horbury, T\BPBI S.%
, O'Brien, H.%
, Morris, J.%
\BCBL {}\ \BBA {} Crabtree, A.%
\end{APACrefauthors}%
\unskip\
\newblock
\APACrefYearMonthDay{2025}{}{}.
\newblock
\APACrefbtitle {Real-time prediction of geomagnetic storms using Solar Orbiter as a far upstream solar wind monitor.} {Real-time prediction of geomagnetic storms using solar orbiter as a far upstream solar wind monitor.}
\newblock
\begin{APACrefURL} \url{https://arxiv.org/abs/2508.13892} \end{APACrefURL}
\PrintBackRefs{\CurrentBib}

\bibitem [\protect \citeauthoryear {%
Dessler%
\ \BBA {} Parker%
}{%
Dessler%
\ \BBA {} Parker%
}{%
{\protect \APACyear {1959}}%
}]{%
desslerparker1959}
\APACinsertmetastar {%
desslerparker1959}%
\begin{APACrefauthors}%
Dessler, A\BPBI J.%
\BCBT {}\ \BBA {} Parker, E\BPBI N.%
\end{APACrefauthors}%
\unskip\
\newblock
\APACrefYearMonthDay{1959}{}{}.
\newblock
{\BBOQ}\APACrefatitle {Hydromagnetic theory of geomagnetic storms} {Hydromagnetic theory of geomagnetic storms}.{\BBCQ}
\newblock
\APACjournalVolNumPages{Journal of Geophysical Research (1896-1977)}{64}{12}{2239-2252}.
\newblock
\begin{APACrefURL} \url{https://agupubs.onlinelibrary.wiley.com/doi/abs/10.1029/JZ064i012p02239} \end{APACrefURL}
\newblock
\begin{APACrefDOI} \doi{https://doi.org/10.1029/JZ064i012p02239} \end{APACrefDOI}
\PrintBackRefs{\CurrentBib}

\bibitem [\protect \citeauthoryear {%
Echer%
, Gonzalez%
\BCBL {}\ \BBA {} Tsurutani%
}{%
Echer%
\ \protect \BOthers {.}}{%
{\protect \APACyear {2008}}%
}]{%
echer2008}
\APACinsertmetastar {%
echer2008}%
\begin{APACrefauthors}%
Echer, E.%
, Gonzalez, W\BPBI D.%
\BCBL {}\ \BBA {} Tsurutani, B\BPBI T.%
\end{APACrefauthors}%
\unskip\
\newblock
\APACrefYearMonthDay{2008}{}{}.
\newblock
{\BBOQ}\APACrefatitle {Interplanetary conditions leading to superintense geomagnetic storms (Dst $\le$ -250 nT) during solar cycle 23} {Interplanetary conditions leading to superintense geomagnetic storms (dst $\le$ -250 nt) during solar cycle 23}.{\BBCQ}
\newblock
\APACjournalVolNumPages{Geophysical Research Letters}{35}{6}{}.
\newblock
\begin{APACrefURL} \url{https://agupubs.onlinelibrary.wiley.com/doi/abs/10.1029/2007GL031755} \end{APACrefURL}
\newblock
\begin{APACrefDOI} \doi{https://doi.org/10.1029/2007GL031755} \end{APACrefDOI}
\PrintBackRefs{\CurrentBib}

\bibitem [\protect \citeauthoryear {%
{Galvin}%
\ \protect \BOthers {.}}{%
{Galvin}%
\ \protect \BOthers {.}}{%
{\protect \APACyear {2008}}%
}]{%
galvin2008plastic}
\APACinsertmetastar {%
galvin2008plastic}%
\begin{APACrefauthors}%
{Galvin}, A\BPBI B.%
, {Kistler}, L\BPBI M.%
, {Popecki}, M\BPBI A.%
, {Farrugia}, C\BPBI J.%
, {Simunac}, K\BPBI D\BPBI C.%
, {Ellis}, L.%
\BDBL {}{Steinfeld}, D.%
\end{APACrefauthors}%
\unskip\
\newblock
\APACrefYearMonthDay{2008}{{\APACmonth{04}}}{}.
\newblock
{\BBOQ}\APACrefatitle {{The Plasma and Suprathermal Ion Composition (PLASTIC) Investigation on the STEREO Observatories}} {{The Plasma and Suprathermal Ion Composition (PLASTIC) Investigation on the STEREO Observatories}}.{\BBCQ}
\newblock
\APACjournalVolNumPages{\ssr}{136}{1-4}{437-486}.
\newblock
\begin{APACrefDOI} \doi{10.1007/s11214-007-9296-x} \end{APACrefDOI}
\PrintBackRefs{\CurrentBib}

\bibitem [\protect \citeauthoryear {%
Gonzalez%
\ \protect \BOthers {.}}{%
Gonzalez%
\ \protect \BOthers {.}}{%
{\protect \APACyear {1998}}%
}]{%
gonzalez1998}
\APACinsertmetastar {%
gonzalez1998}%
\begin{APACrefauthors}%
Gonzalez, W\BPBI D.%
, de Gonzalez, A\BPBI L\BPBI C.%
, Dal~Lago, A.%
, Tsurutani, B\BPBI T.%
, Arballo, J\BPBI K.%
, Lakhina, G\BPBI K.%
\BDBL {}Wu, S\BHBI T.%
\end{APACrefauthors}%
\unskip\
\newblock
\APACrefYearMonthDay{1998}{}{}.
\newblock
{\BBOQ}\APACrefatitle {Magnetic cloud field intensities and solar wind velocities} {Magnetic cloud field intensities and solar wind velocities}.{\BBCQ}
\newblock
\APACjournalVolNumPages{Geophysical Research Letters}{25}{7}{963-966}.
\newblock
\begin{APACrefURL} \url{https://agupubs.onlinelibrary.wiley.com/doi/abs/10.1029/98GL00703} \end{APACrefURL}
\newblock
\begin{APACrefDOI} \doi{https://doi.org/10.1029/98GL00703} \end{APACrefDOI}
\PrintBackRefs{\CurrentBib}

\bibitem [\protect \citeauthoryear {%
Gonzalez%
\ \protect \BOthers {.}}{%
Gonzalez%
\ \protect \BOthers {.}}{%
{\protect \APACyear {1994}}%
}]{%
gonzalez1994}
\APACinsertmetastar {%
gonzalez1994}%
\begin{APACrefauthors}%
Gonzalez, W\BPBI D.%
, Joselyn, J\BPBI A.%
, Kamide, Y.%
, Kroehl, H\BPBI W.%
, Rostoker, G.%
, Tsurutani, B\BPBI T.%
\BCBL {}\ \BBA {} Vasyliunas, V\BPBI M.%
\end{APACrefauthors}%
\unskip\
\newblock
\APACrefYearMonthDay{1994}{}{}.
\newblock
{\BBOQ}\APACrefatitle {What is a geomagnetic storm?} {What is a geomagnetic storm?}{\BBCQ}
\newblock
\APACjournalVolNumPages{Journal of Geophysical Research: Space Physics}{99}{A4}{5771-5792}.
\newblock
\begin{APACrefURL} \url{https://agupubs.onlinelibrary.wiley.com/doi/abs/10.1029/93JA02867} \end{APACrefURL}
\newblock
\begin{APACrefDOI} \doi{https://doi.org/10.1029/93JA02867} \end{APACrefDOI}
\PrintBackRefs{\CurrentBib}

\bibitem [\protect \citeauthoryear {%
{Good}%
\ \BBA {} {Forsyth}%
}{%
{Good}%
\ \BBA {} {Forsyth}%
}{%
{\protect \APACyear {2016}}%
}]{%
good2016interplanetary}
\APACinsertmetastar {%
good2016interplanetary}%
\begin{APACrefauthors}%
{Good}, S\BPBI W.%
\BCBT {}\ \BBA {} {Forsyth}, R\BPBI J.%
\end{APACrefauthors}%
\unskip\
\newblock
\APACrefYearMonthDay{2016}{{\APACmonth{01}}}{}.
\newblock
{\BBOQ}\APACrefatitle {{Interplanetary Coronal Mass Ejections Observed by MESSENGER and Venus Express}} {{Interplanetary Coronal Mass Ejections Observed by MESSENGER and Venus Express}}.{\BBCQ}
\newblock
\APACjournalVolNumPages{\solphys}{291}{1}{239-263}.
\newblock
\begin{APACrefDOI} \doi{10.1007/s11207-015-0828-3} \end{APACrefDOI}
\PrintBackRefs{\CurrentBib}

\bibitem [\protect \citeauthoryear {%
Gopalswamy%
\ \protect \BOthers {.}}{%
Gopalswamy%
\ \protect \BOthers {.}}{%
{\protect \APACyear {2022}}%
}]{%
gopalswamy2022}
\APACinsertmetastar {%
gopalswamy2022}%
\begin{APACrefauthors}%
Gopalswamy, N.%
, Yashiro, S.%
, Akiyama, S.%
, Xie, H.%
, Mäkelä, P.%
, Fok, M\BHBI C.%
\BCBL {}\ \BBA {} Ferradas, C\BPBI P.%
\end{APACrefauthors}%
\unskip\
\newblock
\APACrefYearMonthDay{2022}{}{}.
\newblock
{\BBOQ}\APACrefatitle {What Is Unusual About the Third Largest Geomagnetic Storm of Solar Cycle 24?} {What is unusual about the third largest geomagnetic storm of solar cycle 24?}{\BBCQ}
\newblock
\APACjournalVolNumPages{Journal of Geophysical Research: Space Physics}{127}{8}{e2022JA030404}.
\newblock
\begin{APACrefURL} \url{https://agupubs.onlinelibrary.wiley.com/doi/abs/10.1029/2022JA030404} \end{APACrefURL}
\newblock
\APACrefnote{e2022JA030404 2022JA030404}
\newblock
\begin{APACrefDOI} \doi{https://doi.org/10.1029/2022JA030404} \end{APACrefDOI}
\PrintBackRefs{\CurrentBib}

\bibitem [\protect \citeauthoryear {%
{Gosling}%
, {McComas}%
, {Phillips}%
\BCBL {}\ \BBA {} {Bame}%
}{%
{Gosling}%
\ \protect \BOthers {.}}{%
{\protect \APACyear {1991}}%
}]{%
gosling1991geo}
\APACinsertmetastar {%
gosling1991geo}%
\begin{APACrefauthors}%
{Gosling}, J\BPBI T.%
, {McComas}, D\BPBI J.%
, {Phillips}, J\BPBI L.%
\BCBL {}\ \BBA {} {Bame}, S\BPBI J.%
\end{APACrefauthors}%
\unskip\
\newblock
\APACrefYearMonthDay{1991}{{\APACmonth{05}}}{}.
\newblock
{\BBOQ}\APACrefatitle {{Geomagnetic activity associated with earth passage of interplanetary shock disturbances and coronal mass ejections}} {{Geomagnetic activity associated with earth passage of interplanetary shock disturbances and coronal mass ejections}}.{\BBCQ}
\newblock
\APACjournalVolNumPages{\jgr}{96}{A5}{7831-7839}.
\newblock
\begin{APACrefDOI} \doi{10.1029/91JA00316} \end{APACrefDOI}
\PrintBackRefs{\CurrentBib}

\bibitem [\protect \citeauthoryear {%
Harris%
\ \protect \BOthers {.}}{%
Harris%
\ \protect \BOthers {.}}{%
{\protect \APACyear {2020}}%
}]{%
harris2020numpy}
\APACinsertmetastar {%
harris2020numpy}%
\begin{APACrefauthors}%
Harris, C\BPBI R.%
, Millman, K\BPBI J.%
, van~der Walt, S\BPBI J.%
, Gommers, R.%
, Virtanen, P.%
, Cournapeau, D.%
\BDBL {}Oliphant, T\BPBI E.%
\end{APACrefauthors}%
\unskip\
\newblock
\APACrefYearMonthDay{2020}{{\APACmonth{09}}}{}.
\newblock
{\BBOQ}\APACrefatitle {Array programming with {NumPy}} {Array programming with {NumPy}}.{\BBCQ}
\newblock
\APACjournalVolNumPages{Nature}{585}{7825}{357--362}.
\newblock
\begin{APACrefURL} \url{https://doi.org/10.1038/s41586-020-2649-2} \end{APACrefURL}
\newblock
\begin{APACrefDOI} \doi{10.1038/s41586-020-2649-2} \end{APACrefDOI}
\PrintBackRefs{\CurrentBib}

\bibitem [\protect \citeauthoryear {%
{Henon}%
}{%
{Henon}%
}{%
{\protect \APACyear {1969}}%
}]{%
henon1969}
\APACinsertmetastar {%
henon1969}%
\begin{APACrefauthors}%
{Henon}, M.%
\end{APACrefauthors}%
\unskip\
\newblock
\APACrefYearMonthDay{1969}{{\APACmonth{02}}}{}.
\newblock
{\BBOQ}\APACrefatitle {{Numerical exploration of the restricted problem, V}} {{Numerical exploration of the restricted problem, V}}.{\BBCQ}
\newblock
\APACjournalVolNumPages{\aap}{1}{}{223-238}.
\PrintBackRefs{\CurrentBib}

\bibitem [\protect \citeauthoryear {%
Hunter%
}{%
Hunter%
}{%
{\protect \APACyear {2007}}%
}]{%
hunter2007matplotlib}
\APACinsertmetastar {%
hunter2007matplotlib}%
\begin{APACrefauthors}%
Hunter, J\BPBI D.%
\end{APACrefauthors}%
\unskip\
\newblock
\APACrefYearMonthDay{2007}{}{}.
\newblock
{\BBOQ}\APACrefatitle {Matplotlib: A 2D graphics environment} {Matplotlib: A 2d graphics environment}.{\BBCQ}
\newblock
\APACjournalVolNumPages{Computing in Science \& Engineering}{9}{3}{90--95}.
\newblock
\begin{APACrefDOI} \doi{10.1109/MCSE.2007.55} \end{APACrefDOI}
\PrintBackRefs{\CurrentBib}

\bibitem [\protect \citeauthoryear {%
Iyemori%
}{%
Iyemori%
}{%
{\protect \APACyear {1990}}%
}]{%
iyemori1990}
\APACinsertmetastar {%
iyemori1990}%
\begin{APACrefauthors}%
Iyemori, T.%
\end{APACrefauthors}%
\unskip\
\newblock
\APACrefYearMonthDay{1990}{}{}.
\newblock
{\BBOQ}\APACrefatitle {Storm-Time Magnetospheric Currents Inferred from Mid-Latitude Geomagnetic Field Variations} {Storm-time magnetospheric currents inferred from mid-latitude geomagnetic field variations}.{\BBCQ}
\newblock
\APACjournalVolNumPages{Journal of geomagnetism and geoelectricity}{42}{11}{1249-1265}.
\newblock
\begin{APACrefDOI} \doi{10.5636/jgg.42.1249} \end{APACrefDOI}
\PrintBackRefs{\CurrentBib}

\bibitem [\protect \citeauthoryear {%
{Kaiser}%
\ \protect \BOthers {.}}{%
{Kaiser}%
\ \protect \BOthers {.}}{%
{\protect \APACyear {2008}}%
}]{%
kaiser2008stereo}
\APACinsertmetastar {%
kaiser2008stereo}%
\begin{APACrefauthors}%
{Kaiser}, M\BPBI L.%
, {Kucera}, T\BPBI A.%
, {Davila}, J\BPBI M.%
, {St. Cyr}, O\BPBI C.%
, {Guhathakurta}, M.%
\BCBL {}\ \BBA {} {Christian}, E.%
\end{APACrefauthors}%
\unskip\
\newblock
\APACrefYearMonthDay{2008}{{\APACmonth{04}}}{}.
\newblock
{\BBOQ}\APACrefatitle {{The STEREO Mission: An Introduction}} {{The STEREO Mission: An Introduction}}.{\BBCQ}
\newblock
\APACjournalVolNumPages{\ssr}{136}{1-4}{5-16}.
\newblock
\begin{APACrefDOI} \doi{10.1007/s11214-007-9277-0} \end{APACrefDOI}
\PrintBackRefs{\CurrentBib}

\bibitem [\protect \citeauthoryear {%
{Kay}%
\ \protect \BOthers {.}}{%
{Kay}%
\ \protect \BOthers {.}}{%
{\protect \APACyear {2024}}%
}]{%
kay2024CMEarrivals}
\APACinsertmetastar {%
kay2024CMEarrivals}%
\begin{APACrefauthors}%
{Kay}, C.%
, {Palmerio}, E.%
, {Riley}, P.%
, {Mays}, M\BPBI L.%
, {Nieves-Chinchilla}, T.%
, {Romano}, M.%
\BDBL {}{Chulaki}, A.%
\end{APACrefauthors}%
\unskip\
\newblock
\APACrefYearMonthDay{2024}{{\APACmonth{07}}}{}.
\newblock
{\BBOQ}\APACrefatitle {{Updating Measures of CME Arrival Time Errors}} {{Updating Measures of CME Arrival Time Errors}}.{\BBCQ}
\newblock
\APACjournalVolNumPages{Space Weather}{22}{7}{e2024SW003951}.
\newblock
\begin{APACrefDOI} \doi{10.1029/2024SW003951} \end{APACrefDOI}
\PrintBackRefs{\CurrentBib}

\bibitem [\protect \citeauthoryear {%
{Kilpua}%
, {Koskinen}%
\BCBL {}\ \BBA {} {Pulkkinen}%
}{%
{Kilpua}%
\ \protect \BOthers {.}}{%
{\protect \APACyear {2017}}%
}]{%
kilpua2017coronal}
\APACinsertmetastar {%
kilpua2017coronal}%
\begin{APACrefauthors}%
{Kilpua}, E.%
, {Koskinen}, H\BPBI E\BPBI J.%
\BCBL {}\ \BBA {} {Pulkkinen}, T\BPBI I.%
\end{APACrefauthors}%
\unskip\
\newblock
\APACrefYearMonthDay{2017}{{\APACmonth{11}}}{}.
\newblock
{\BBOQ}\APACrefatitle {{Coronal mass ejections and their sheath regions in interplanetary space}} {{Coronal mass ejections and their sheath regions in interplanetary space}}.{\BBCQ}
\newblock
\APACjournalVolNumPages{\lrsp}{14}{1}{5}.
\newblock
\begin{APACrefDOI} \doi{10.1007/s41116-017-0009-6} \end{APACrefDOI}
\PrintBackRefs{\CurrentBib}

\bibitem [\protect \citeauthoryear {%
Kilpua%
, Lugaz%
, Mays%
\BCBL {}\ \BBA {} Temmer%
}{%
Kilpua%
\ \protect \BOthers {.}}{%
{\protect \APACyear {2019}}%
}]{%
kilpua2019review}
\APACinsertmetastar {%
kilpua2019review}%
\begin{APACrefauthors}%
Kilpua, E.%
, Lugaz, N.%
, Mays, M\BPBI L.%
\BCBL {}\ \BBA {} Temmer, M.%
\end{APACrefauthors}%
\unskip\
\newblock
\APACrefYearMonthDay{2019}{}{}.
\newblock
{\BBOQ}\APACrefatitle {Forecasting the Structure and Orientation of Earthbound Coronal Mass Ejections} {Forecasting the structure and orientation of earthbound coronal mass ejections}.{\BBCQ}
\newblock
\APACjournalVolNumPages{Space Weather}{17}{4}{498-526}.
\newblock
\begin{APACrefURL} \url{https://agupubs.onlinelibrary.wiley.com/doi/abs/10.1029/2018SW001944} \end{APACrefURL}
\newblock
\begin{APACrefDOI} \doi{https://doi.org/10.1029/2018SW001944} \end{APACrefDOI}
\PrintBackRefs{\CurrentBib}

\bibitem [\protect \citeauthoryear {%
King%
\ \BBA {} Papitashvili%
}{%
King%
\ \BBA {} Papitashvili%
}{%
{\protect \APACyear {2005}}%
}]{%
king2025OMNI}
\APACinsertmetastar {%
king2025OMNI}%
\begin{APACrefauthors}%
King, J\BPBI H.%
\BCBT {}\ \BBA {} Papitashvili, N\BPBI E.%
\end{APACrefauthors}%
\unskip\
\newblock
\APACrefYearMonthDay{2005}{}{}.
\newblock
{\BBOQ}\APACrefatitle {Solar wind spatial scales in and comparisons of hourly Wind and ACE plasma and magnetic field data} {Solar wind spatial scales in and comparisons of hourly wind and ace plasma and magnetic field data}.{\BBCQ}
\newblock
\APACjournalVolNumPages{Journal of Geophysical Research: Space Physics}{110}{A2}{}.
\newblock
\begin{APACrefURL} \url{https://agupubs.onlinelibrary.wiley.com/doi/abs/10.1029/2004JA010649} \end{APACrefURL}
\newblock
\begin{APACrefDOI} \doi{https://doi.org/10.1029/2004JA010649} \end{APACrefDOI}
\PrintBackRefs{\CurrentBib}

\bibitem [\protect \citeauthoryear {%
{Klein}%
\ \BBA {} {Burlaga}%
}{%
{Klein}%
\ \BBA {} {Burlaga}%
}{%
{\protect \APACyear {1982}}%
}]{%
klein1982interplanetary}
\APACinsertmetastar {%
klein1982interplanetary}%
\begin{APACrefauthors}%
{Klein}, L\BPBI W.%
\BCBT {}\ \BBA {} {Burlaga}, L\BPBI F.%
\end{APACrefauthors}%
\unskip\
\newblock
\APACrefYearMonthDay{1982}{{\APACmonth{02}}}{}.
\newblock
{\BBOQ}\APACrefatitle {{Interplanetary magnetic clouds at 1 AU}} {{Interplanetary magnetic clouds at 1 AU}}.{\BBCQ}
\newblock
\APACjournalVolNumPages{\jgr}{87}{A2}{613-624}.
\newblock
\begin{APACrefDOI} \doi{10.1029/JA087iA02p00613} \end{APACrefDOI}
\PrintBackRefs{\CurrentBib}

\bibitem [\protect \citeauthoryear {%
Kruchten%
, Seier%
\BCBL {}\ \BBA {} Parmer%
}{%
Kruchten%
\ \protect \BOthers {.}}{%
{\protect \APACyear {2026}}%
}]{%
plotly2026}
\APACinsertmetastar {%
plotly2026}%
\begin{APACrefauthors}%
Kruchten, N.%
, Seier, A.%
\BCBL {}\ \BBA {} Parmer, C.%
\end{APACrefauthors}%
\unskip\
\newblock
\APACrefYearMonthDay{2026}{}{}.
\newblock
\APACrefbtitle {An interactive, open-source, and browser-based graphing library for Python.} {An interactive, open-source, and browser-based graphing library for python.}
\newblock
\begin{APACrefURL} \url{"https://github.com/plotly/plotly.py"} \end{APACrefURL}
\newblock
\begin{APACrefDOI} \doi{10.5281/zenodo.14503524} \end{APACrefDOI}
\PrintBackRefs{\CurrentBib}

\bibitem [\protect \citeauthoryear {%
{Laker}%
\ \protect \BOthers {.}}{%
{Laker}%
\ \protect \BOthers {.}}{%
{\protect \APACyear {2024}}%
}]{%
laker_2024}
\APACinsertmetastar {%
laker_2024}%
\begin{APACrefauthors}%
{Laker}, R.%
, {Horbury}, T\BPBI S.%
, {O'Brien}, H.%
, {Fauchon-Jones}, E\BPBI J.%
, {Angelini}, V.%
, {Fargette}, N.%
\BDBL {}{Dumbovi{\'c}}, M.%
\end{APACrefauthors}%
\unskip\
\newblock
\APACrefYearMonthDay{2024}{{\APACmonth{02}}}{}.
\newblock
{\BBOQ}\APACrefatitle {{Using Solar Orbiter as an Upstream Solar Wind Monitor for Real Time Space Weather Predictions}} {{Using Solar Orbiter as an Upstream Solar Wind Monitor for Real Time Space Weather Predictions}}.{\BBCQ}
\newblock
\APACjournalVolNumPages{Space Weather}{22}{2}{e2023SW003628}.
\newblock
\begin{APACrefDOI} \doi{10.1029/2023SW003628} \end{APACrefDOI}
\PrintBackRefs{\CurrentBib}

\bibitem [\protect \citeauthoryear {%
Liemohn%
\ \protect \BOthers {.}}{%
Liemohn%
\ \protect \BOthers {.}}{%
{\protect \APACyear {2018}}%
}]{%
Liemohn2018}
\APACinsertmetastar {%
Liemohn2018}%
\begin{APACrefauthors}%
Liemohn, M\BPBI W.%
, McCollough, J\BPBI P.%
, Jordanova, V\BPBI K.%
, Ngwira, C\BPBI M.%
, Morley, S\BPBI K.%
, Cid, C.%
\BDBL {}Vasile, R.%
\end{APACrefauthors}%
\unskip\
\newblock
\APACrefYearMonthDay{2018}{}{}.
\newblock
{\BBOQ}\APACrefatitle {Model Evaluation Guidelines for Geomagnetic Index Predictions} {Model evaluation guidelines for geomagnetic index predictions}.{\BBCQ}
\newblock
\APACjournalVolNumPages{Space Weather}{16}{12}{2079-2102}.
\newblock
\begin{APACrefURL} \url{https://agupubs.onlinelibrary.wiley.com/doi/abs/10.1029/2018SW002067} \end{APACrefURL}
\newblock
\begin{APACrefDOI} \doi{https://doi.org/10.1029/2018SW002067} \end{APACrefDOI}
\PrintBackRefs{\CurrentBib}

\bibitem [\protect \citeauthoryear {%
{Lindsay}%
, {Russell}%
\BCBL {}\ \BBA {} {Luhmann}%
}{%
{Lindsay}%
\ \protect \BOthers {.}}{%
{\protect \APACyear {1999}}%
}]{%
lindsay1999dst}
\APACinsertmetastar {%
lindsay1999dst}%
\begin{APACrefauthors}%
{Lindsay}, G\BPBI M.%
, {Russell}, C\BPBI T.%
\BCBL {}\ \BBA {} {Luhmann}, J\BPBI G.%
\end{APACrefauthors}%
\unskip\
\newblock
\APACrefYearMonthDay{1999}{{\APACmonth{05}}}{}.
\newblock
{\BBOQ}\APACrefatitle {{Predictability of Dst index based upon solar wind conditions monitored inside 1 AU}} {{Predictability of Dst index based upon solar wind conditions monitored inside 1 AU}}.{\BBCQ}
\newblock
\APACjournalVolNumPages{\jgr}{104}{A5}{10335-10344}.
\newblock
\begin{APACrefDOI} \doi{10.1029/1999JA900010} \end{APACrefDOI}
\PrintBackRefs{\CurrentBib}

\bibitem [\protect \citeauthoryear {%
Liu%
, Hu%
, Zhao%
, Chen%
\BCBL {}\ \BBA {} Wang%
}{%
Liu%
\ \protect \BOthers {.}}{%
{\protect \APACyear {2024}}%
}]{%
liu2024may}
\APACinsertmetastar {%
liu2024may}%
\begin{APACrefauthors}%
Liu, Y\BPBI D.%
, Hu, H.%
, Zhao, X.%
, Chen, C.%
\BCBL {}\ \BBA {} Wang, R.%
\end{APACrefauthors}%
\unskip\
\newblock
\APACrefYearMonthDay{2024}{}{}.
\newblock
\APACrefbtitle {A Pileup of Coronal Mass Ejections Produced the Largest Geomagnetic Storm in Two Decades.} {A pileup of coronal mass ejections produced the largest geomagnetic storm in two decades.}
\newblock
\begin{APACrefURL} \url{https://arxiv.org/abs/2409.11492} \end{APACrefURL}
\PrintBackRefs{\CurrentBib}

\bibitem [\protect \citeauthoryear {%
Loto'aniu%
\ \protect \BOthers {.}}{%
Loto'aniu%
\ \protect \BOthers {.}}{%
{\protect \APACyear {2022}}%
}]{%
lotoaniu2022}
\APACinsertmetastar {%
lotoaniu2022}%
\begin{APACrefauthors}%
Loto'aniu, P\BPBI T\BPBI M.%
, Romich, K.%
, Rowland, W.%
, Codrescu, S.%
, Biesecker, D.%
, Johnson, J.%
\BDBL {}Stevens, M.%
\end{APACrefauthors}%
\unskip\
\newblock
\APACrefYearMonthDay{2022}{}{}.
\newblock
{\BBOQ}\APACrefatitle {Validation of the DSCOVR Spacecraft Mission Space Weather Solar Wind Products} {Validation of the dscovr spacecraft mission space weather solar wind products}.{\BBCQ}
\newblock
\APACjournalVolNumPages{Space Weather}{20}{10}{e2022SW003085}.
\newblock
\begin{APACrefURL} \url{https://agupubs.onlinelibrary.wiley.com/doi/abs/10.1029/2022SW003085} \end{APACrefURL}
\newblock
\APACrefnote{e2022SW003085 2022SW003085}
\newblock
\begin{APACrefDOI} \doi{https://doi.org/10.1029/2022SW003085} \end{APACrefDOI}
\PrintBackRefs{\CurrentBib}

\bibitem [\protect \citeauthoryear {%
Lugaz%
\ \protect \BOthers {.}}{%
Lugaz%
\ \protect \BOthers {.}}{%
{\protect \APACyear {2025}}%
}]{%
lugaz2025subL1}
\APACinsertmetastar {%
lugaz2025subL1}%
\begin{APACrefauthors}%
Lugaz, N.%
, Al-Haddad, N.%
, Zhuang, B.%
, Möstl, C.%
, Davies, E\BPBI E.%
, Farrugia, C\BPBI J.%
\BDBL {}Galvin, A\BPBI B.%
\end{APACrefauthors}%
\unskip\
\newblock
\APACrefYearMonthDay{2025}{}{}.
\newblock
{\BBOQ}\APACrefatitle {The Need for a Sub-L1 Space Weather Research Mission: Current Knowledge Gaps on Coronal Mass Ejections} {The need for a sub-l1 space weather research mission: Current knowledge gaps on coronal mass ejections}.{\BBCQ}
\newblock
\APACjournalVolNumPages{Space Weather}{23}{2}{e2024SW004189}.
\newblock
\begin{APACrefURL} \url{https://agupubs.onlinelibrary.wiley.com/doi/abs/10.1029/2024SW004189} \end{APACrefURL}
\newblock
\APACrefnote{e2024SW004189 2024SW004189}
\newblock
\begin{APACrefDOI} \doi{https://doi.org/10.1029/2024SW004189} \end{APACrefDOI}
\PrintBackRefs{\CurrentBib}

\bibitem [\protect \citeauthoryear {%
Lugaz%
, Lee%
\BCBL {}\ \protect \BOthers {.}}{%
Lugaz%
, Lee%
\BCBL {}\ \protect \BOthers {.}}{%
{\protect \APACyear {2024}}%
}]{%
Lugaz2024mission}
\APACinsertmetastar {%
Lugaz2024mission}%
\begin{APACrefauthors}%
Lugaz, N.%
, Lee, C\BPBI O.%
, Al-Haddad, N.%
, Lillis, R\BPBI J.%
, Jian, L\BPBI K.%
, Curtis, D\BPBI W.%
\BDBL {}Nieves-Chinchilla, T.%
\end{APACrefauthors}%
\unskip\
\newblock
\APACrefYearMonthDay{2024}{Sep}{20}.
\newblock
{\BBOQ}\APACrefatitle {The Need for Near-Earth Multi-Spacecraft Heliospheric Measurements and an Explorer Mission to Investigate Interplanetary Structures and Transients in the Near-Earth Heliosphere} {The need for near-earth multi-spacecraft heliospheric measurements and an explorer mission to investigate interplanetary structures and transients in the near-earth heliosphere}.{\BBCQ}
\newblock
\APACjournalVolNumPages{Space Science Reviews}{220}{7}{73}.
\newblock
\begin{APACrefURL} \url{https://doi.org/10.1007/s11214-024-01108-8} \end{APACrefURL}
\newblock
\begin{APACrefDOI} \doi{10.1007/s11214-024-01108-8} \end{APACrefDOI}
\PrintBackRefs{\CurrentBib}

\bibitem [\protect \citeauthoryear {%
Lugaz%
, Zhuang%
\BCBL {}\ \protect \BOthers {.}}{%
Lugaz%
, Zhuang%
\BCBL {}\ \protect \BOthers {.}}{%
{\protect \APACyear {2024}}%
}]{%
lugaz2024MEwidth}
\APACinsertmetastar {%
lugaz2024MEwidth}%
\begin{APACrefauthors}%
Lugaz, N.%
, Zhuang, B.%
, Scolini, C.%
, Al-Haddad, N.%
, Farrugia, C\BPBI J.%
, Winslow, R\BPBI M.%
\BDBL {}Galvin, A\BPBI B.%
\end{APACrefauthors}%
\unskip\
\newblock
\APACrefYearMonthDay{2024}{feb}{}.
\newblock
{\BBOQ}\APACrefatitle {The Width of Magnetic Ejecta Measured near 1 au: Lessons from STEREO-A Measurements in 2021–2022} {The width of magnetic ejecta measured near 1 au: Lessons from stereo-a measurements in 2021–2022}.{\BBCQ}
\newblock
\APACjournalVolNumPages{The Astrophysical Journal}{962}{2}{193}.
\newblock
\begin{APACrefURL} \url{https://dx.doi.org/10.3847/1538-4357/ad17b9} \end{APACrefURL}
\newblock
\begin{APACrefDOI} \doi{10.3847/1538-4357/ad17b9} \end{APACrefDOI}
\PrintBackRefs{\CurrentBib}

\bibitem [\protect \citeauthoryear {%
Luhmann%
\ \protect \BOthers {.}}{%
Luhmann%
\ \protect \BOthers {.}}{%
{\protect \APACyear {2008}}%
}]{%
luhmann2008impact}
\APACinsertmetastar {%
luhmann2008impact}%
\begin{APACrefauthors}%
Luhmann, J\BPBI G.%
, Curtis, D\BPBI W.%
, Schroeder, P.%
, McCauley, J.%
, Lin, R\BPBI P.%
, Larson, D\BPBI E.%
\BDBL {}Gosling, J\BPBI T.%
\end{APACrefauthors}%
\unskip\
\newblock
\APACrefYearMonthDay{2008}{Apr}{01}.
\newblock
{\BBOQ}\APACrefatitle {STEREO IMPACT Investigation Goals, Measurements, and Data Products Overview} {Stereo impact investigation goals, measurements, and data products overview}.{\BBCQ}
\newblock
\APACjournalVolNumPages{Space Science Reviews}{136}{1}{117-184}.
\newblock
\begin{APACrefURL} \url{https://doi.org/10.1007/s11214-007-9170-x} \end{APACrefURL}
\newblock
\begin{APACrefDOI} \doi{10.1007/s11214-007-9170-x} \end{APACrefDOI}
\PrintBackRefs{\CurrentBib}

\bibitem [\protect \citeauthoryear {%
{Manchester}%
\ \protect \BOthers {.}}{%
{Manchester}%
\ \protect \BOthers {.}}{%
{\protect \APACyear {2017}}%
}]{%
manchester2017physical}
\APACinsertmetastar {%
manchester2017physical}%
\begin{APACrefauthors}%
{Manchester}, W.%
, {Kilpua}, E\BPBI K\BPBI J.%
, {Liu}, Y\BPBI D.%
, {Lugaz}, N.%
, {Riley}, P.%
, {T{\"o}r{\"o}k}, T.%
\BCBL {}\ \BBA {} {Vr{\v{s}}nak}, B.%
\end{APACrefauthors}%
\unskip\
\newblock
\APACrefYearMonthDay{2017}{{\APACmonth{11}}}{}.
\newblock
{\BBOQ}\APACrefatitle {{The Physical Processes of CME/ICME Evolution}} {{The Physical Processes of CME/ICME Evolution}}.{\BBCQ}
\newblock
\APACjournalVolNumPages{\ssr}{212}{3-4}{1159-1219}.
\newblock
\begin{APACrefDOI} \doi{10.1007/s11214-017-0394-0} \end{APACrefDOI}
\PrintBackRefs{\CurrentBib}

\bibitem [\protect \citeauthoryear {%
{McComas}%
\ \protect \BOthers {.}}{%
{McComas}%
\ \protect \BOthers {.}}{%
{\protect \APACyear {1998}}%
}]{%
mccomas1998solar}
\APACinsertmetastar {%
mccomas1998solar}%
\begin{APACrefauthors}%
{McComas}, D\BPBI J.%
, {Bame}, S\BPBI J.%
, {Barker}, P.%
, {Feldman}, W\BPBI C.%
, {Phillips}, J\BPBI L.%
, {Riley}, P.%
\BCBL {}\ \BBA {} {Griffee}, J\BPBI W.%
\end{APACrefauthors}%
\unskip\
\newblock
\APACrefYearMonthDay{1998}{{\APACmonth{07}}}{}.
\newblock
{\BBOQ}\APACrefatitle {{Solar Wind Electron Proton Alpha Monitor (SWEPAM) for the Advanced Composition Explorer}} {{Solar Wind Electron Proton Alpha Monitor (SWEPAM) for the Advanced Composition Explorer}}.{\BBCQ}
\newblock
\APACjournalVolNumPages{\ssr}{86}{}{563-612}.
\newblock
\begin{APACrefDOI} \doi{10.1023/A:1005040232597} \end{APACrefDOI}
\PrintBackRefs{\CurrentBib}

\bibitem [\protect \citeauthoryear {%
{Milo{\v{s}}i{\'c}}%
, {Temmer}%
, {Heinemann}%
, {Hofmeister}%
\BCBL {}\ \BBA {} {Asvestari}%
}{%
{Milo{\v{s}}i{\'c}}%
\ \protect \BOthers {.}}{%
{\protect \APACyear {2025}}%
}]{%
milosic2025}
\APACinsertmetastar {%
milosic2025}%
\begin{APACrefauthors}%
{Milo{\v{s}}i{\'c}}, D.%
, {Temmer}, M.%
, {Heinemann}, S\BPBI G.%
, {Hofmeister}, S.%
\BCBL {}\ \BBA {} {Asvestari}, E.%
\end{APACrefauthors}%
\unskip\
\newblock
\APACrefYearMonthDay{2025}{{\APACmonth{07}}}{}.
\newblock
{\BBOQ}\APACrefatitle {{Case study on the evolution of corotating interaction regions for the ``smiley coronal holes'': 0.3 to 1 AU}} {{Case study on the evolution of corotating interaction regions for the ``smiley coronal holes'': 0.3 to 1 AU}}.{\BBCQ}
\newblock
\APACjournalVolNumPages{\aap}{699}{}{A267}.
\newblock
\begin{APACrefDOI} \doi{10.1051/0004-6361/202453096} \end{APACrefDOI}
\PrintBackRefs{\CurrentBib}

\bibitem [\protect \citeauthoryear {%
Morley%
}{%
Morley%
}{%
{\protect \APACyear {2020}}%
}]{%
morley2020}
\APACinsertmetastar {%
morley2020}%
\begin{APACrefauthors}%
Morley, S\BPBI K.%
\end{APACrefauthors}%
\unskip\
\newblock
\APACrefYearMonthDay{2020}{}{}.
\newblock
{\BBOQ}\APACrefatitle {Challenges and Opportunities in Magnetospheric Space Weather Prediction} {Challenges and opportunities in magnetospheric space weather prediction}.{\BBCQ}
\newblock
\APACjournalVolNumPages{Space Weather}{18}{3}{e2018SW002108}.
\newblock
\begin{APACrefURL} \url{https://agupubs.onlinelibrary.wiley.com/doi/abs/10.1029/2018SW002108} \end{APACrefURL}
\newblock
\APACrefnote{e2018SW002108 10.1029/2018SW002108}
\newblock
\begin{APACrefDOI} \doi{https://doi.org/10.1029/2018SW002108} \end{APACrefDOI}
\PrintBackRefs{\CurrentBib}

\bibitem [\protect \citeauthoryear {%
M{\"o}stl%
, Davies%
\BCBL {}\ \BBA {} Weiler%
}{%
M{\"o}stl%
\ \protect \BOthers {.}}{%
{\protect \APACyear {2025}}%
}]{%
moestl_icme_figshare}
\APACinsertmetastar {%
moestl_icme_figshare}%
\begin{APACrefauthors}%
M{\"o}stl, C.%
, Davies, E.%
\BCBL {}\ \BBA {} Weiler, E.%
\end{APACrefauthors}%
\unskip\
\newblock
\APACrefYearMonthDay{2025}{}{}.
\newblock
\APACrefbtitle {{HELIO4CAST Interplanetary Coronal Mass Ejection Catalog v2.3}.} {{HELIO4CAST Interplanetary Coronal Mass Ejection Catalog v2.3}.}
\newblock
\APACaddressPublisher{}{figshare}.
\newblock
\begin{APACrefURL} \url{https://doi.org/10.6084/m9.figshare.6356420.v24} \end{APACrefURL}
\newblock
\begin{APACrefDOI} \doi{10.6084/m9.figshare.6356420.v24} \end{APACrefDOI}
\PrintBackRefs{\CurrentBib}

\bibitem [\protect \citeauthoryear {%
Möstl%
\ \protect \BOthers {.}}{%
Möstl%
\ \protect \BOthers {.}}{%
{\protect \APACyear {2026}}%
}]{%
moestl2026icmecat}
\APACinsertmetastar {%
moestl2026icmecat}%
\begin{APACrefauthors}%
Möstl, C.%
, Davies, E\BPBI E.%
, Weiler, E.%
, Amerstorfer, U\BPBI V.%
, Weiss, A\BPBI J.%
, Rüdisser, H\BPBI T.%
\BDBL {}Heyner, D.%
\end{APACrefauthors}%
\unskip\
\newblock
\APACrefYearMonthDay{2026}{}{}.
\newblock
\APACrefbtitle {On the magnetic field evolution of interplanetary coronal mass ejections from 0.07 to 5.4 au.} {On the magnetic field evolution of interplanetary coronal mass ejections from 0.07 to 5.4 au.}
\newblock
\begin{APACrefURL} \url{https://arxiv.org/abs/2512.04730} \end{APACrefURL}
\PrintBackRefs{\CurrentBib}

\bibitem [\protect \citeauthoryear {%
Niehof%
, Morley%
, Welling%
\BCBL {}\ \BBA {} Larsen%
}{%
Niehof%
\ \protect \BOthers {.}}{%
{\protect \APACyear {2022}}%
}]{%
niehof2022spacepy}
\APACinsertmetastar {%
niehof2022spacepy}%
\begin{APACrefauthors}%
Niehof, J\BPBI T.%
, Morley, S\BPBI K.%
, Welling, D\BPBI T.%
\BCBL {}\ \BBA {} Larsen, B\BPBI A.%
\end{APACrefauthors}%
\unskip\
\newblock
\APACrefYearMonthDay{2022}{}{}.
\newblock
{\BBOQ}\APACrefatitle {The SpacePy space science package at 12 years} {The spacepy space science package at 12 years}.{\BBCQ}
\newblock
\APACjournalVolNumPages{Frontiers in Astronomy and Space Sciences}{9}{}{}.
\newblock
\begin{APACrefDOI} \doi{10.3389/fspas.2022.1023612} \end{APACrefDOI}
\PrintBackRefs{\CurrentBib}

\bibitem [\protect \citeauthoryear {%
{Nose}%
, {Iyemori}%
, {Sugiura}%
\BCBL {}\ \BBA {} {Kamei}%
}{%
{Nose}%
\ \protect \BOthers {.}}{%
{\protect \APACyear {2015}}%
}]{%
kyoto_dst}
\APACinsertmetastar {%
kyoto_dst}%
\begin{APACrefauthors}%
{Nose}, M.%
, {Iyemori}, T.%
, {Sugiura}, M.%
\BCBL {}\ \BBA {} {Kamei}, T.%
\end{APACrefauthors}%
\unskip\
\newblock
\APACrefYearMonthDay{2015}{}{}.
\newblock
\APACrefbtitle {Geomagnetic Dst index.} {Geomagnetic dst index.}
\newblock
\APACaddressPublisher{}{World Data Center for Geomagnetism, Kyoto}.
\newblock
\begin{APACrefURL} \url{https://isds-datadoi.nict.go.jp/wds/10.17593__14515-74000.html} \end{APACrefURL}
\newblock
\begin{APACrefDOI} \doi{10.17593/14515-74000} \end{APACrefDOI}
\PrintBackRefs{\CurrentBib}

\bibitem [\protect \citeauthoryear {%
{Odstrcil}%
}{%
{Odstrcil}%
}{%
{\protect \APACyear {2003}}%
}]{%
odstrcil2003enlil}
\APACinsertmetastar {%
odstrcil2003enlil}%
\begin{APACrefauthors}%
{Odstrcil}, D.%
\end{APACrefauthors}%
\unskip\
\newblock
\APACrefYearMonthDay{2003}{{\APACmonth{08}}}{}.
\newblock
{\BBOQ}\APACrefatitle {{Modeling 3-D solar wind structure}} {{Modeling 3-D solar wind structure}}.{\BBCQ}
\newblock
\APACjournalVolNumPages{Advances in Space Research}{32}{4}{497-506}.
\newblock
\begin{APACrefDOI} \doi{10.1016/S0273-1177(03)00332-6} \end{APACrefDOI}
\PrintBackRefs{\CurrentBib}

\bibitem [\protect \citeauthoryear {%
Palmerio%
}{%
Palmerio%
}{%
{\protect \APACyear {2025}}%
}]{%
palmerio2025_commentary_weiler}
\APACinsertmetastar {%
palmerio2025_commentary_weiler}%
\begin{APACrefauthors}%
Palmerio, E.%
\end{APACrefauthors}%
\unskip\
\newblock
\APACrefYearMonthDay{2025}{}{}.
\newblock
{\BBOQ}\APACrefatitle {Monitoring the Solar Wind Before It Reaches L1} {Monitoring the solar wind before it reaches l1}.{\BBCQ}
\newblock
\APACjournalVolNumPages{Space Weather}{23}{11}{e2025SW004452}.
\newblock
\begin{APACrefURL} \url{https://agupubs.onlinelibrary.wiley.com/doi/abs/10.1029/2025SW004452} \end{APACrefURL}
\newblock
\APACrefnote{e2025SW004452 2025SW004452}
\newblock
\begin{APACrefDOI} \doi{https://doi.org/10.1029/2025SW004452} \end{APACrefDOI}
\PrintBackRefs{\CurrentBib}

\bibitem [\protect \citeauthoryear {%
pandas~development team%
}{%
pandas~development team%
}{%
{\protect \APACyear {2020}}%
}]{%
reback2020pandas}
\APACinsertmetastar {%
reback2020pandas}%
\begin{APACrefauthors}%
pandas~development team, T.%
\end{APACrefauthors}%
\unskip\
\newblock
\APACrefYearMonthDay{2020}{{\APACmonth{02}}}{}.
\newblock
\APACrefbtitle {pandas-dev/pandas: Pandas.} {pandas-dev/pandas: Pandas.}
\newblock
\APACaddressPublisher{}{Zenodo}.
\newblock
\begin{APACrefURL} \url{https://doi.org/10.5281/zenodo.3509134} \end{APACrefURL}
\newblock
\begin{APACrefDOI} \doi{10.5281/zenodo.3509134} \end{APACrefDOI}
\PrintBackRefs{\CurrentBib}

\bibitem [\protect \citeauthoryear {%
Papitashvili%
\ \BBA {} King%
}{%
Papitashvili%
\ \BBA {} King%
}{%
{\protect \APACyear {2020}}%
}]{%
omnimin}
\APACinsertmetastar {%
omnimin}%
\begin{APACrefauthors}%
Papitashvili, N\BPBI E.%
\BCBT {}\ \BBA {} King, J\BPBI H.%
\end{APACrefauthors}%
\unskip\
\newblock
\APACrefYearMonthDay{2020}{}{}.
\newblock
\APACrefbtitle {OMNI 1-min Data.} {Omni 1-min data.}
\newblock
\APACaddressPublisher{}{NASA Space Physics Data Facility}.
\newblock
\begin{APACrefURL} \url{https://spdf.gsfc.nasa.gov/pub/data/omni/high_res_omni/} \end{APACrefURL}
\newblock
\begin{APACrefDOI} \doi{10.48322/45bb-8792} \end{APACrefDOI}
\PrintBackRefs{\CurrentBib}

\bibitem [\protect \citeauthoryear {%
Pokharia%
, Prasad%
, Bhoj%
\BCBL {}\ \BBA {} Mathpal%
}{%
Pokharia%
\ \protect \BOthers {.}}{%
{\protect \APACyear {2018}}%
}]{%
pokharia2018}
\APACinsertmetastar {%
pokharia2018}%
\begin{APACrefauthors}%
Pokharia, M.%
, Prasad, L.%
, Bhoj, C.%
\BCBL {}\ \BBA {} Mathpal, C.%
\end{APACrefauthors}%
\unskip\
\newblock
\APACrefYearMonthDay{2018}{}{}.
\newblock
{\BBOQ}\APACrefatitle {Study of Geomagnetic Storms and Solar and Interplanetary Parameters for Solar Cycles 22 and 24} {Study of geomagnetic storms and solar and interplanetary parameters for solar cycles 22 and 24}.{\BBCQ}
\newblock
\APACjournalVolNumPages{Solar Physics}{293}{9}{126}.
\newblock
\begin{APACrefURL} \url{https://doi.org/10.1007/s11207-018-1345-y} \end{APACrefURL}
\newblock
\begin{APACrefDOI} \doi{10.1007/s11207-018-1345-y} \end{APACrefDOI}
\PrintBackRefs{\CurrentBib}

\bibitem [\protect \citeauthoryear {%
{Pomoell}%
\ \BBA {} {Poedts}%
}{%
{Pomoell}%
\ \BBA {} {Poedts}%
}{%
{\protect \APACyear {2018}}%
}]{%
pomoell2018euhforia}
\APACinsertmetastar {%
pomoell2018euhforia}%
\begin{APACrefauthors}%
{Pomoell}, J.%
\BCBT {}\ \BBA {} {Poedts}, S.%
\end{APACrefauthors}%
\unskip\
\newblock
\APACrefYearMonthDay{2018}{{\APACmonth{06}}}{}.
\newblock
{\BBOQ}\APACrefatitle {{EUHFORIA: European heliospheric forecasting information asset}} {{EUHFORIA: European heliospheric forecasting information asset}}.{\BBCQ}
\newblock
\APACjournalVolNumPages{Journal of Space Weather and Space Climate}{8}{}{A35}.
\newblock
\begin{APACrefDOI} \doi{10.1051/swsc/2018020} \end{APACrefDOI}
\PrintBackRefs{\CurrentBib}

\bibitem [\protect \citeauthoryear {%
Richardson%
\ \BBA {} Paularena%
}{%
Richardson%
\ \BBA {} Paularena%
}{%
{\protect \APACyear {1998}}%
}]{%
richardson1998}
\APACinsertmetastar {%
richardson1998}%
\begin{APACrefauthors}%
Richardson, J\BPBI D.%
\BCBT {}\ \BBA {} Paularena, K\BPBI I.%
\end{APACrefauthors}%
\unskip\
\newblock
\APACrefYearMonthDay{1998}{}{}.
\newblock
{\BBOQ}\APACrefatitle {The orientation of plasma structure in the solar wind} {The orientation of plasma structure in the solar wind}.{\BBCQ}
\newblock
\APACjournalVolNumPages{Geophysical Research Letters}{25}{12}{2097-2100}.
\newblock
\begin{APACrefURL} \url{https://agupubs.onlinelibrary.wiley.com/doi/abs/10.1029/98GL01520} \end{APACrefURL}
\newblock
\begin{APACrefDOI} \doi{https://doi.org/10.1029/98GL01520} \end{APACrefDOI}
\PrintBackRefs{\CurrentBib}

\bibitem [\protect \citeauthoryear {%
{Riley}%
\ \BBA {} {Lionello}%
}{%
{Riley}%
\ \BBA {} {Lionello}%
}{%
{\protect \APACyear {2011}}%
}]{%
riley2011mapping}
\APACinsertmetastar {%
riley2011mapping}%
\begin{APACrefauthors}%
{Riley}, P.%
\BCBT {}\ \BBA {} {Lionello}, R.%
\end{APACrefauthors}%
\unskip\
\newblock
\APACrefYearMonthDay{2011}{{\APACmonth{06}}}{}.
\newblock
{\BBOQ}\APACrefatitle {{Mapping Solar Wind Streams from the Sun to 1 AU: A Comparison of Techniques}} {{Mapping Solar Wind Streams from the Sun to 1 AU: A Comparison of Techniques}}.{\BBCQ}
\newblock
\APACjournalVolNumPages{\solphys}{270}{}{575-592}.
\newblock
\begin{APACrefDOI} \doi{10.1007/s11207-011-9766-x} \end{APACrefDOI}
\PrintBackRefs{\CurrentBib}

\bibitem [\protect \citeauthoryear {%
{Riley}%
\ \protect \BOthers {.}}{%
{Riley}%
\ \protect \BOthers {.}}{%
{\protect \APACyear {2018}}%
}]{%
riley2018forecasting}
\APACinsertmetastar {%
riley2018forecasting}%
\begin{APACrefauthors}%
{Riley}, P.%
, {Mays}, M\BPBI L.%
, {Andries}, J.%
, {Amerstorfer}, T.%
, {Biesecker}, D.%
, {Delouille}, V.%
\BDBL {}{Zhao}, X.%
\end{APACrefauthors}%
\unskip\
\newblock
\APACrefYearMonthDay{2018}{{\APACmonth{09}}}{}.
\newblock
{\BBOQ}\APACrefatitle {{Forecasting the Arrival Time of Coronal Mass Ejections: Analysis of the CCMC CME Scoreboard}} {{Forecasting the Arrival Time of Coronal Mass Ejections: Analysis of the CCMC CME Scoreboard}}.{\BBCQ}
\newblock
\APACjournalVolNumPages{Space Weather}{16}{9}{1245-1260}.
\newblock
\begin{APACrefDOI} \doi{10.1029/2018SW001962} \end{APACrefDOI}
\PrintBackRefs{\CurrentBib}

\bibitem [\protect \citeauthoryear {%
Rüdisser%
, Nguyen%
, Le~Louëdec%
, Davies%
\BCBL {}\ \BBA {} Möstl%
}{%
Rüdisser%
\ \protect \BOthers {.}}{%
{\protect \APACyear {2026}}%
}]{%
ruedisser2025ARCANE}
\APACinsertmetastar {%
ruedisser2025ARCANE}%
\begin{APACrefauthors}%
Rüdisser, H\BPBI T.%
, Nguyen, G.%
, Le~Louëdec, J.%
, Davies, E\BPBI E.%
\BCBL {}\ \BBA {} Möstl, C.%
\end{APACrefauthors}%
\unskip\
\newblock
\APACrefYearMonthDay{2026}{}{}.
\newblock
{\BBOQ}\APACrefatitle {ARCANE–Early Detection of Interplanetary Coronal Mass Ejections} {Arcane–early detection of interplanetary coronal mass ejections}.{\BBCQ}
\newblock
\APACjournalVolNumPages{Space Weather}{24}{2}{e2025SW004537}.
\newblock
\begin{APACrefURL} \url{https://agupubs.onlinelibrary.wiley.com/doi/abs/10.1029/2025SW004537} \end{APACrefURL}
\newblock
\APACrefnote{e2025SW004537 2025SW004537}
\newblock
\begin{APACrefDOI} \doi{https://doi.org/10.1029/2025SW004537} \end{APACrefDOI}
\PrintBackRefs{\CurrentBib}

\bibitem [\protect \citeauthoryear {%
{Sckopke}%
}{%
{Sckopke}%
}{%
{\protect \APACyear {1966}}%
}]{%
sckopke1966}
\APACinsertmetastar {%
sckopke1966}%
\begin{APACrefauthors}%
{Sckopke}, N.%
\end{APACrefauthors}%
\unskip\
\newblock
\APACrefYearMonthDay{1966}{{\APACmonth{07}}}{}.
\newblock
{\BBOQ}\APACrefatitle {{A general relation between the energy of trapped particles and the disturbance field near the Earth}} {{A general relation between the energy of trapped particles and the disturbance field near the Earth}}.{\BBCQ}
\newblock
\APACjournalVolNumPages{\jgr}{71}{13}{3125-3130}.
\newblock
\begin{APACrefDOI} \doi{10.1029/JZ071i013p03125} \end{APACrefDOI}
\PrintBackRefs{\CurrentBib}

\bibitem [\protect \citeauthoryear {%
{Smith}%
\ \protect \BOthers {.}}{%
{Smith}%
\ \protect \BOthers {.}}{%
{\protect \APACyear {1998}}%
}]{%
smith1998ACEmag}
\APACinsertmetastar {%
smith1998ACEmag}%
\begin{APACrefauthors}%
{Smith}, C\BPBI W.%
, {L'Heureux}, J.%
, {Ness}, N\BPBI F.%
, {Acu{\~n}a}, M\BPBI H.%
, {Burlaga}, L\BPBI F.%
\BCBL {}\ \BBA {} {Scheifele}, J.%
\end{APACrefauthors}%
\unskip\
\newblock
\APACrefYearMonthDay{1998}{{\APACmonth{07}}}{}.
\newblock
{\BBOQ}\APACrefatitle {{The ACE Magnetic Fields Experiment}} {{The ACE Magnetic Fields Experiment}}.{\BBCQ}
\newblock
\APACjournalVolNumPages{\ssr}{86}{}{613-632}.
\newblock
\begin{APACrefDOI} \doi{10.1023/A:1005092216668} \end{APACrefDOI}
\PrintBackRefs{\CurrentBib}

\bibitem [\protect \citeauthoryear {%
{Stone}%
\ \protect \BOthers {.}}{%
{Stone}%
\ \protect \BOthers {.}}{%
{\protect \APACyear {1998}}%
{\protect \APACexlab {{\protect \BCnt {1}}}}}]{%
stone1998ace}
\APACinsertmetastar {%
stone1998ace}%
\begin{APACrefauthors}%
{Stone}, E\BPBI C.%
, {Frandsen}, A\BPBI M.%
, {Mewaldt}, R\BPBI A.%
, {Christian}, E\BPBI R.%
, {Margolies}, D.%
, {Ormes}, J\BPBI F.%
\BCBL {}\ \BBA {} {Snow}, F.%
\end{APACrefauthors}%
\unskip\
\newblock
\APACrefYearMonthDay{1998{\protect \BCnt {1}}}{{\APACmonth{07}}}{}.
\newblock
{\BBOQ}\APACrefatitle {{The Advanced Composition Explorer}} {{The Advanced Composition Explorer}}.{\BBCQ}
\newblock
\APACjournalVolNumPages{\ssr}{86}{}{1-22}.
\newblock
\begin{APACrefDOI} \doi{10.1023/A:1005082526237} \end{APACrefDOI}
\PrintBackRefs{\CurrentBib}

\bibitem [\protect \citeauthoryear {%
{Stone}%
\ \protect \BOthers {.}}{%
{Stone}%
\ \protect \BOthers {.}}{%
{\protect \APACyear {1998}}%
{\protect \APACexlab {{\protect \BCnt {2}}}}}]{%
stone1998advanced}
\APACinsertmetastar {%
stone1998advanced}%
\begin{APACrefauthors}%
{Stone}, E\BPBI C.%
, {Frandsen}, A\BPBI M.%
, {Mewaldt}, R\BPBI A.%
, {Christian}, E\BPBI R.%
, {Margolies}, D.%
, {Ormes}, J\BPBI F.%
\BCBL {}\ \BBA {} {Snow}, F.%
\end{APACrefauthors}%
\unskip\
\newblock
\APACrefYearMonthDay{1998{\protect \BCnt {2}}}{{\APACmonth{07}}}{}.
\newblock
{\BBOQ}\APACrefatitle {{The Advanced Composition Explorer}} {{The Advanced Composition Explorer}}.{\BBCQ}
\newblock
\APACjournalVolNumPages{\ssr}{86}{}{1-22}.
\newblock
\begin{APACrefDOI} \doi{10.1023/A:1005082526237} \end{APACrefDOI}
\PrintBackRefs{\CurrentBib}

\bibitem [\protect \citeauthoryear {%
Sugiura%
}{%
Sugiura%
}{%
{\protect \APACyear {1964}}%
}]{%
sugiura1964}
\APACinsertmetastar {%
sugiura1964}%
\begin{APACrefauthors}%
Sugiura, M.%
\end{APACrefauthors}%
\unskip\
\newblock
\APACrefYearMonthDay{1964}{}{}.
\newblock
{\BBOQ}\APACrefatitle {Hourly values of equatorial Dst for the IGY} {Hourly values of equatorial dst for the igy}.{\BBCQ}
\newblock
\APACjournalVolNumPages{Annals of the International Geophysical Year}{}{}{}.
\PrintBackRefs{\CurrentBib}

\bibitem [\protect \citeauthoryear {%
{SunPy Community}%
\ \protect \BOthers {.}}{%
{SunPy Community}%
\ \protect \BOthers {.}}{%
{\protect \APACyear {2020}}%
}]{%
sunpy_2020}
\APACinsertmetastar {%
sunpy_2020}%
\begin{APACrefauthors}%
{SunPy Community}%
, {Barnes}, W\BPBI T.%
, {Bobra}, M\BPBI G.%
, {Christe}, S\BPBI D.%
, {Freij}, N.%
, {Hayes}, L\BPBI A.%
\BDBL {}{Dang}, T\BPBI K.%
\end{APACrefauthors}%
\unskip\
\newblock
\APACrefYearMonthDay{2020}{{\APACmonth{02}}}{}.
\newblock
{\BBOQ}\APACrefatitle {{The SunPy Project: Open Source Development and Status of the Version 1.0 Core Package}} {{The SunPy Project: Open Source Development and Status of the Version 1.0 Core Package}}.{\BBCQ}
\newblock
\APACjournalVolNumPages{\apj}{890}{1}{68}.
\newblock
\begin{APACrefURL} \url{https://zenodo.org/records/13743565} \end{APACrefURL}
\newblock
\begin{APACrefDOI} \doi{10.3847/1538-4357/ab4f7a} \end{APACrefDOI}
\PrintBackRefs{\CurrentBib}

\bibitem [\protect \citeauthoryear {%
Temerin%
\ \BBA {} Li%
}{%
Temerin%
\ \BBA {} Li%
}{%
{\protect \APACyear {2002}}%
}]{%
temerin2002model}
\APACinsertmetastar {%
temerin2002model}%
\begin{APACrefauthors}%
Temerin, M.%
\BCBT {}\ \BBA {} Li, X.%
\end{APACrefauthors}%
\unskip\
\newblock
\APACrefYearMonthDay{2002}{}{}.
\newblock
{\BBOQ}\APACrefatitle {A new model for the prediction of Dst on the basis of the solar wind} {A new model for the prediction of dst on the basis of the solar wind}.{\BBCQ}
\newblock
\APACjournalVolNumPages{Journal of Geophysical Research: Space Physics}{107}{A12}{SMP 31-1-SMP 31-8}.
\newblock
\begin{APACrefURL} \url{https://agupubs.onlinelibrary.wiley.com/doi/abs/10.1029/2001JA007532} \end{APACrefURL}
\newblock
\begin{APACrefDOI} \doi{https://doi.org/10.1029/2001JA007532} \end{APACrefDOI}
\PrintBackRefs{\CurrentBib}

\bibitem [\protect \citeauthoryear {%
{Temerin}%
\ \BBA {} {Li}%
}{%
{Temerin}%
\ \BBA {} {Li}%
}{%
{\protect \APACyear {2006}}%
}]{%
Temerin2006}
\APACinsertmetastar {%
Temerin2006}%
\begin{APACrefauthors}%
{Temerin}, M.%
\BCBT {}\ \BBA {} {Li}, X.%
\end{APACrefauthors}%
\unskip\
\newblock
\APACrefYearMonthDay{2006}{{\APACmonth{04}}}{}.
\newblock
{\BBOQ}\APACrefatitle {{Dst model for 1995-2002}} {{Dst model for 1995-2002}}.{\BBCQ}
\newblock
\APACjournalVolNumPages{\jgr (Space Physics)}{111}{A4}{A04221}.
\newblock
\begin{APACrefDOI} \doi{10.1029/2005JA011257} \end{APACrefDOI}
\PrintBackRefs{\CurrentBib}

\bibitem [\protect \citeauthoryear {%
{Verbeke}%
, {Baratashvili}%
\BCBL {}\ \BBA {} {Poedts}%
}{%
{Verbeke}%
\ \protect \BOthers {.}}{%
{\protect \APACyear {2022}}%
}]{%
verbeke2022}
\APACinsertmetastar {%
verbeke2022}%
\begin{APACrefauthors}%
{Verbeke}, C.%
, {Baratashvili}, T.%
\BCBL {}\ \BBA {} {Poedts}, S.%
\end{APACrefauthors}%
\unskip\
\newblock
\APACrefYearMonthDay{2022}{{\APACmonth{06}}}{}.
\newblock
{\BBOQ}\APACrefatitle {{ICARUS, a new inner heliospheric model with a flexible grid}} {{ICARUS, a new inner heliospheric model with a flexible grid}}.{\BBCQ}
\newblock
\APACjournalVolNumPages{\aap}{662}{}{A50}.
\newblock
\begin{APACrefDOI} \doi{10.1051/0004-6361/202141981} \end{APACrefDOI}
\PrintBackRefs{\CurrentBib}

\bibitem [\protect \citeauthoryear {%
Virtanen%
\ \protect \BOthers {.}}{%
Virtanen%
\ \protect \BOthers {.}}{%
{\protect \APACyear {2020}}%
}]{%
scipy2020}
\APACinsertmetastar {%
scipy2020}%
\begin{APACrefauthors}%
Virtanen, P.%
, Gommers, R.%
, Oliphant, T\BPBI E.%
, Haberland, M.%
, Reddy, T.%
, Cournapeau, D.%
\BDBL {}{SciPy 1.0 Contributors}%
\end{APACrefauthors}%
\unskip\
\newblock
\APACrefYearMonthDay{2020}{}{}.
\newblock
{\BBOQ}\APACrefatitle {{{SciPy} 1.0: Fundamental Algorithms for Scientific Computing in Python}} {{{SciPy} 1.0: Fundamental Algorithms for Scientific Computing in Python}}.{\BBCQ}
\newblock
\APACjournalVolNumPages{Nature Methods}{17}{}{261--272}.
\newblock
\begin{APACrefDOI} \doi{10.1038/s41592-019-0686-2} \end{APACrefDOI}
\PrintBackRefs{\CurrentBib}

\bibitem [\protect \citeauthoryear {%
{Vourlidas}%
, {Patsourakos}%
\BCBL {}\ \BBA {} {Savani}%
}{%
{Vourlidas}%
\ \protect \BOthers {.}}{%
{\protect \APACyear {2019}}%
}]{%
vourlidas2019review}
\APACinsertmetastar {%
vourlidas2019review}%
\begin{APACrefauthors}%
{Vourlidas}, A.%
, {Patsourakos}, S.%
\BCBL {}\ \BBA {} {Savani}, N\BPBI P.%
\end{APACrefauthors}%
\unskip\
\newblock
\APACrefYearMonthDay{2019}{{\APACmonth{07}}}{}.
\newblock
{\BBOQ}\APACrefatitle {{Predicting the geoeffective properties of coronal mass ejections: current status, open issues and path forward}} {{Predicting the geoeffective properties of coronal mass ejections: current status, open issues and path forward}}.{\BBCQ}
\newblock
\APACjournalVolNumPages{Philosophical Transactions of the Royal Society of London Series A}{377}{2148}{20180096}.
\newblock
\begin{APACrefDOI} \doi{10.1098/rsta.2018.0096} \end{APACrefDOI}
\PrintBackRefs{\CurrentBib}

\bibitem [\protect \citeauthoryear {%
{Wang}%
, {Ye}%
\BCBL {}\ \BBA {} {Wang}%
}{%
{Wang}%
\ \protect \BOthers {.}}{%
{\protect \APACyear {2003}}%
}]{%
wang2003}
\APACinsertmetastar {%
wang2003}%
\begin{APACrefauthors}%
{Wang}, Y\BPBI M.%
, {Ye}, P\BPBI Z.%
\BCBL {}\ \BBA {} {Wang}, S.%
\end{APACrefauthors}%
\unskip\
\newblock
\APACrefYearMonthDay{2003}{{\APACmonth{10}}}{}.
\newblock
{\BBOQ}\APACrefatitle {{Multiple magnetic clouds: Several examples during March-April 2001}} {{Multiple magnetic clouds: Several examples during March-April 2001}}.{\BBCQ}
\newblock
\APACjournalVolNumPages{Journal of Geophysical Research (Space Physics)}{108}{A10}{1370}.
\newblock
\begin{APACrefDOI} \doi{10.1029/2003JA009850} \end{APACrefDOI}
\PrintBackRefs{\CurrentBib}

\bibitem [\protect \citeauthoryear {%
Wang%
, Ye%
, Wang%
, Zhou%
\BCBL {}\ \BBA {} Wang%
}{%
Wang%
\ \protect \BOthers {.}}{%
{\protect \APACyear {2002}}%
}]{%
wang2002}
\APACinsertmetastar {%
wang2002}%
\begin{APACrefauthors}%
Wang, Y\BPBI M.%
, Ye, P\BPBI Z.%
, Wang, S.%
, Zhou, G\BPBI P.%
\BCBL {}\ \BBA {} Wang, J\BPBI X.%
\end{APACrefauthors}%
\unskip\
\newblock
\APACrefYearMonthDay{2002}{}{}.
\newblock
{\BBOQ}\APACrefatitle {A statistical study on the geoeffectiveness of Earth-directed coronal mass ejections from March 1997 to December 2000} {A statistical study on the geoeffectiveness of earth-directed coronal mass ejections from march 1997 to december 2000}.{\BBCQ}
\newblock
\APACjournalVolNumPages{Journal of Geophysical Research: Space Physics}{107}{A11}{SSH 2-1-SSH 2-9}.
\newblock
\begin{APACrefURL} \url{https://agupubs.onlinelibrary.wiley.com/doi/abs/10.1029/2002JA009244} \end{APACrefURL}
\newblock
\begin{APACrefDOI} \doi{https://doi.org/10.1029/2002JA009244} \end{APACrefDOI}
\PrintBackRefs{\CurrentBib}

\bibitem [\protect \citeauthoryear {%
Wanliss%
\ \BBA {} Showalter%
}{%
Wanliss%
\ \BBA {} Showalter%
}{%
{\protect \APACyear {2006}}%
}]{%
wanliss2006}
\APACinsertmetastar {%
wanliss2006}%
\begin{APACrefauthors}%
Wanliss, J\BPBI A.%
\BCBT {}\ \BBA {} Showalter, K\BPBI M.%
\end{APACrefauthors}%
\unskip\
\newblock
\APACrefYearMonthDay{2006}{}{}.
\newblock
{\BBOQ}\APACrefatitle {High-resolution global storm index: Dst versus SYM-H} {High-resolution global storm index: Dst versus sym-h}.{\BBCQ}
\newblock
\APACjournalVolNumPages{Journal of Geophysical Research: Space Physics}{111}{A2}{}.
\newblock
\begin{APACrefURL} \url{https://agupubs.onlinelibrary.wiley.com/doi/abs/10.1029/2005JA011034} \end{APACrefURL}
\newblock
\begin{APACrefDOI} \doi{https://doi.org/10.1029/2005JA011034} \end{APACrefDOI}
\PrintBackRefs{\CurrentBib}

\bibitem [\protect \citeauthoryear {%
Weiler%
, Davies%
\BCBL {}\ \BBA {} M{\"o}stl%
}{%
Weiler%
\ \protect \BOthers {.}}{%
{\protect \APACyear {2026}}%
}]{%
Weiler2026_figshare}
\APACinsertmetastar {%
Weiler2026_figshare}%
\begin{APACrefauthors}%
Weiler, E.%
, Davies, E.%
\BCBL {}\ \BBA {} M{\"o}stl, C.%
\end{APACrefauthors}%
\unskip\
\newblock
\APACrefYearMonthDay{2026}{}{}.
\newblock
\APACrefbtitle {{Statistical Sub-L1 Analysis}.} {{Statistical Sub-L1 Analysis}.}
\newblock
\APACaddressPublisher{}{figshare}.
\newblock
\begin{APACrefURL} \url{https://doi.org/10.6084/m9.figshare.31410306.v2} \end{APACrefURL}
\newblock
\begin{APACrefDOI} \doi{10.6084/m9.figshare.31410306.v2} \end{APACrefDOI}
\PrintBackRefs{\CurrentBib}

\bibitem [\protect \citeauthoryear {%
Weiler%
\ \protect \BOthers {.}}{%
Weiler%
\ \protect \BOthers {.}}{%
{\protect \APACyear {2025}}%
}]{%
weiler2025}
\APACinsertmetastar {%
weiler2025}%
\begin{APACrefauthors}%
Weiler, E.%
, Möstl, C.%
, Davies, E\BPBI E.%
, Veronig, A\BPBI M.%
, Amerstorfer, U\BPBI V.%
, Amerstorfer, T.%
\BDBL {}Reiss, M.%
\end{APACrefauthors}%
\unskip\
\newblock
\APACrefYearMonthDay{2025}{}{}.
\newblock
{\BBOQ}\APACrefatitle {First Observations of a Geomagnetic Superstorm With a Sub-L1 Monitor} {First observations of a geomagnetic superstorm with a sub-l1 monitor}.{\BBCQ}
\newblock
\APACjournalVolNumPages{Space Weather}{23}{3}{e2024SW004260}.
\newblock
\begin{APACrefURL} \url{https://agupubs.onlinelibrary.wiley.com/doi/abs/10.1029/2024SW004260} \end{APACrefURL}
\newblock
\APACrefnote{e2024SW004260 2024SW004260}
\newblock
\begin{APACrefDOI} \doi{https://doi.org/10.1029/2024SW004260} \end{APACrefDOI}
\PrintBackRefs{\CurrentBib}

\bibitem [\protect \citeauthoryear {%
Zhang%
, Dere%
, Howard%
\BCBL {}\ \BBA {} Bothmer%
}{%
Zhang%
\ \protect \BOthers {.}}{%
{\protect \APACyear {2003}}%
}]{%
zhang2003}
\APACinsertmetastar {%
zhang2003}%
\begin{APACrefauthors}%
Zhang, J.%
, Dere, K\BPBI P.%
, Howard, R\BPBI A.%
\BCBL {}\ \BBA {} Bothmer, V.%
\end{APACrefauthors}%
\unskip\
\newblock
\APACrefYearMonthDay{2003}{jan}{}.
\newblock
{\BBOQ}\APACrefatitle {Identification of Solar Sources of Major Geomagnetic Storms between 1996 and 2000} {Identification of solar sources of major geomagnetic storms between 1996 and 2000}.{\BBCQ}
\newblock
\APACjournalVolNumPages{The Astrophysical Journal}{582}{1}{520}.
\newblock
\begin{APACrefURL} \url{https://doi.org/10.1086/344611} \end{APACrefURL}
\newblock
\begin{APACrefDOI} \doi{10.1086/344611} \end{APACrefDOI}
\PrintBackRefs{\CurrentBib}

\bibitem [\protect \citeauthoryear {%
{Zhang}%
\ \protect \BOthers {.}}{%
{Zhang}%
\ \protect \BOthers {.}}{%
{\protect \APACyear {2007}}%
}]{%
zhang2007}
\APACinsertmetastar {%
zhang2007}%
\begin{APACrefauthors}%
{Zhang}, J.%
, {Richardson}, I\BPBI G.%
, {Webb}, D\BPBI F.%
, {Gopalswamy}, N.%
, {Huttunen}, E.%
, {Kasper}, J\BPBI C.%
\BDBL {}{Zhukov}, A\BPBI N.%
\end{APACrefauthors}%
\unskip\
\newblock
\APACrefYearMonthDay{2007}{{\APACmonth{10}}}{}.
\newblock
{\BBOQ}\APACrefatitle {{Solar and interplanetary sources of major geomagnetic storms (Dst <= -100 nT) during 1996-2005}} {{Solar and interplanetary sources of major geomagnetic storms (Dst <= -100 nT) during 1996-2005}}.{\BBCQ}
\newblock
\APACjournalVolNumPages{Journal of Geophysical Research (Space Physics)}{112}{A10}{A10102}.
\newblock
\begin{APACrefDOI} \doi{10.1029/2007JA012321} \end{APACrefDOI}
\PrintBackRefs{\CurrentBib}

\bibitem [\protect \citeauthoryear {%
{Zurbuchen}%
\ \BBA {} {Richardson}%
}{%
{Zurbuchen}%
\ \BBA {} {Richardson}%
}{%
{\protect \APACyear {2006}}%
}]{%
zurbuchen2006situ}
\APACinsertmetastar {%
zurbuchen2006situ}%
\begin{APACrefauthors}%
{Zurbuchen}, T\BPBI H.%
\BCBT {}\ \BBA {} {Richardson}, I\BPBI G.%
\end{APACrefauthors}%
\unskip\
\newblock
\APACrefYearMonthDay{2006}{{\APACmonth{03}}}{}.
\newblock
{\BBOQ}\APACrefatitle {{In-Situ Solar Wind and Magnetic Field Signatures of Interplanetary Coronal Mass Ejections}} {{In-Situ Solar Wind and Magnetic Field Signatures of Interplanetary Coronal Mass Ejections}}.{\BBCQ}
\newblock
\APACjournalVolNumPages{\ssr}{123}{1-3}{31-43}.
\newblock
\begin{APACrefDOI} \doi{10.1007/s11214-006-9010-4} \end{APACrefDOI}
\PrintBackRefs{\CurrentBib}

\bibitem [\protect \citeauthoryear {%
Zwickl%
\ \protect \BOthers {.}}{%
Zwickl%
\ \protect \BOthers {.}}{%
{\protect \APACyear {1998}}%
}]{%
zwickl1998noaartsw}
\APACinsertmetastar {%
zwickl1998noaartsw}%
\begin{APACrefauthors}%
Zwickl, R\BPBI D.%
, Doggett, K\BPBI A.%
, Sahm, S.%
, Barrett, W\BPBI P.%
, Grubb, R\BPBI N.%
, Detman, T\BPBI R.%
\BDBL {}Maruyama, T.%
\end{APACrefauthors}%
\unskip\
\newblock
\APACrefYearMonthDay{1998}{{\APACmonth{07}}}{}.
\newblock
{\BBOQ}\APACrefatitle {The {{NOAA Real-Time Solar-Wind}} ({{RTSW}}) {{System}} Using {{ACE Data}}} {The {{NOAA Real-Time Solar-Wind}} ({{RTSW}}) {{System}} using {{ACE Data}}}.{\BBCQ}
\newblock
\APACjournalVolNumPages{Space Science Reviews}{86}{}{633--648}.
\newblock
\begin{APACrefDOI} \doi{10.1023/A:1005044300738} \end{APACrefDOI}
\PrintBackRefs{\CurrentBib}

\end{thebibliography}


\end{document}